\documentclass[pra,twocolumn]{revtex4}
\usepackage{graphicx}
\usepackage{bm}
\usepackage{bbm}
\usepackage{amsmath}

\renewcommand{\Re}{\operatorname{Re}}
\renewcommand{\Im}{\operatorname{Im}}
\newcommand{\cc}{\text{c.c.}}
\newcommand{\Hc}{\text{H.c.}}

\DeclareMathOperator{\tr}{tr}

\newcommand{\StateJ}{\Xi}
\newcommand{\StateF}{\Upsilon}
\newcommand{\StateSingle}{\chi}

\begin{document}

\title{Dark-time decay of the retrieval efficiency of light stored as a Rydberg excitation in a noninteracting ultracold gas}

\author{Steffen Schmidt-Eberle}
\author{Thomas Stolz}
\author{Gerhard Rempe}
\author{Stephan D\"urr}
\affiliation{Max-Planck-Institut f\"{u}r Quantenoptik, Hans-Kopfermann-Str.\ 1, 85748 Garching, Germany}

\begin{abstract}
We study the dark-time decay of the retrieval efficiency for light stored in a Rydberg state in an ultracold gas of $^{87}$Rb atoms based on electromagnetically induced transparency (EIT). Using low atomic density to avoid dephasing caused by atom-atom interactions, we measure a $1/e$ time of 30 $\mu$s for the $80S$ state in free expansion. One of the dominant limitations is the combination of photon recoil and thermal atomic motion at 0.2 $\mu$K. If the 1064-nm dipole trap is left on, then the $1/e$ time is reduced to 13 $\mu$s, in agreement with a model taking differential light shifts and gravitational sag into account. To characterize how coherent the retrieved light is, we overlap it with reference light and measure the visibility $V$ of the resulting interference pattern, obtaining $V> 90\%$ for short dark time. Our experimental work is accompanied by a detailed model for the dark-time decay of the retrieval efficiency of light stored in atomic ensembles. The model is generally applicable for photon storage in Dicke states, such as in EIT with $\Lambda$-type or ladder-type level schemes and in Duan-Lukin-Cirac-Zoller single-photon sources. The model includes a treatment of the dephasing caused by thermal atomic motion combined with net photon recoil, as well as the influence of trapping potentials. It takes into account that the signal light field is typically not a plane wave. The model maps the retrieval efficiency to single-atom properties and shows that the retrieval efficiency is related to the decay of fringe visibility in Ramsey spectroscopy and to the spatial first-order coherence function of the gas.
\end{abstract}

\maketitle

\section{Introduction}

Decay of coherence is a major performance-limiting factor in photonic quantum memories \cite{Lvovsky:09}. Among the various schemes for such memories, the focus of the present work is on storage and retrieval of light \cite{Fleischhauer:00, Fleischhauer:02, Liu:01, Phillips:01} in an ultracold atomic ensemble based on electromagnetically induced transparency (EIT) \cite{Fleischhauer:05}. The storage produces a spin wave in the atomic ensemble. Various physical effects cause the coherence of the spin wave to decay as a function of the dark time between storage and retrieval. As a result, the efficiency of the retrieval decays. Over the years, various techniques have been used to extend the decay time. For ground-state EIT in ultracold atomic gases, a $1/e$ time of 16 s was reached a few years ago \cite{Dudin:13}.

Aiming at the creation of strong interactions between photons, EIT has been combined with Rydberg states \cite{Friedler:05, Pritchard:10, Firstenberg:16}. Coherence in Rydberg EIT suffers from the fact that the sensitivity of a Rydberg atom to interactions with surrounding ground-state atoms increases with atomic density and with increasing principal quantum number \cite{Balewski:13, Baur:14, Baur:phd, Mirgorodskiy:17}. In addition, when the atoms are held in a red-detuned optical dipole trap, the differential light shifts between the ground and Rydberg state are large. To avoid light shifts, many experiments are carried out in free expansion. In recent experiments on Rydberg EIT, the $1/e$ decay time has been pushed to 12 $\mu$s for the $45S$ state in free expansion \cite{Mirgorodskiy:17} and to approximately 20 $\mu$s for the $65S$ state in a magic-wavelength optical lattice specifically designed to fight decoherence \cite{Lampen:18}.

We recently demonstrated a photon-photon quantum gate based on Rydberg interactions \cite{Tiarks:19}. The gate required a fairly high atomic density of $2\times10^{12}$ cm$^{-3}$ because light had to accumulate a $\pi$ phase shift in a single pass through the blockade volume. Correspondingly, atom-atom interactions caused a decay of coherence on a time scale of a few microseconds, which was the key limiting factor for the overall performance of the gate. Much better performance of a photon-photon gate should be achievable using Rydberg blockade in an atomic ensemble inside an optical resonator of moderate finesse (Rydberg cavity gate) \cite{Hao:15, Das:16, Wade:16} because, among other things, light passing many times through a blockade volume can accumulate a large effect even at low atomic density, where decoherence should be much slower.

Here, we present an experiment in which we extend the $1/e$ decay time of Rydberg EIT retrieval from the $80S$ state in $^{87}$Rb to 30 $\mu$s in free expansion. This is achieved by operating at a low peak atomic density of $5\times10^{10}$ cm$^{-3}$ to make atom-atom interactions negligible and at a low temperature of 0.2 $\mu$K to reduce the effect of net photon recoil. We combine this with an experimental study of the dependence of the decay time on temperature. In addition, we overlap the retrieved light with a local oscillator and measure a visibility above 90\%. This is a considerable improvement over the 66\% visibility which we previously measured at a peak density of $2\times10^{12}$ cm$^{-3}$ \cite{Tiarks:19}. If the dipole trap with a wavelength of 1064 nm is left on during the experiment, the $1/e$ decay time is reduced to 13 $\mu$s.

Our experimental work is accompanied by a theoretical analysis of the decay of the EIT retrieval efficiency for light stored in a gas of noninteracting atoms. The model is applicable to EIT-based storage in $\Lambda$-type or ladder-type level schemes. The retrieval efficiency can be calculated from single-particle properties. We show that the resulting expressions are analogous to expressions for the decay of the fringe visibility in Ramsey spectroscopy as a function of dark time. We also note that the EIT retrieval decay is equivalent to the decay of the read efficiency in single-photon sources based on the inherently probabilistic Duan-Lukin-Cirac-Zoller (DLCZ) protocol \cite{Duan:01}. In addition, we show that when applying the model to photon recoil during storage combined with thermal atomic motion, the decay of retrieval efficiency can be regarded as a direct experimental probe of the spatial first-order coherence function of the gas.

We apply this model to study several scenarios, first, photon recoil during storage combined with thermal atomic motion \cite{Ginsberg:07, Zhao:Pan:08, Jenkins:12, Baur:phd}, second, a harmonic differential light-shift potential \cite{Kuhr:05, Yang:11, Jenkins:12, Afek:17, Lampen:18}, and, third, release from a harmonic trap \cite{Jenkins:12}. According to the model, photon recoil combined with thermal motion is one of the dominant limitations for the observed $1/e$ time of 30 $\mu$s. This model also agrees well with the temperature dependence of the decay time measured here. The observed reduction of the $1/e$ time to 13 $\mu$s in the dipole trap is explained by the model when taking into account the differential light shift and the gravitational sag of the atomic cloud in the trapping potential. The model predicts that producing the same harmonic trapping potential with 532-nm light should make the light-shift contribution to the decay irrelevant compared to the presently observed 30 $\mu$s.

This paper is organized as follows. In Sec.\ \ref{sec-exper} we present experimental results and compare with results from a model. This model is introduced in Sec.\ \ref{sec-model} and applied to several experimentally relevant situations in Sec.\ \ref{sec-applications}.

\section{Experiment}

\label{sec-exper}

This section presents experimental results of the decay of the retrieval efficiency during the dark time. It begins with a description of the experimental setup and procedure in Sec.\ \ref{sec-exper-procedure}, which is followed by Sec.\ \ref{sec-exper-recoil} presenting an experimental study of the decay of the retrieval efficiency in free expansion. The results agree largely with results from a model taking into account photon recoil combined with thermal atomic motion. In Sec.\ \ref{sec-exper-trap} we observe that the decay becomes faster if the optical dipole trap is left on during the dark time. In Sec.\ \ref{sec-exper-visibility} we investigate to which degree the retrieved light is coherent.

\subsection{Experimental Procedure}

\label{sec-exper-procedure}

Our experiment is based on a double magneto-optical trap (MOT) system. $^{87}$Rb atoms are collected in a first MOT from a background vapor and then transferred to a second MOT at much lower background pressure. After polarization-gradient cooling and optical pumping, the atoms are transferred into a Ioffe-Pritchard magnetic trap. Radio-frequency (rf) induced evaporative cooling in the magnetic trap allows it to reach sub-microkelvin temperatures. The atoms are finally transferred into an optical dipole trap. In the past, we produced BECs in this apparatus \cite{ernst:98}. In recent years, however, we have used the apparatus  to study Rydberg EIT, for which the high atomic density in a BEC of typically $10^{14}$ cm$^{-3}$ is disadvantageous because atom-atom interactions would produce fast dephasing, see e.g.\ Ref.\ \cite{Baur:14}. To avoid this, we operate at much lower atomic densities. Technically, we achieve this by deliberately loading much fewer atoms than possible. As a result, the typical temperatures of a few hundred nanokelvin are far above quantum degeneracy.

\begin{figure}[!t]
\includegraphics[scale=1]{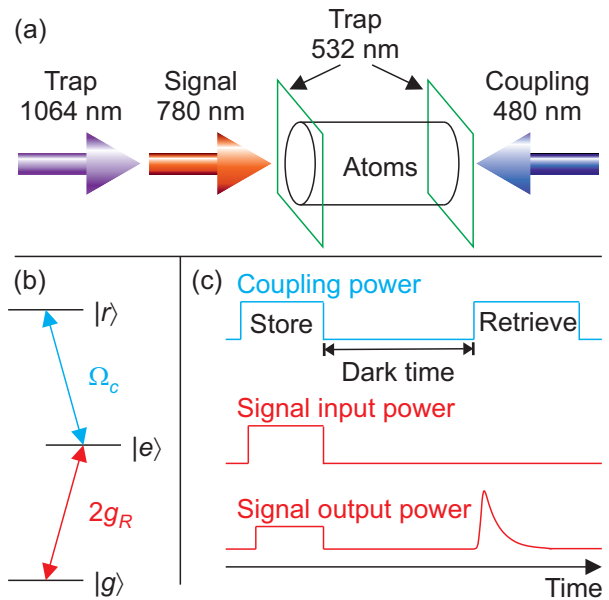}
\caption{(a) Scheme of the beam geometry. Trapping light at 1064 nm creates a radially confining optical dipole trap for the atomic ensemble. 532 nm light sheets provide a box-like longitudinal optical dipole potential. 780 nm signal light copropagates with the 1064 nm trapping light. 480 nm EIT coupling light counterpropagates the signal light to minimize the net photon recoil in the two-photon transition $|g\rangle \leftrightarrow |r\rangle$. (b) Scheme of the atomic levels and transitions. A weak signal light field with vacuum Rabi frequency $2g_R$ drives the transition $|g\rangle \leftrightarrow |e\rangle$. Strong EIT coupling light with Rabi frequency $\Omega_c$ drives the transition $|e\rangle \leftrightarrow |r\rangle$. (c) Scheme of the timing sequence. The storage pulse consists of incoming signal and coupling light. Storage of the signal light is achieved by switching off the coupling light. Because of imperfections, some signal light leaks through the atomic ensemble, thus appearing at the output with some group delay. The storage is followed by a dark time, after which the coupling light is switched back on. This causes the stored signal light to be retrieved.}
\label{fig-schemes}
\end{figure}

The dipole trap is made of a horizontally propagating light beam with a wavelength of 1064 nm and a beam waist ($1/e^2$ radius of intensity) of 140 $\mu$m, see Fig.\ \ref{fig-schemes}(a). At a power of 3.7 W, the trap depth is estimated to be $k_B\times 18$ $\mu$K, where we used the dynamical polarizability of the ground state at 1064 nm of 687.3 a.u.\ \cite{Arora:12}. Here, $k_B$ is the Boltzmann constant and one atomic unit is $1.649\times 10^{-41}$ J/(V/m)$^2$. This beam provides a radial confinement with a harmonic trapping frequency of $\omega/2\pi= 96$ Hz estimated from the beam waist and power. This agrees fairly well with the measured value of 87(8) Hz. The axial confinement resulting from the divergence of the 1064 nm beam is estimated to be below 0.1 Hz, which is negligible. The gravitational acceleration of $g=9.8$ m/s$^2$ causes a gravitational sag of $x_{g,s}= g/\omega^2= 27$ $\mu$m. In the radial direction the one-dimensional (1d) root-mean-square (rms) radius is $\sigma_x= \sqrt{k_BT/m\omega^2}$, where $m$ is the atomic mass, yielding e.g.\ $\sigma_x= 7$ $\mu$m for $T= 0.2$ $\mu$K.

In the axial direction, a box-like potential is produced using two light sheets at a wavelength of 532 nm with beam waists of 55 $\mu$m along gravity and 14 $\mu$m orthogonal thereto, similar to Ref.\ \cite{Tiarks:19}. Adjusting the power of the 532 nm light sheets allows for evaporative cooling in the dipole trap without changing the radial confinement provided by the 1064 nm light. The resulting temperature depends on the power of the light sheets. The separation of the centers of the light sheets is typically 0.43 mm. Combining this number with the temperature and with the dynamical polarizability of $-250$ a.u.\ \cite{Saffman:05} for the ground state at 532 nm, we can estimate the length of the medium $L$ (full width at half maximum).

For example, for $T= 0.2$ $\mu$K and a typical power of 25 mW for each light sheet we estimate $L= 0.40$ mm. Combining this with an atom number of e.g.\ $N= 1\times10^4$ yields a peak atomic density of $\varrho_0= 8\times 10^{10}$ cm$^{-3}$ and a peak phase space density of $\varrho_0\lambda_{dB}^3= 6\times10^{-3}$, where $\lambda_{dB}= \hbar \sqrt{2\pi/mk_BT}$ is the thermal de Broglie wavelength. As $\varrho_0\lambda_{dB}^3\ll 1$, the temperature is far above quantum degeneracy.

The atomic sample is prepared in the stretched spin state $|g\rangle= |5S_{1/2},F{=}m_F{=}2\rangle$ of the atomic ground state, where $F,m_F$ are the hyperfine quantum numbers. The quantization axis is chosen along the wave vector of the 1064-nm light beam. A magnetic field of 24 $\mu$T applied along the quantization axis preserves the spin preparation of the sample.

The sample can be illuminated with an EIT signal-light beam with a beam waist of $w=8$ $\mu$m and a wavelength of $\lambda_{eg}= 780.24$ nm, resonantly driving the $|g\rangle \leftrightarrow |e\rangle$ transition, where $|e\rangle= |5P_{3/2},F{=}m_F{=}3\rangle$. A scheme of the atomic levels and transitions is shown Fig.\ \ref{fig-schemes}(b). The EIT signal-light beam copropagates with the 1064 nm dipole trapping beam. In addition, the sample can be illuminated with an EIT coupling-light beam with a beam waist of 29 $\mu$m and a wavelength of $\lambda_{re}= 480$ nm, resonantly driving the $|e\rangle \leftrightarrow |r\rangle$ transition, where $|r\rangle= |nS_{1/2},F{=}m_F{=}2\rangle$ is a Rydberg state with principal quantum number $n$. The EIT coupling-light beam counterpropagates the EIT signal-light beam to minimize the net photon recoil of the two-photon transition from $|g\rangle$ to $|r\rangle$. Note that the states $|g\rangle$ and $|r\rangle$ experience, to a good approximation, the same linear Zeeman effect.

We use the following timing sequence, schematically shown in Fig.\ \ref{fig-schemes}(c) to achieve EIT-based storage and retrieval of the signal light. First, the EIT coupling light is turned on. Next, a pulse of EIT signal light is sent onto the sample. The incoming EIT signal pulse has a rectangular temporal shape. Unless otherwise noted, its duration is 0.5 $\mu$s. Because of EIT, the signal light becomes a Rydberg polariton when inside the sample. The pulse experiences a much-reduced group velocity, which causes a drastic spatial compression of the pulse in the longitudinal direction upon entering the sample \cite{Fleischhauer:05}. When the signal pulse is inside the medium, the coupling light is switched off. Hence, the signal light is converted into a stationary Rydberg excitation \cite{Fleischhauer:00, Fleischhauer:02}. After a variable dark time $t$, the coupling light is switched back on. This couples the population in state $|r\rangle$ to the state $|e\rangle$ from where spontaneous emission into state $|g\rangle$ can occur. There is interference between the light emitted from the large number of atoms in the ensemble. Ideally, the interference is such that the signal light pulse resumes propagation with an unchanged form of the longitudinal and transverse wave packet. This retrieval can, to a good approximation, be regarded as the time-reversed process of the storage.

In practice, various physical effects can cause deviations from this ideal retrieval scenario. While a possible change in the longitudinal wave packet could, in principle, be compensated by shaping the temporal profile of the coupling light pulse during retrieval, a change in the transverse profile is typically hard to compensate. Hence, the fraction $\eta$ of the incoming light which is emitted into the original transverse mode is an important figure of merit. $\eta$ is the combined efficiency of the storage-and-retrieval process. For brevity, we refer to $\eta$ as the retrieval efficiency throughout this work. To measure $\eta$, we focus the light emitted from the atomic sample into a single-mode optical fiber and measure the light intensity behind the fiber. In the absence of the atomic ensemble, we achieve a fiber-coupling efficiency of 45\% for coupling the signal light, impinging at the position where the atoms would usually be, into the optical fiber. In our present work, we are not interested in the retrieval efficiency $\eta$ at short dark time. Instead, we focus on the decay of the retrieval efficiency $\eta$ as a function of the dark time $t$ between storage and retrieval.

The incoming EIT signal pulse is derived from a continuous-wave laser using an acousto-optical modulator for pulse shaping. To a good approximation, it can be modeled as a coherent state with a Poissonian photon number distribution. The average number of photons is approximately one. Hence, the probability of having more than one incoming photon is not negligible. Nonetheless, the probability that two stored Rydberg excitations interact with each other during the dark time is negligible. This is partly because the storage efficiency is fairly low, typically between 10 and 20\%, and partly because the spatial compression of the EIT signal pulse inside the medium is moderate, giving it a length of several hundreds of micrometers, which is large compared to the radius over which the van der Waals interaction between two stationary Rydberg excitations is noticeable.

To avoid decay of $\eta$ resulting from the differential light shifts created by the dipole trapping potential, we switch the dipole trap off before sending the EIT signal light pulse into the sample. Hence, each storage-and-retrieval experiment takes place during free expansion. As the preparation of the atomic sample is time consuming, we perform many repetitions of the experiment on the same atomic sample. To do so, we recapture the atomic ensemble by switching the dipole trap back on 35 $\mu$s after switching it off. This is well after the retrieval is over.

The number of repetitions which we can perform on one sample is limited by heating caused, firstly, by spontaneous emission of 780 nm photons which were absorbed because of imperfect EIT and, secondly and most importantly, by periodically switching the dipole trapping light on and off. We choose to perform 1000 repetitions of the experiment for each atomic sample. During the course of these 1000 repetitions for one atomic sample, we typically observe a 20\% increase in temperature. Because of spontaneous evaporation over the 532-nm light sheet barriers, this is accompanied by a 20\% decrease in atom number. We separate these repetitions by 1 ms from one another. On one hand, this gives possibly left-behind Rydberg excitations time to decay spontaneously to the ground state. On the other hand, this means that the 35 $\mu$s during which the trap is off have a negligible time-averaged effect. So, the 1000 repetitions take a total time of 1 s, after which we prepare a new atomic sample, which takes between 13 and 19 s depending on the choice of the final temperature. Clearly, it would be desirable to avoid the periodic release and recapture because without it, the heating would be much reduced, allowing it to perform typically ten times as many repetitions of the experiment before needing to prepare a new atomic sample. This would increase the time-averaged data acquisition rate by an order of magnitude when choosing the separation between repetitions to 100 $\mu$s as in Ref.\ \cite{Tiarks:19}.

Compared to our previous experiments \cite{Tiarks:19}, we operate at lower atomic density which increases the group velocity for the EIT signal light pulse. Hence, a lower EIT coupling Rabi frequency would be needed to achieve a similar group velocity. This is partly compensated by the longer medium for which a somewhat higher group velocity is needed for optimal storage efficiency. We choose the power of the EIT coupling beam to be between 5 and 25 mW depending on atomic density and principal quantum number. The EIT coupling Rabi frequency ranges between 3 and 12 MHz. The repulsive potential which the 480 nm EIT coupling light causes for the ground-state atoms has negligible effect because this light is on typically only for 0.3\% of the time, similar to Refs.\ \cite{Baur:14, Tiarks:19}.

\subsection{Decay Caused by Photon Recoil Combined with Thermal Motion}

\label{sec-exper-recoil}

An important mechanism which causes decay of the retrieval efficiency is thermal atomic motion combined with the net photon recoil $\hbar\bm k_R$ transferred during storage on the two-photon transition from $|g\rangle$ to $|r\rangle$. According to Eq.\ \eqref{tau-R}, this mechanism is expected to cause a decay of the retrieval efficiency governed by
\begin{align}
\label{experiment-tau-R}
\frac{\eta(t)}{\eta(0)}
= e^{-t^2/\tau_R^2}
,&&
\tau_R
= \frac{1}{k_R} \sqrt{\frac{m}{k_BT}}
.\end{align}
Hence, $\eta(t)$ is expected to exhibit a Gaussian decay as a function of dark time $t$ with $1/e$ time $\tau_R$.

\begin{figure}[!t]
\includegraphics[width=\columnwidth]{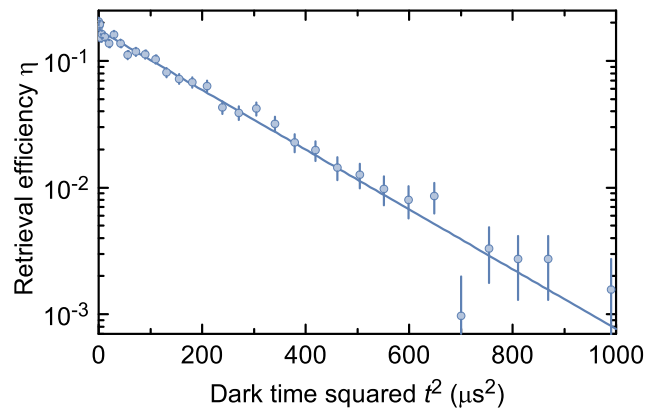}
\caption{Retrieval efficiency $\eta$ as a function of the square of the dark time $t$ for the $70S$ state at $T= 2.0$ $\mu$K in free expansion. The data cover two orders of magnitude in $\eta$. The line is a fit of a Gaussian decay according to Eq.\ \eqref{experiment-tau-R}, which agrees well with the experimental data. With a logarithmic vertical axis and $t^2$ on the horizontal axis, the fit curve is a straight line.}
\label{fig-eta-vs-time}
\end{figure}

To test the prediction of a Gaussian decay, we consider the measurement of $\eta$ as a function of $t$ displayed in Fig.\ \ref{fig-eta-vs-time}. The data span from $t=0.7$ to $31.5$ $\mu$s. The line is a fit of the Gaussian decay from Eq.\ \eqref{experiment-tau-R} to the data. The vertical axis in the figure is logarithmic and the horizontal axis shows $t^2$. With these axes, the Gaussian decay becomes a straight line. The fit agrees well with the data. To analyze the data in Fig.\ \ref{fig-eta-vs-time} from a different perspective, we fit $\eta(t)= \eta(0)\exp[-(t/\tau)^p]$ with free fit parameters $p$, $\tau$, and $\eta(0)$ to these data, obtaining the best-fit value $p=2.0(1)$. Again, this shows that a Gaussian models the situation well.

For comparison, Ref.\ \cite{Ginsberg:07} observed an exponential decay of the retrieval efficiency for ground-state EIT in an uncondensed cloud of sodium atoms with a $1/e$ time which is a factor of 0.7 shorter than the expectation from Eq.\ \eqref{experiment-tau-R}. It is difficult to explain this in hindsight without performing additional experimental tests on that setup.

\begin{figure}[!t]
\includegraphics[width=\columnwidth]{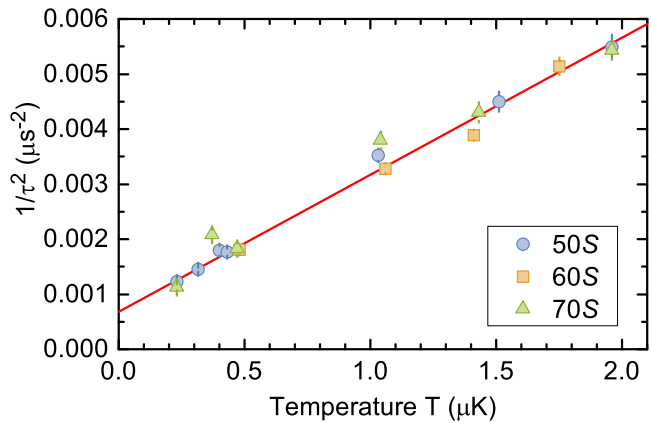}
\caption{Inverse decay time squared $1/\tau^2$ of the retrieval efficiency in free expansion as a function of temperature $T$. Data points with different symbols correspond to different principal quantum numbers $n$. There is no discernable dependence on $n$ in the parameter range studied here. According to Eq.\ \eqref{experiment-tau-R} all data in this figure are expected to fall onto a straight line through the origin. A straight-line fit to all data yields a slope which agrees well with the expectation from Eq.\ \eqref{experiment-tau-R}. For small $T$, the fit reveals an additional decay mechanism of presently unclear origin. Extrapolating the straight line to $T= 0$ yields $\tau=38(2)$ $\mu$s. The in-trap peak atomic density is $\varrho_0 \leq 1.7\times10^{11}$ cm$^{-3}$ for all these data so that atom-atom collisions are negligible.}
\label{fig-tau-vs-T}
\end{figure}

To study how the $1/e$ time $\tau$ extracted from the fit depends on temperature $T$, we record a series of data sets similar to the one shown in Fig.\ \ref{fig-eta-vs-time} and extract $\tau$ by fitting the Gaussian of Eq.\ \eqref{experiment-tau-R} to each data set. The resulting values of $\tau$ are shown in Fig.\ \ref{fig-tau-vs-T}. The data cover an order of magnitude in $T$. Different symbols represent different principal quantum numbers. The rightmost data point for the $70S$ state represents the data set from Fig.\ \ref{fig-eta-vs-time}.

To change the temperature, we varied the power of each light sheet between 25 and 320 mW. We estimate $L$ between 0.39 and 0.40 mm and $\varrho_0$ between $5\times 10^{10}$ and $1.7\times 10^{11}$ cm$^{-3}$ for all data in Figs.\ \ref{fig-eta-vs-time} and \ref{fig-tau-vs-T}. By varying the atomic density in additional measurements not shown here, we experimentally verified that for these principal quantum numbers, a density dependence of $\tau$ appears only for noticeably higher density, which means that atom-atom interactions have negligible effect in Figs.\ \ref{fig-eta-vs-time} and \ref{fig-tau-vs-T}.

As Fig.\ \ref{fig-tau-vs-T} shows $1/\tau^2$ versus $T$, Eq.\ \eqref{experiment-tau-R} predicts that all data should fall onto a straight line through the origin. The line in Fig.\ \ref{fig-tau-vs-T} is a straight-line fit to the data. The fit agrees well with the data. The slope can be expressed in terms of a best-fit value for the wavelength of the spin wave $\lambda_R= 2\pi/k_R= 1.23(3)$ $\mu$m. This agrees well with the $\lambda_R= (\lambda_{re}^{-1}-\lambda_{eg}^{-1})^{-1}= 1.25$ $\mu$m expected in the counterpropagating geometry of our experiment. Hence, for large enough $T$ the temperature dependence of $\tau$ agrees well with the prediction from Eq.\ \eqref{experiment-tau-R}.

In the limit $T\to 0$, however, the fit extrapolates to $\tau= 38(2)$ $\mu$s instead of $\tau\to \infty$ expected from Eq.\ \eqref{experiment-tau-R}. This indicates that there is an additional decay mechanism becoming relevant at low temperature. The physical origin thereof is presently unclear \cite{supplemental:model:experiment}.

\subsection{Decay Caused by Dipole-Trapping Light}

\label{sec-exper-trap}

Another important mechanism which causes decay of the retrieval efficiency is a spatially inhomogeneous difference of the potentials experienced by states $|g\rangle$ and $|r\rangle$. The 1064 nm dipole trapping light creates an attractive potential for the ground state with dynamical polarizability $\alpha_g= 687.3$ a.u., see above. For the Rydberg state, the dynamical polarizability is well approximated by that of a free electron, see e.g.\ Ref.\ \cite{Saffman:05}, yielding $\alpha_r= -550$ a.u.\ at 1064 nm. If this light is left on during the experiment, this will cause a dark-time decay of $\eta$ because the differential light shift depends on the atomic position.

\begin{figure}[!t]
\includegraphics[width=\columnwidth]{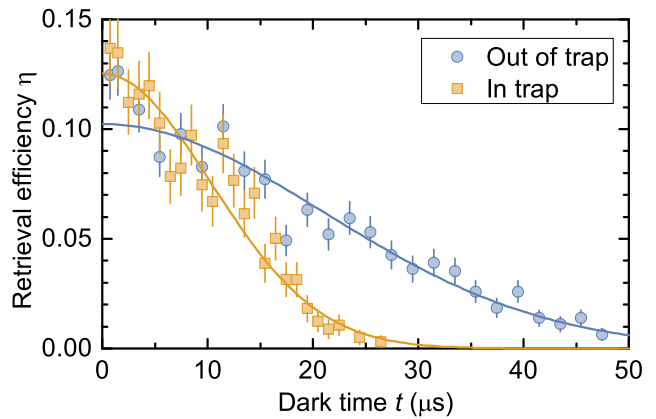}
\caption{Retrieval efficiency as a function of the dark time for the $80S$ state. Blue circles represent data taken after release from the trap. Orange squares represent data taken in the 1064-nm dipole trap. According to Eqs.\ \eqref{experiment-tau-R} and \eqref{experiment-tau-F} we expect a Gaussian decay in either case. Gaussian fits (lines) yield $1/e$ times of $\tau= 30(1)$ $\mu$s and $\tau= 12.5(6)$ $\mu$s with the trap off and on, respectively. Data were taken at a low temperature of 0.2 $\mu$K to keep the dephasing rate caused by photon recoil combined with thermal motion small and at a low in-trap peak atomic density of $5\times10^{10}$ cm$^{-3}$ to avoid dephasing caused by atom-atom interactions.}
\label{fig-trap}
\end{figure}

To study the size of this effect, we recorded the experimental data shown in Fig.\ \ref{fig-trap}. While the blue circles represent data taken in free expansion after switching the trap off, the orange squares represent data taken in the dipole trap. A Gaussian fit according to Eq.\ \eqref{experiment-tau-R} yields a $1/e$ decay time of $\tau= 30(1)$ $\mu$s for the free-expansion data. Clearly, the in-trap data decay much faster.

To model the in-trap decay time, we approximate the radial confinement produced by the 1064-nm light as harmonic, assume that the sample is axially homogeneous and neglect the presence of the 532-nm light sheets. According to Eqs.\ \eqref{tau-kappa} and \eqref{eta-tau-kappa}, this yields an algebraic decay
\begin{align}
\label{experiment-tau-kappa}
\frac{\eta(t)}{\eta(0)}
= \frac{1}{1+{t^2}/{\tau_\kappa^2}}
,&&
\tau_\kappa
= \frac{4\hbar}{w_r^2|\kappa_g-\kappa_r|}
,\end{align}
where $\kappa_g$ and $\kappa_r$ are the spring constants of the harmonic potentials $V_g(\bm x)= \frac12\kappa_g(x^2+y^2)$ and $V_r(\bm x)= \frac12\kappa_r(x^2+y^2)$ experienced by atoms in states $|g\rangle$ and $|r\rangle$, respectively, and the radial sizes are $w= 8$ $\mu$m for the signal beam waist, $2\sigma_x= \sqrt{4k_BT/m\omega^2}= 14$ $\mu$m for the ground-state atom cloud, and, according to Eq.\ \eqref{w-r-tau-F}
\begin{align}
\label{experiment-w-r-tau-F}
w_r
= \left( \frac{1}{4\sigma_x^2} +\frac{1}{w^2}\right)^{-1/2}
\end{align}
for that part of the atom cloud which was transferred into state $|r\rangle$ during storage. Experiments on EIT-based storage and retrieval are typically operated in the regime $w\ll 2\sigma_x$ because otherwise some part of the light would transversely miss the atomic ensemble, resulting in low storage efficiency. Hence, typically $w_r\approx w$.

As the differential potential $V_r-V_g$ depends quadratically on the radial position in this model, it imprints a phase onto the retrieved light which depends quadratically on the radial position. This is equivalent to inserting a lens. Hence, the dark-time decay of the amount of light coupled into the single-mode fiber is not caused by dephasing in the sense that the phase evolution of different atoms would fluctuate as a result of a fluctuating external parameter. Instead, this dark-time decay is caused by changing the focussing of the retrieved light. For fixed dark time, one could compensate this, in principle, by changing the alignment of the single-mode fiber.

For the parameters of Fig.\ \ref{fig-trap}, this model predicts a $1/e$ time of $\tau_\kappa\sqrt{e-1}= 120$ $\mu$s. This would suggest that this additional mechanism for trap-induced decay should be negligible compared to the 30 $\mu$s decay time from the free-expansion data. However, the experimental in-trap data clearly decay much faster.

This discrepancy is resolved when taking into account gravitational sag, i.e.\ the fact that gravity shifts the equilibrium position of the atomic cloud away from the center of the dipole trapping beam. As a result, the differential potential $V_r-V_g$ between states $|r\rangle$ and $|g\rangle$ varies linearly in position when moving away from the cloud center, whereas in the absence of gravitational sag, it would vary quadratically. Hence, gravitational sag causes the finite-size atomic cloud to sample larger values of the differential potential.

In our experiment, the radius of the atomic cloud $\sigma_x= 7$ $\mu$m at $T= 0.2$ $\mu$m is much smaller than the gravitational sag $x_{g,s}= 27$ $\mu$m. Hence, the curvature of the potential becomes negligible and according to Eqs.\ \eqref{w-r-tau-F} and \eqref{eta-tau-F} we expect $\eta$ to decay as
\begin{align}
\label{experiment-tau-F}
\frac{\eta(t)}{\eta(0)}
= \exp\left(-\frac{t^2}{\tau_F^2} \right)
,&&
\tau_F
= \frac{2\hbar}{w_r|F|}
,\end{align}
where $\bm F= -\nabla (V_r-V_g)$ is the differential force, to be taken at the cloud center. Note that taking into account the finite size $w_r$ of the atomic cloud transferred to state $|r\rangle$ is crucial here, because in the limit $w_r\to \infty$, Eq.\ \eqref{experiment-tau-F} yields $\tau_F\to 0$. The plausibility of this is discussed in Sec.\ \ref{sec-quadratic}.

As the differential potential depends linearly on the position along gravity in this model, it imprints a phase onto the retrieved light which depends linearly on the position along gravity. This is equivalent to inserting a prism. Hence, the dark-time decay of the amount of light coupled into the single-mode fiber is not caused by dephasing but by changing the direction of the wave vector of the retrieved light beam. For fixed dark time, one could compensate this, in principle, by changing the alignment of the single-mode fiber.

A fit of the Gaussian model Eq.\ \eqref{experiment-tau-F} to the in-trap data in Fig.\ \ref{fig-trap} agrees well with the data. It yields a best-fit value for the $1/e$ time of $\tau_F= 12.5(6)$ $\mu$s, in good agreement with the prediction 11.8 $\mu$s from Eq.\ \eqref{experiment-tau-F}. In principle, the faster decay of $\eta$ in the presence of the dipole trap could also be caused by photoionization of the Rydberg state by the trapping light. Quantitatively, however, photoionization is expected to be much slower than the time scale observed here \cite{supplemental:model:experiment}.

While release and recapture solves the problem of the reduced in-trap decay time, it causes heating which drastically reduces the time-averaged data acquisition rate, as discussed above. This limitation could be overcome by operating in blue-detuned or magic-wavelength dipole traps \cite{Saffman:05, Zhang:11:trap, Li:13, Piotrowicz:13, Lampen:18}. To obtain a quantitative estimate for the parameters of our experiment, we consider
\begin{align}
\tau_F
= \frac{2\hbar}{mgw_r|1-\frac{\alpha_r}{\alpha_g}|}
\end{align}
from Eq.\ \eqref{tau-F-tau-kappa-alpha}. Hence, apart from $w_r$, the only relevant quantity here is $|1-\frac{\alpha_r}{\alpha_g}|$. For 1064 nm, the above-quoted values of the polarizabilities yield $|1-\frac{\alpha_r}{\alpha_g}|= 1.8$. For 532 nm, however, the polarizabilities are $\alpha_g= -250$ a.u., see above, and $\alpha_r= -140$ a.u.\ estimated from a free electron. This yields $|1-\frac{\alpha_r}{\alpha_g}|= 0.45$. Hence, if we replaced the 1064-nm dipole trap by a 532-nm hollow-beam dipole trap, we would expect an approximately four-fold increase of $\tau_F$ to 50 $\mu$s. Hence, $\tau_F$ would have negligible effect compared to the observed 30 $\mu$s decay time.

Note that similarly if $\kappa_g$ is unchanged then according to Eq.\ \eqref{tau-F-tau-kappa-alpha} $\tau_\kappa$ will also improve by a factor of approximately 4 when making the transition to a 532 nm trap, meaning that $\tau_\kappa$ remains irrelevant.

\subsection{Visibility}

\label{sec-exper-visibility}

In addition to the efficiency, the retrieved light has another crucial property, namely the degree to which it is coherent. To quantify this, one can overlap the light with a reference beam, vary the phase of the reference beam, and quote the fringe visibility $V= (I_\text{max}-I_\text{min})/(I_\text{max}+I_\text{min})$ of the resulting sinusoidal interference pattern, where $I_\text{max}$ and $I_\text{min}$ denote the maximum and minimum of the intensity. To characterize the coherence of the retrieved light, one will of course quote the value of $V$ for a parameter setting in which the powers of the two light fields are balanced.

As directed retrieval is a coherent phenomenon, $\eta$ is also some measure of coherence in the atomic system at the time of retrieval. Hence, one might wonder whether they react identically to experimental imperfections. However, that does not have to be the case. To give an example of a mechanism onto which they react quite differently, we consider shot-to-shot fluctuations of the energy of the Rydberg state. This would cause shot-to-shot fluctuations of the phase of the retrieved light. When taking an ensemble average over many shots, these phase fluctuations would yield a decay of $V$ as a function of the dark time $t$. But the same fluctuations would have no effect whatsoever on $\eta$.

In our experiment, we measure $V$ with a slightly different technique. We overlap the lefthand circularly polarized signal light beam with a copropagating righthand circularly polarized reference light beam at the same frequency. Polarization tomography reveals the normalized Stokes vector. Here, $V$ is the length of the projection of the normalized Stokes vector onto the plane which contains all linear polarizations, see e.g.\ Ref.\ \cite{Tiarks:16}. Again, the powers of the signal and reference light must be balanced to avoid underestimating the degree to which the retrieved light is coherent. In the absence of atoms, we measure a visibility of $V_0= 97.3(6)\%$. This is caused e.g.\ by imperfections in the polarization tomography, in balancing the beam powers, and in the active stabilization of the differential phase between signal and reference light. $V_0$ sets the technical detection limit of our present measurement.

\begin{figure}[!t]
\includegraphics[width=\columnwidth]{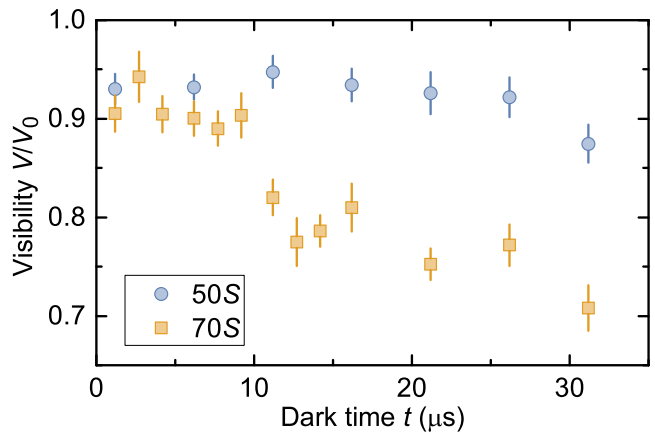}
\caption{Dependence of the visibility $V$ on the dark time $t$. $V$ characterizes how coherent the retrieved light is. It is measured by overlapping the retrieved light with reference light. For short dark time, $V$ reaches values above 90\%. There is a discernable decay as a function of $t$ for the 70$S$ data, but not for the 50$S$ data. Data are normalized with respect to the technical detection limit $V_0$ in the present setup. The in-trap peak atomic density is roughly $2\times10^{11}$ cm$^{-3}$ for all data in this figure.}
\label{fig-visibility}
\end{figure}

Figure \ref{fig-visibility} shows $V$ as a function of dark time $t$ for storage in Rydberg states 50$S$ and 70$S$ after release from the dipole trap. The values are normalized to the technical detection limit $V_0$, which is not related to the physics in the atomic system. The data in Fig.\ \ref{fig-visibility} were taken at in-trap peak atomic densities between $\varrho_0= 1.7\times10^{11}$ and $2.4\times10^{11}$ cm$^{-3}$, temperatures between 0.3 and 0.4 $\mu$K, and $L= 0.39$ mm.

The input signal pulse is rectangular and has a duration of 4.5 $\mu$s out of which only a small fraction near the end of the pulse is stored. This somewhat exaggerated length of the input pulse provides ample time for possible transients to decay. Such transients may result from switching on the pulse. The retrieved pulse has an approximately exponentially decaying shape with a $1/e$ time of typically 0.8 $\mu$s, suggesting that the stored part of the input pulse might have had a similar length. As mentioned above, we overlap the lefthand circularly polarized retrieved light with righthand circularly polarized reference light. For simplicity, we use a rectangular pulse shape for the reference light. We process data only in a time interval with a duration of typically 0.5 $\mu$s, because outside this interval, the beam powers would be poorly balanced.

For short dark time, $V$ reaches values above 90\% in Fig.\ \ref{fig-visibility}. This is a big improvement over the 66(2)\% which we reported in Ref.\ \cite{Tiarks:19} for a measurement for $t=4.5$ $\mu$s, $\varrho_0= 2\times10^{12}$ cm$^{-3}$, and storage in state 69$S$. The much lower atomic density in the present measurement is crucial for this improvement \cite{supplemental:model:experiment}.

For storage in the 50$S$ state, we observe no discernible decay of $V(t)$ in the time interval studied here. Measuring for much longer times would become cumbersome because there would be only a small retrieved signal. For storage in the 70$S$ state, there clearly is a decay of $V(t)$ but not a very fast one. As the decay depends on principal quantum number and as we confirmed in an additional measurement that it does not improve when lowering the atomic density \cite{supplemental:model:experiment}, the most likely explanation for the observed decay of $V(t)$ seems to be a fluctuating Stark shift of the Rydberg state caused by fluctuating stray electric fields.

\section{Model}

\label{sec-model}

Here, we develop a model for the dark-time decay of the efficiency $\eta$ in EIT-based storage and retrieval of light. Our model ignores loss of photons during storage and during the propagation of light inside the medium. Instead, it focusses on the decay of the efficiency as a function of the dark time between storage and retrieval. We start with a brief description of dark polaritons in Sec.\ \ref{sec-polariton}. This is a straightforward generalization of a similar treatment \cite{Fleischhauer:02} which did not take photon recoil into account. It sets the stage for the following discussion. In Sec.\ \ref{sec-storage}, we use this formalism to derive an expression for the efficiency $\eta$ for a separable initial state, which may be pure or mixed. The result simplifies if the initial state is uncorrelated, as discussed in Sec.\ \ref{sec-uncorrelated}. The results can be simplified even further in the frequently encountered situation, in which the Hamiltonian for ground-state atoms is identical before and after the storage, as discussed in Sec.\ \ref{sec-Hg-rho}. The treatment up to that point assumes for simplicity that the mode of the incoming EIT signal field $u(\bm x)$ is a plane wave. A generalization beyond this assumption is discussed in Sec.\ \ref{sec-beyond-plane-waves}. The relation to DLCZ sources and to Ramsey spectroscopy is discussed in Sec.\ \ref{sec-DLCZ-Ramsey}. A generalization to entangled initial states is discussed in appendix \ref{sec-app-entangled}.

\subsection{Dark Polaritons}

\label{sec-polariton}

We consider EIT-based storage and retrieval in an ensemble of noninteracting, identical, three-level atoms with ladder-type energy level scheme. A straightforward generalization to $\Lambda$-type energy level schemes is discussed in appendix \ref{sec-app-Hamiltonian}. The internal state of an atom has a basis of energy eigenstates which in order of ascending energies are $|g\rangle$, $|e\rangle$, and $|r\rangle$. In addition, the atom has an external state describing its center-of-mass motion. We assume that a signal (coupling) light field is resonant with the $|g\rangle \leftrightarrow |e\rangle$ ($|e\rangle \leftrightarrow |r\rangle$) transition, see Fig.\ \ref{fig-schemes}(b). The coupling light is assumed to be a plane wave $e^{i\bm k_c\cdot\bm x}$ with wave vector $\bm k_c$. We consider only a single mode of the EIT signal light with mode function $u(\bm x)$ normalized to $\int_{\mathcal V} d^3x |u(\bm x)|^2= 1$, where $\mathcal V$ is the quantization volume. We abbreviate $v(\bm x)= u(\bm x)e^{i\bm k_c\cdot\bm x}$. This is normalized to $\int_{\mathcal V} d^3x |v(\bm x)|^2=1$. For simplicity, the following description assumes that $u(\bm x)$ is a plane wave $u(\bm x)= e^{i\bm k_s\cdot\bm x}/\sqrt{\mathcal V}$ with wave vector $\bm k_s$. A generalization beyond this assumption is discussed in Sec.\ \ref{sec-beyond-plane-waves}. $\hbar\bm k_R$ with $\bm k_R= \bm k_s+\bm k_c$ is the recoil momentum transferred to an atom in the two-photon transition from $|g\rangle$ to $|r\rangle$.

While the initial state $|g\rangle$ and final state $|r\rangle$ of the storage process are assumed to be long lived, the intermediate state $|e\rangle$ is subject to decay to state $|g\rangle$ with rate coefficient $\Gamma_e$ accompanied by emission of a photon. The desired part of these emissions produces photons in mode $u(\bm x)$. The remaining part produces photons in spatial modes orthogonal to $u(\bm x)$. We refer to the latter process as spontaneous emission.

We define the number of excitations as the number of signal photons in mode $u(\bm x)$ plus the number of atoms in states $|e\rangle$ and $|r\rangle$. In the absence of spontaneous emission, the number of excitations is conserved. As a result, the subspaces of Hilbert space describing states with a given number of excitations are invariant under time evolution as long as spontaneous emission is ignored. In the presence of spontaneous emission, the number of excitations can only decrease. In our experiment, the number of stored excitations is typically less than one. Hence, we restrict our model to the subspaces with one or zero excitations. The zero-excitation subspace is trivial, so we focus on the single-excitation subspace.

Hence, EIT-based storage starts with one photon in the mode $u(\bm x)$ and all atoms in internal state $|g\rangle$. The external initial state, however, is often a mixed state. The initial $N$-atom state can generally be described by a density matrix $\rho_{N,\text{in}}$, where the subscript $_\text{in}$ indicates the initial state before storage. For simplicity, we restrict the discussion here to situations in which $\rho_{N,\text{in}}$ is separable. The model is easily extended to entangled initial states, see appendix \ref{sec-app-entangled}. As the density matrix $\rho_{N,\text{in}}$ is separable it can be written as, see e.g.\ Ref.\ \cite{Lewenstein:00},
\begin{align}
\label{rho-N}
\rho_{N,\text{in}}
= \sum_n P_n |\Psi_{g,n,\text{in}}\rangle\langle\Psi_{g,n,\text{in}}|
,\end{align}
where each of the initial $N$-atom pure states $|\Psi_{g,n,\text{in}}\rangle$ is separable, i.e\ a product of $N$ single-atom states. The probabilities $P_n$ fulfill $P_n\geq0$ and $\sum_n P_n= 1$. Obviously, the storage-and-retrieval efficiency for the initial state $\rho_{N,\text{in}}$ of Eq.\ \eqref{rho-N} is
\begin{align}
\label{eta-eta-n}
\eta(t)
= \sum_n P_n\eta_n(t)
,\end{align}
where $t$ is the dark time and $\eta_n(t)$ denotes the storage-and-retrieval efficiency which would be obtained if the initial state was the pure state $|\Psi_{g,n,\text{in}}\rangle$.

To calculate $\eta_n(t)$ we note that, as stated above, the pure state $|\Psi_{g,n,\text{in}}\rangle$ is assumed to be a product state with all atoms in internal state $|g\rangle$. Hence, it has the form $|\Psi_{g,n,\text{in}}\rangle= \bigotimes_{i=1}^N |\psi_{g,n,i}(0),g_{i}\rangle$, where $|\psi_{g,n,i}(0)\rangle$ and $|g_{i}\rangle$ are the initial external and internal states of the $i$th atom, respectively. The argument $(0)$ in $|\psi_{g,n,i}(0)\rangle$ refers to zero dark time. Throughout this paper, $N$-atom states (external single-atom states) are represented by uppercase Greek letters (lowercase Greek letters).

It is easy to show that the three-dimensional (3d) subspace of Hilbert space spanned by the orthonormal set of $N$-atom states
\begin{subequations}
\begin{align}
&
\label{Psi-g-n(0)}
|\Psi_{g,n,\text{in}},1_s\rangle
= |1_s\rangle\bigotimes_{i=1}^N |\psi_{g,n,i}(0),g_{i}\rangle
,\\
&
|\Psi_{e,n}(0)\rangle
= \frac1{\sqrt N} \sum_{i=1}^N |\psi_{e,n,i}(0),e_{i}\rangle
\bigotimes_{\substack{i'=1\\ i'\neq i}}^N |\psi_{g,n,i'}(0),g_{i'}\rangle
,\\
&
\label{Psi-r-n(0)}
|\Psi_{r,n}(0)\rangle
= \frac1{\sqrt N} \sum_{i=1}^N |\psi_{r,n,i}(0),r_{i}\rangle
\bigotimes_{\substack{i'=1\\ i'\neq i}}^N |\psi_{g,n,i'}(0),g_{i'}\rangle
\end{align}
\label{3d-subspace}%
\end{subequations}
is invariant under application of the atom-light interaction Hamiltonian $\mathcal V_{al}$ detailed in appendix \ref{sec-app-Hamiltonian}. Here, $|1_s\rangle$ is the single-photon Fock state of the mode $u(\bm x)$ of the signal light. The external single-atom states $|\psi_{e,n,i}(0)\rangle$ and $|\psi_{r,n,i}(0)\rangle$ have the position representations
\begin{subequations}
\begin{align}
\psi_{e,n,i}(\bm x,0)
&
= \sqrt{\mathcal V}u(\bm x)\psi_{g,n,i}(\bm x,0)
,\\
\label{psi-r-n-i(0)}
\psi_{r,n,i}(\bm x,0)
&
= \sqrt{\mathcal V}v(\bm x)\psi_{g,n,i}(\bm x,0)
\end{align}
\label{psi-e-n-i(0)-psi-r-n-i(0)}%
\end{subequations}
and are properly normalized because $u(\bm x)$ and $v(\bm x)$ are properly normalized plane waves. The $N$-atom states $|\Psi_{e,n}(0)\rangle$ and $|\Psi_{r,n}(0)\rangle$ are singly excited Dicke states, in which we omitted multiplication with the vacuum state $|0_s\rangle$ of the signal mode for brevity.

For later use, we introduce an operator $R_i^\dag$ acting on the external degree of the $i$th atom with position representation
\begin{align}
\label{R-def}
R_i^\dag(\bm x)
= \sqrt{\mathcal V}v(\bm x)
.\end{align}
Hence Eq.\ \eqref{psi-r-n-i(0)} can be rewritten as
\begin{align}
\label{psi-r-R-psi-g}
|\psi_{r,n,i}(0)\rangle
= R_i^\dag |\psi_{g,n,i}(0)\rangle
.\end{align}
As $v(\bm x)$ is a plane wave, $R_i^\dag$ is unitary. A related single-particle operator $S_{r,i}^\dag= R_i^\dag \otimes|r_i\rangle\langle g_i|$ acting onto the external and internal degrees of freedom of the $i$th atom is studied in Ref.\ \cite{supplemental:model:experiment}.

It is easy to show that with respect to the orthonormal basis $(|\Psi_{g,n,\text{in}},1_s\rangle, |\Psi_{e,n}(0)\rangle, |\Psi_{r,n}(0)\rangle)$ of the 3d invariant subspace, the atom-light interaction Hamiltonian $\mathcal V_{al}$ has the matrix representation
\begin{align}
\label{V-al-matrix}
\mathcal V_{al}
=
\frac\hbar2
\begin{pmatrix}
0 & 2g_R \sqrt N  & 0 \\
2g_R \sqrt N & 0 & \Omega_c \\
0 & \Omega_c & 0 \\
\end{pmatrix}
,\end{align}
where $2g_R$ is the vacuum Rabi frequency of the signal mode and $\Omega_c$ the Rabi frequency of the EIT coupling light. The factor $\sqrt N$ comes about because $g_R$ causes transitions between a product state and a Dicke state.

Obviously, the $N$-atom state
\begin{align}
|\Psi_{d,n}(\vartheta)\rangle
= \cos\vartheta |\Psi_{g,n,\text{in}},1_s\rangle -\sin\vartheta |\Psi_{r,n}(0)\rangle
\end{align}
with mixing angle $\vartheta$ given by
\begin{align}
\tan\vartheta
= \frac{2g_R\sqrt N}{\Omega_c}
\end{align}
is an eigenstate of $\mathcal V_{al}$. Hence, for signal light in vacuum with $\Omega_c\neq 0$, one obtains $\vartheta= 0$. If we were to extend our formalism to a signal light pulse instead of a plane wave, then $\vartheta= 0$ would be the initial value before the pulse enters the medium and the final value after the retrieved pulse left the medium. Conversely, for a pulse inside the medium with $\Omega_c= 0$ during the dark time between storage and retrieval, one would obtain $\vartheta= \pi/2$.

The state $|\Psi_{d,n}\rangle$ is dark in the sense that it shows no spontaneous emission when spontaneous emission into modes orthogonal to $u(\bm x)$ is added to the model, because $\langle e_i|\Psi_{d,n}\rangle= 0$ for all $i$. For $0<\vartheta<\pi/2$ the state $|\Psi_{d,n}\rangle$ describes a superposition of a photon and a copropagating atomic excitation, which is why it is called dark polariton. For $\vartheta= 0$, however, it describes a single photon and for $\vartheta= \pi/2$ it describes the Dicke state $|\Psi_{r,n}(0)\rangle$, which is commonly referred to as a spin wave.

The other two eigenstates of $\mathcal V_{al}$ in Eq.\ \eqref{V-al-matrix} are bright states because they rapidly decay by spontaneous emission into modes orthogonal to $u(\bm x)$. In our model, these bright states which couple to $u(\bm x)$ never become populated because we will assume below that the population adiabatically follows the dark state $|\Psi_{d,n}\rangle$.

\subsection{EIT-Based Storage and Retrieval}

\label{sec-storage}

In an experiment, a signal light pulse of finite duration is stored in an atomic medium of finite length. Aiming at large storage efficiency would entail a nontrivial treatment of the longitudinal wave function of the signal light pulse, particularly when entering and leaving the medium. In addition, one would need to address the question of whether the complete light pulse fits into the medium longitudinally. If that is not the case, this will lead to leakage of signal light through the medium during the time of storage. Furthermore, there is the issue of residual absorption because of imperfect EIT while the pulse propagates inside the medium before storage and after retrieval. The residual absorption can be caused e.g.\ by dephasing or by the nonzero frequency width of the light pulse which results from its finite duration. Such issues have been addressed in the literature in detail, see e.g.\ Ref.\ \cite{Gorshkov:07:PRL}. We do not attempt to model leakage and residual absorption here because these issues tend to affect the efficiency in a way which is independent of the dark time. In Eq.\ \eqref{eta-P-Psi-r}, leakage and residual absorption will be subsumed in an empirical correction factor $\eta_0$.

To describe storage, we use a model quite similar to Refs.\ \cite{Fleischhauer:00, Fleischhauer:02}. In this model, the medium is homogeneous along the $z$ axis with a quantization length $L_z$. Hence, the signal-light pulse cannot enter or leave the medium and no spatial pulse compression occurs, which drastically simplifies the model. The experimental initial situation with a light pulse outside the medium is modelled by starting with a very large value of $\Omega_c$, which results in $\vartheta\approx 0$. Along with this, the system is assumed to be prepared in the dark state $|\Psi_{d,n}\rangle= |\Psi_{g,n,\text{in}},1_s\rangle$. Next, $\Omega_c$ is ramped to 0. We assume that this ramp is slow enough that the population, to a good approximation, adiabatically follows the dark state. The time evolution during the ramp, which can be fairly complicated in general, is thus simply modelled as adiabatic following, much like in Ref.\ \cite{Fleischhauer:02}. After this ramp, $\vartheta= \pi/2$ so that the dark state has evolved into $|\Psi_{d,n}\rangle= |\Psi_{r,n}(0)\rangle$, which means that storage in the form of a spin wave has been achieved. The assumption of adiabatic following means that no spontaneous emission into modes orthogonal to $u(\bm x)$ occurs during storage.

As an aside, we note that in an experiment, the rotation of $\vartheta$ occurs almost exclusively when the pulse enters the medium. When the pulse is inside the homogeneous part of the medium, typically $\vartheta\approx \pi/2$. The actual temporal ramp of $\Omega_c$ in the experiment changes $\vartheta$ only by a small amount, bringing it all the way to $\pi/2$.

Now, we deviate from Refs.\ \cite{Fleischhauer:00, Fleischhauer:02}. We further simplify the model by assuming that $u(\bm x)$ is a plane wave. As a result, keeping track of the longitudinal properties of the signal light becomes trivial.

We now turn to the dark time $t$ between storage and retrieval. As the Hamiltonian is time independent during the dark time, it yields a time-evolution operator of the simple form $\mathcal U_d(t)= e^{-i\mathcal H_dt/\hbar}$, where $\mathcal H_d$ is the $N$-atom dark-time Hamiltonian. For simplicity, we restrict our model to a situation in which $\mathcal H_d= \sum_{i=1}^N H_{d,i}$ is a sum of single-atom Hamiltonians $H_{d,i}$ of the form
\begin{align}
\label{H-d-i}
H_{d,i}
= H_{g,i} \otimes |g_{i}\rangle\langle g_{i}|+H_{r,i}\otimes |r_{i}\rangle\langle r_{i}|
,\end{align}
where $H_{g,i}$ and $H_{r,i}$ are operators acting on the external state of the $i$th atom. Hence, the atoms are noninteracting and the internal state of each atom is unchanged during the dark time. We abbreviate
\begin{subequations}
\begin{align}
U_{g,i}(t)
&
= e^{-iH_{g,i}t/\hbar}
,&
|\psi_{g,n,i}(t)\rangle
&
= U_{g,i} |\psi_{g,n,i}(0)\rangle
,\\
U_{r,i}(t)
&
= e^{-iH_{r,i}t/\hbar}
,&
|\psi_{r,n,i}(t)\rangle
&
= U_{r,i} |\psi_{r,n,i}(0)\rangle
.\end{align}
\label{psi-g-n-i(t)-psi-r-n-i(t)}%
\end{subequations}
Throughout this paper, $N$-atom (single-atom) operators other than the density matrix are represented by uppercase calligraphic (italic) letters.

Hence, the $N$-atom state at the end of the dark time reads
\begin{align}
\label{Psi-r-n(t)}
|\Psi_{r,n}(t)\rangle
&
= \mathcal U_d(t) |\Psi_{r,n}(0)\rangle
\\ & \notag
= \frac1{\sqrt N} \sum_{i=1}^N |\psi_{r,n,i}(t),r_{i}\rangle
\bigotimes_{\substack{i'=1\\ i'\neq i}}^N |\psi_{g,n,i'}(t),g_{i'}\rangle
.\end{align}
Typically, the term $\mathcal H_d$ in the Hamiltonian will also be present during storage and retrieval. But for simplicity, we assume that the dynamics during storage and retrieval are dominated by $\mathcal V_{al}$ so that $\mathcal H_d$ has negligible effect during storage and retrieval.

After the dark time, the EIT coupling light is turned back on for retrieval. Ideally, this will cause directed retrieval of the photon into the original spatial mode $u(\bm x)$ as a result of interference of light emitted from the large number $N$ of atoms \cite{Fleischhauer:00, Fleischhauer:02}. The write-read efficiency $\eta$ is the ratio of the average number of photons retrieved into the original mode $u(\bm x)$ divided by the average number of incoming photons before storage. Again, we neglect a variety of experimental complications, e.g.\ the fact that after the finite-duration signal light pulse resumes propagation, it experiences some residual absorption before leaving the finite-length medium. Instead, we model the retrieval process by assuming that $\Omega_c$ is slowly ramped back up from zero, where $\vartheta= \pi/2$, to a very large value of $\Omega_c$, finally resulting in $\vartheta\approx 0$. We study the final number of photons in the plane-wave mode $u(\bm x)$ which remains inside the homogeneous medium and does not experience absorption in our model because it is monochromatic and meets the two-photon resonance condition.

Hence, much like the storage process, we model the retrieval as an adiabatic passage, but now with $\vartheta$ evolving back from $\pi/2$ to 0. For zero dark time, the considerations can be restricted to the 3d invariant subspace given by Eq.\ \eqref{3d-subspace} and the retrieval is simply the time reversed process of the storage. For nonzero dark time, however, things become more complicated. If retrieval is successful, then by definition the excitation reappears in the mode $u(\bm x)$ of the signal light, which obviously implies that all atoms are finally in internal state $|g\rangle$. But for nonzero dark time it is not immediately clear, what the final external $N$-atom state will be.

To include this aspect in our calculation, let $\mathcal H_\text{ext}$ denote the Hilbert space containing all external $N$-atom states. Note that the fact that $\mathcal H_\text{ext}$ is a Hilbert space implies that it contains product states and entangled states. Let $W_u$ denote the subspace of spin-wave states obtained when applying storage with mode function $u(\bm x)$ to $\mathcal H_\text{ext}$. Let us temporarily assume that the state after the dark time $|\Psi_{r,n}(t)\rangle$ is an element of $W_u$. Hence, there exists an $N$-atom state $|\Psi_f\rangle$ which would turn into $|\Psi_{r,n}(t)\rangle$ upon storage. As we treat the retrieval as the time reversed version of storage, it is now clear that the state $|\Psi_{r,n}(t)\rangle$ causes retrieval with 100\% efficiency into mode $u(\bm x)$ with final atomic state $|\Psi_f\rangle$. In this way, we found the possibly nontrivial final $N$-atom state $|\Psi_f\rangle$.

Now we turn to a general state $|\Psi_{r,n}(t)\rangle$ which does not have to be an element of $W_u$. We use $\mathcal P_u$ to denote the orthogonal projector onto the subspace $W_u$ and use this to decompose this state into $\mathcal P_u|\Psi_{r,n}(t)\rangle$ and
$(\mathbbm 1-\mathcal P_u) |\Psi_{r,n}(t)\rangle$. As $\mathcal P_u|\Psi_{r,n}(t)\rangle$ is an element of $W_u$ it causes retrieval into mode $u(\bm x)$ with perfect efficiency, as explained above. Conversely, as $(\mathbbm 1-\mathcal P_u) |\Psi_{r,n}(t)\rangle$ is orthogonal to $W_u$ it does not couple to the mode $u(\bm x)$ for reasons discussed in appendix \ref{sec-app-retrieval}. Hence, the efficiency is
\begin{align}
\label{eta-P-Psi-r}
\eta_n(t)
= \eta_0 \lVert \mathcal P_u|\Psi_{r,n}(t)\rangle \rVert^2
,\end{align}
where $\lVert...\rVert$ denotes the norm of a vector and we included a constant factor $\eta_0$ with $0\leq \eta_0\leq 1$ which serves to represent imperfections during storage and retrieval, such as leakage, residual absorption, and imperfections in the adiabaticity when rotating $\vartheta$. We assume that $\eta_0$ is independent of $n$. In the following we always consider $N\gg 1$, because this is a necessary condition for making the directed emission dominate over spontaneous emission in random directions.

As detailed in appendix \ref{sec-app-retrieval}, Eq.\ \eqref{eta-P-Psi-r} with $N\gg 1$ yields for a separable pure initial state
\begin{align}
\label{eta-n-Q-plane-wave}
\frac{\eta_n(t)}{\eta_0}
= \frac1{N^2} \left|\sum_{i=1}^N Q_{n,i}(t) \right|^2
\end{align}
with
\begin{align}
\label{Q-def}
Q_{n,i}(t)
&
= \langle\psi_{g,n,i}(t)|R_i|\psi_{r,n,i}(t)\rangle
\notag \\ &
= \langle\psi_{g,n,i}(0)|U_{g,i}^\dag(t) R_i U_{r,i}(t)R_i^\dag|\psi_{g,n,i}(0)\rangle
.\end{align}
Note that Eq.\ \eqref{psi-r-R-psi-g} combined with $\langle\psi_{r,n,i}(0)|\psi_{r,n,i}(0)\rangle= 1$ implies $Q_{n,i}(0)= 1$ so that $\eta_n(0)/\eta_0= 1$. In addition, the unitarity of $R_i$ implies $|Q_{n,i}(t)|^2\leq 1$ for all times so that $\eta_n(t)/\eta_0\leq 1$ for all times.

Equation \eqref{eta-n-Q-plane-wave} for a separable pure initial state can equivalently be written as
\begin{align}
\label{eta-n-fidelity-Phi-Psi-r}
\frac{\eta_n(t)}{\eta_0}
= \left|\langle \Phi_n(t)|\Psi_{r,n}(t)\rangle\right|^2
\end{align}
with a properly-normalized Dicke state
\begin{align}
\label{Phi-n}
|\Phi_n(t)\rangle
= \frac1{\sqrt N} \sum_{i=1}^N \left(R_i^\dag|\psi_{g,n,i}(t),r_{i}\rangle \right)
\bigotimes_{\substack{i'=1\\ i'\neq i}}^N |\psi_{g,n,i'}(t),g_{i'}\rangle
\end{align}
which turns out to be an element of $W_u$. In the language of quantum information processing, $\eta_n(t)/\eta_0$ in Eq.\ \eqref{eta-n-fidelity-Phi-Psi-r} is the fidelity \cite{Jozsa:94} of the states $|\Phi_n(t)\rangle$ and $|\Psi_{r,n}(t)\rangle$.

The $N$-atom state $|\Psi_{r,n}(t)\rangle$ is obtained from the state $|\Psi_{g,n,\text{in}},1_s\rangle$ by storage followed by dark-time propagation, whereas the $N$-atom state $|\Phi_n(t)\rangle$ would be obtained if the temporal order were reversed, namely if the dark-time propagation were followed by storage. Equation \eqref{Q-def} features analogous quantities on the single-particle level, because $|\psi_{r,n,i}(t)\rangle$ is obtained from $|\psi_{g,n,i}(0)\rangle$ by storage followed by dark-time propagation, whereas $R_i^\dag|\psi_{g,n,i}(t)\rangle$ would be obtained by dark-time propagation followed by storage.

Equation \eqref{eta-n-Q-plane-wave} is immediately applicable to a gas of noninteracting bosons at $T= 0$, because in that case one obtains a pure BEC so that all atoms initially occupy the same single-particle wave function. In this situation, we drop the indices $n,i$ from the notation and obtain
\begin{align}
\label{eta-BEC-plane-wave}
\frac{\eta_\text{BEC}(t)}{\eta_0}
= |Q(t)|^2
.\end{align}
More generally, inserting Eq.\ \eqref{eta-n-Q-plane-wave} into Eq.\ \eqref{eta-eta-n}, one immediately finds $\eta(t)$ for the arbitrary separable initial state of Eq.\ \eqref{rho-N}, which may be mixed.

\subsection{Uncorrelated Initial State}

\label{sec-uncorrelated}

We now concentrate on a special case, which is experimentally relevant and allows for further simplifications, finally giving a simple expression. Specifically, we assume that the initial $N$-atom density matrix
\begin{align}
\label{rho-factorize}
\rho_{N,\text{in}}
= \underbrace{\widetilde\rho_\text{in} \otimes \widetilde\rho_\text{in} \otimes \dots \otimes \widetilde\rho_\text{in}}_{N \text{ times}}
\end{align}
is a tensor product of $N$ identical copies of a single-atom density matrix $\widetilde\rho_\text{in}$. As all atoms are initially in internal state $|g\rangle$, we obtain $\widetilde\rho_\text{in}= \rho_\text{in}\otimes |g\rangle\langle g|$, where $\rho_\text{in}$ describes only the external state of a single atom. Note that $\rho_\text{in}$ can be diagonalized as
\begin{align}
\label{rho-single-atom-diagonalized}
\rho_\text{in}
= \sum_n p_n |\psi_{g,n}(0)\rangle\langle\psi_{g,n}(0)|
\end{align}
with probabilities $p_n$. Inserting this into Eq.\ \eqref{rho-factorize} shows that $\rho_{N,\text{in}}$ is separable so that the above formalism is applicable.

As we assumed that the particles are identical, Eq.\ \eqref{rho-factorize} holds if and only if all particles are uncorrelated, which is the case, e.g.\ if $\rho_{N,\text{in}}$ describes a noninteracting gas of identical particles in thermal equilibrium at a temperature far above quantum degeneracy. In that case, $p_n$ is given by Eq.\ \eqref{Gibbs-measure} and the external single-atom states $|\psi_{g,n}(0)\rangle$ are the eigenstates of the Hamiltonian before storage. Note that a noninteracting pure BEC at $T=0$ is also an example of the uncorrelated initial state in Eq.\ \eqref{rho-factorize}. In that case $\rho_\text{in}$ is a pure states.

In addition, as we assumed that the particles are identical, we obtain $H_{g,i}= H_{g,1}$ and $H_{r,i}= H_{r,1}$ for all $i$. As a result, the problem factorizes and all properties of the $N$-atom problem can be expressed in terms of properties of only the first particle. In this situation, we drop the index $i$ from the notation, writing e.g.\ $H_g= H_{g,1}$ and $Q_n(t)= Q_{n,1}(t)$.

The fact that all properties of the $N$-atom problem can be expressed in terms of the properties of only the first particle drastically simplifies the problem. A straightforward calculation based on Eq.\ \eqref{eta-n-Q-plane-wave} yields for an uncorrelated initial state and $N\gg1$
\begin{align}
\label{eta-C-plane-wave}
\frac{\eta(t)}{\eta_0}
= |C(t)|^2
\end{align}
with
\begin{align}
\label{C-def}
C(t)
= \sum_n p_n Q_n(t)
.\end{align}
This is a complex number, which we call the coherence. Note that $Q_n(0)= 1$ for all $n$ implies
\begin{align}
\label{C(0)=1-plane-wave}
C(0)
= 1
\end{align}
as long as $u(\bm x)$ is a plane wave. Inserting $Q_n(t)$ from Eq.\ \eqref{Q-def} yields
\begin{align}
\label{C-URUR}
C(t)
&
= \sum_n p_n \langle\psi_{g,n}(0)|U_g^\dag(t) RU_r(t)R^\dag|\psi_{g,n}(0)\rangle
\notag\\ &
= \tr[\rho_\text{in} U_g^\dag(t) RU_r(t)R^\dag]
.\end{align}

Before proceeding, we note for later use that if
\begin{align}
\label{R=1-Hg=Hr}
R
= \mathbbm 1
&&
\text{and}
&&
H_g
= H_r
,\end{align}
then Eq.\ \eqref{Q-def} obviously yields $Q_n(t)= 1$ for all $n,t$ and we obtain $\eta(t)/\eta_0= 1$ for all $t$.

In addition, we note for later use that there is a number of situations, in which the problem separates in Cartesian coordinates. Specifically, if the time-evolution operator fulfills $U_g(\bm p,\bm x)= U_{g,x}(p_x,x) \linebreak[1] U_{g,y}(p_y,y) \linebreak[1] U_{g,z}(p_z,z)$ and if an analogous statement holds for $U_r(\bm p,\bm x)$, $v(\bm x)$, and $\psi_{g,n}(\bm x,0)$ for all $n$ and if the probabilities fulfill $p_n= p_{n_x} \linebreak[1] p_{n_y} \linebreak[1] p_{n_z}$ for all $n$, then according to Eq.\ \eqref{Q-def} $Q_{n_x,n_y,n_z}(\bm x)= Q_{n_x,x}(x) \linebreak[1] Q_{n_y,y}(y) \linebreak[1] Q_{n_z,z}(z)$ and according to Eqs.\ \eqref{eta-C-plane-wave} and \eqref{C-def}
\begin{align}
\label{separate}
\frac{\eta(t)}{\eta_0}
= \frac{\eta_x(t)}{\eta_{x,0}} \frac{\eta_y(t)}{\eta_{y,0}} \frac{\eta_z(t)}{\eta_{z,0}}
.\end{align}

\subsection{Same Ground-State Hamiltonian before and during the Dark Time}

\label{sec-Hg-rho}

To further simplify the model, we assume that the initial $N$-atom density matrix $\rho_{N,\text{in}}$ commutes with the dark-time Hamiltonian $\mathcal H_g= \sum_{i=1}^N H_{g,i}$ for $N$ atoms in internal state $|g\rangle$
\begin{align}
\label{Hg-rho-commute}
[\mathcal H_g,\rho_{N,\text{in}}]
= 0
.\end{align}
This equation holds e.g.\ in the frequently encountered situation in which $\rho_{N,\text{in}}$ is in thermal equilibrium before storage and the ground-state Hamiltonian $\mathcal H_g$ is identical before storage and during the dark time.

Eq.\ \eqref{Hg-rho-commute} implies that all the $|\psi_{g,n,i}(0)\rangle$ can be chosen such that they are eigenstates of $H_{g,i}$. Let the corresponding eigenvalues be denoted as $E_{g,n,i}$. This yields
\begin{align}
|\psi_{g,n,i}(t)\rangle
= e^{-iE_{g,n,i}t/\hbar} |\psi_{g,n,i}(0)\rangle
\end{align}
so that using Eq.\ \eqref{R-def}, we find that Eq.\ \eqref{Q-def} simplifies to
\begin{align}
\label{Q-n-commuting}
Q_{n,i}(t)
= e^{iE_{g,n,i}t/\hbar} \langle\psi_{r,n,i}(0)|\psi_{r,n,i}(t)\rangle
.\end{align}
The corresponding Eq.\ \eqref{eta-n-fidelity-Phi-Psi-r} for $N$-atom states simplifies to
\begin{align}
\label{eta-n-fidelity-Psi-r(0)-Psi-r(t)}
\frac{\eta_n(t)}{\eta_0}
= \left|\langle \Psi_{r,n}(0)|\Psi_{r,n}(t)\rangle\right|^2
.\end{align}
The last equation has been used previously e.g.\ in Refs.\ \cite{Zhao:Pan:08, Baur:phd, Mirgorodskiy:17} without much justification.

A particularly simple example is obtained if additionally the $|\psi_{r,n,i}(0)\rangle$ for all $n$ are eigenstates of the dark-time Hamiltonian $H_{r,i}$ with eigenvalues $E_{r,n,i}$. This yields
\begin{align}
\label{psi-r-n(t)-simple}
|\psi_{r,n,i}(t)\rangle
= e^{-iE_{r,n,i}t/\hbar} |\psi_{r,n,i}(0)\rangle
\end{align}
so that Eq.\ \eqref{Q-n-commuting} simplifies to
\begin{align}
\label{Q-n-simple}
Q_{n,i}(t)
= e^{i(E_{g,n,i}-E_{r,n,i})t/\hbar} \langle\psi_{r,n,i}(0)|\psi_{r,n,i}(0)\rangle
.\end{align}
As long as $u(\bm x)$ is a plane wave $\langle\psi_{r,n,i}(0)|\psi_{r,n,i}(0)\rangle= 1$, as discussed above.

\subsection{Beyond a Plane-Wave Signal Light Field}

\label{sec-beyond-plane-waves}

The formalism discussed so far can be extended to situations in which the mode function $u(\bm x)$ of the signal light is not a plane wave, as detailed in appendix \ref{sec-app-beyond-plane-waves}. Here, we briefly summarize the central results of that treatment. This treatment is experimentally relevant because typically $w_r\approx w$, as pointed out in the context of Eq.\ \eqref{experiment-w-r-tau-F}. Hence, the finite signal beam waist $w$ is crucial for correctly modeling the radius $w_r$ of the part of the atom cloud which was transferred into state $|r\rangle$ during storage.

The quantities $|\psi_{g,n,i}(t)\rangle$, $|\psi_{r,n,i}(t)\rangle$, $R_i^\dag$, $Q_{n,i}(t)$, and $C(t)$ are still defined by Eqs.\ \eqref{psi-e-n-i(0)-psi-r-n-i(0)}, \eqref{R-def}, \eqref{psi-g-n-i(t)-psi-r-n-i(t)}, \eqref{Q-def}, and \eqref{C-def}. In addition to imprinting the phase factor which represents the net photon recoil, the operator $R_i^\dag$ now also imprints the finite beam waist $w$ onto the part of the atom cloud which was transferred into state $|r\rangle$ during storage. But now the operator $R_i^\dag$ is no longer unitary and the single-particle state $R_i^\dag|\psi_{g,n,i}(t)\rangle$ which appears in the definition \eqref{Q-def} of $Q_{n,i}(t)$ is no longer properly normalized. Instead, calculating its norm squared yields a dimensionless real number
\begin{align}
\label{M-n-i(t)}
M_{n,i}(t)
&
= \langle\psi_{g,n,i}(t)|R_iR_i^\dag|\psi_{g,n,i}(t)\rangle
\notag \\ &
= \mathcal V \int_{\mathcal V} d^3x |u(\bm x) \psi_{g,n,i}(\bm x,t)|^2
,\end{align}
which describes how well the mode $u(\bm x)$ overlaps with the atomic wave function $\psi_{g,n,i}(\bm x,t)$. Note that combination with Eq.\ \eqref{psi-r-R-psi-g} yields $\langle\psi_{r,n,i}(0)|\psi_{r,n,i}(0)\rangle= M_{n,i}(0)$.

As shown in appendix \ref{sec-app-beyond-plane-waves}, Eq.\ \eqref{eta-BEC-plane-wave} generalizes to
\begin{align}
\label{eta-BEC}
\frac{\eta_\text{BEC}(t)}{\eta_0}
= \frac{|Q(t)|^2}{M(0)M(t)}
.\end{align}
We turn to the uncorrelated state of Eq.\ \eqref{rho-factorize}. As shown in appendix \ref{sec-app-beyond-plane-waves}, Eq.\ \eqref{eta-C-plane-wave} generalizes to
\begin{align}
\label{eta-C}
\frac{\eta(t)}{\eta_0}
= \frac{|C(t)|^2}{\mu(0)\mu(t)}
,\end{align}
where
\begin{align}
\label{mu-def}
\mu(t)
= \sum_n p_n M_n(t)
= \mathcal V \int_{\mathcal V} d^3x |u(\bm x)|^2 \varrho_g(\bm x,t)
\end{align}
is the average of all the $M_n(t)$ and
\begin{align}
\label{varrho}
\varrho_g(\bm x,t)
= \sum_n p_n |\psi_{g,n}(\bm x,t)|^2
\end{align}
is the spatial density distribution of a single atom, normalized to $\int_{\mathcal V} d^3x \varrho_g(\bm x,t)= 1$. Note that $\mu(0)= C(0)$ according to appendix \ref{sec-app-beyond-plane-waves}.

If Eq.\ \eqref{Hg-rho-commute} holds, typically because the Hamiltonian is identical before and after storage, then according to appendix \ref{sec-app-beyond-plane-waves}, we obtain $\mu(t)= \mu(0)$ so that Eq.\ \eqref{eta-C} simplifies to
\begin{align}
\label{eta-C-no-mu}
\frac{\eta(t)}{\eta_0}
= \left| \frac{C(t)}{C(0)}\right|^2
\end{align}
and Eqs.\ \eqref{Q-n-commuting} and \eqref{Q-n-simple} remain unchanged, now with $\langle\psi_{r,n}(0)|\psi_{r,n}(0)\rangle= M_n(0)$ according to the text below Eq.\ \eqref{M-n-i(t)}.

\subsection{Relation to DLCZ Sources and Ramsey Spectroscopy}

\label{sec-DLCZ-Ramsey}

The central results of Secs.\ \ref{sec-storage}--\ref{sec-beyond-plane-waves} also apply to single-photon sources based on the DLCZ protocol \cite{Duan:01}. This is because the write pulse of a DLCZ source, while using a somewhat different mechanism, prepares the Dicke state $|\Psi_{r,n}(0)\rangle$ of Eq.\ \eqref{Psi-r-n(0)}. The subsequent time evolution during the dark time and during the DLCZ read pulse is largely identical to the dark time and the retrieval in EIT-based storage and retrieval. Hence, processes which cause the efficiency to decay as a function of dark time affect DLCZ sources and EIT-based storage and retrieval in the same way.

Furthermore, as shown in Ref.\ \cite{supplemental:model:experiment}, an appropriately designed Ramsey experiment with the purely kinetic dark-time Hamiltonian of Eq.\ \eqref{H-kin} has fringe visibility
\begin{align}
\label{V-C}
V(t)
= \left|\frac{C_1(t)}{C_1(0)}\right|
\end{align}
with $C_1(t)$ defined in Ref.\ \cite{supplemental:model:experiment}. This equation holds for a plane-wave signal beam $u(\bm x)$ and an arbitrary initial state. Alternatively, it also holds if the pulse area of the Ramsey pulses is small and $\rho_{N,\text{in}}$ commutes with the dark-time Hamiltonian. For an uncorrelated initial state or for Ramsey spectroscopy performed on a single atom $C_1(t)$ becomes identical to $C(t)$ from Eq.\ \eqref{C-URUR}.

For an uncorrelated initial state with a plane-wave signal beam, Eqs.\ \eqref{eta-C-plane-wave} and \eqref{C(0)=1-plane-wave} hold and combination with Eq.\ \eqref{V-C} and $C_1(t)= C(t)$ yields
\begin{align}
\label{eta-V}
\frac{\eta(t)}{\eta_0}
= V^2(t)
.\end{align}
Alternatively, if the initial state is uncorrelated, and $\rho_{N,\text{in}}$ commutes with the dark-time Hamiltonian, then Eq.\ \eqref{eta-C-no-mu} holds and combination with Eq.\ \eqref{V-C} and $C_1(t)= C(t)$ yields Eq.\ \eqref{eta-V}. Hence, in both of these situations, the analysis of the processes which cause the visibility in Ramsey spectroscopy and the efficiency in EIT-based storage and retrieval to decay as a function of dark time are equivalent. Studying whether this equivalence holds for other initial states is beyond the present scope.

\section{Applications of the Model}

\label{sec-applications}

In this section, we apply the above model to a few selected situations. The first situation, discussed in Sec.\ \ref{sec-recoil}, deals with the decay of $\eta$ caused by photon recoil during storage combined with thermal atomic motion. This situation is closely related to the spatial first-order coherence function of the gas, as pointed out in Sec.\ \ref{sec-g1}. In Sec.\ \ref{sec-Raman-Nath} we use the Raman-Nath approximation to derive an expression for the decay of $\eta$ resulting if atoms in states $|g\rangle$ and $|r\rangle$ experience different potentials $V_g(\bm x)$ and $V_r(\bm x)$ during the dark time. In Sec.\ \ref{sec-quadratic}, we apply this expression to a situation in which both $V_g(\bm x)$ and $V_r(\bm x)$ are harmonic and gravitational sag is taken into account.

\subsection{Photon Recoil and Thermal Motion}

\label{sec-recoil}

Here we discuss the decay of $\eta$ caused by nonzero total photon recoil $\hbar\bm k_R$ during storage combined with thermal atomic motion at a temperature $T$ far above quantum degeneracy. For simplicity, we assume that the EIT signal light mode $u(\bm x)$ is a plane wave. According to Eq.\ \eqref{C-def}, we only need to consider single-particle properties. The $|\psi_{g,n}(0)\rangle$ are the eigenstates of the Hamiltonian $H_g$ before storage. The probability of occupying the $n$th single-particle state is
\begin{align}
\label{Gibbs-measure}
p_n
= \frac1Z e^{-\beta E_{g,n}}
,\end{align}
where the normalization constant $Z= \sum_n e^{-\beta E_{g,n}}$ is the canonical partition function and $\beta= 1/k_BT$.

We assume that the single-atom Hamiltonian before and after storage is
\begin{align}
\label{H-kin}
H_g
= H_r
= \frac{\bm p^2}{2m}
,\end{align}
where $\bm p$ and $m$ are the momentum and the mass of the atom, i.e.\ the single-particle potentials vanish before and after storage
$V_g(\bm x)
= V_r(\bm x)
= 0
.$
Hence, the $|\psi_{g,n}(0)\rangle$ have the position representation
\begin{align}
\label{psi-g-plane-wave}
\psi_{g,n}(\bm x,0)
= \frac{e^{i\bm k_n\cdot\bm x}}{\sqrt{\mathcal V}}
\end{align}
with wave vectors $\bm k_n$ meeting periodic boundary conditions. The external states $\psi_{r,n}(\bm x,0)= e^{i(\bm k_n+\bm k_R)\cdot\bm x}/\sqrt{\mathcal V}$ created during storage are eigenstates of the dark-time Hamiltonian so that Eqs.\ \eqref{C-def} and \eqref{Q-n-simple} apply with
\begin{align}
E_{g,n}
= \frac{\hbar^2 k_n^2}{2m}
,&&
E_{r,n}
= \frac{\hbar^2 (\bm k_n+\bm k_R)^2}{2m}
.\end{align}

In Eq.\ \eqref{C-def}, $\sum_n p_n$ expresses the thermal average. For high enough temperature $T$ or for large enough quantization volume $\mathcal V$, we approximate the parameter $\bm k_n$ as continuous with probability density
\begin{align}
p(\bm k)
= \frac{e^{-k^2/2\sigma_k^2}}{(2\pi\sigma_k^2)^{3/2}}
,&&
\sigma_k= \frac{\sqrt{mk_BT}}\hbar= \frac{\sqrt{2\pi}}{\lambda_{dB}}
.\end{align}
Equation \eqref{C-def} with $\sum_n p_n$ approximated as $\int d^3k p(\bm k)$ yields $C(t)= e^{-i\hbar k_R^2t/2m} e^{-t^2/2\tau_R^2}$ so that Eq.\ \eqref{eta-C-plane-wave} yields (see also Refs.\ \cite{Zhao:Pan:08, Jenkins:12, Baur:phd})
\begin{align}
\label{tau-R}
\frac{\eta(t)}{\eta_0}
= e^{-t^2/\tau_R^2}
,&&
\tau_R
= \frac{1}{k_R\sigma_v}
,\end{align}
where $\sigma_v= \hbar\sigma_k/m= \sqrt{k_BT/m}$ is the 1d rms width of the thermal velocity distribution. $\eta(t)/\eta_0$ displays a Gaussian decay with $1/e$ time $\tau_R$. The expression for $\tau_R$ can be interpreted as the condition that the typical distance $\sigma_v \tau_R$ travelled because of thermal motion  equals the reduced wavelength $\lambda_R/2\pi= 1/k_R$ of the spin wave.

\subsection{Relation to the Spatial Coherence Function}

\label{sec-g1}

Alternatively, $\tau_R$ in Eq.\ \eqref{tau-R} can be written as (see also Ref.\ \cite{Baur:phd})
\begin{align}
\label{tau-R-v-R}
\tau_R
= \frac{\lambda_{dB}}{v_R\sqrt{2\pi}}
,\end{align}
where $\bm v_R= \hbar\bm k_R/m$ is the recoil velocity associated with $\bm k_R$. This can be interpreted as the condition that the distance $v_R \tau_R$ travelled because of the photon recoil equals the coherence length $l_c= \lambda_{dB}/\sqrt{2\pi}= 1/\sigma_k$ of the gas. The latter is obtained from the spatial first-order coherence function $g^{(1)}(r)$ which has the property \cite{naraschewski:99}
\begin{align}
|g^{(1)}(r)|^2
= e^{-2\pi r^2/\lambda_{dB}^2}
\end{align}
for a homogeneous, noninteracting gas with $T$ far above quantum degeneracy.

The appearance of $g^{(1)}(r)$ in this problem is not a coincidence. In fact, as shown in Ref.\ \cite{supplemental:model:experiment} the situation considered here with the purely kinetic dark-time Hamiltonian of Eq.\ \eqref{H-kin} and a plane-wave signal beam $u(\bm x)$ yields
\begin{align}
\label{eta-g1}
\frac{\eta(t)}{\eta_0}
= |g^{(1)}(v_R t)|^2
.\end{align}
As discussed in Ref.\ \cite{supplemental:model:experiment}, this holds for an uncorrelated state, such as a noninteracting pure BEC at $T=0$ or a thermalized gas with $T$ far above quantum degeneracy. In addition it holds for arbitrary separable pure initial states. For other initial states, there might be deviations from this relation, as alluded to in Ref.\ \cite{supplemental:model:experiment}, but a detailed study of such deviations is beyond the present scope. The idea that there is some relation between $\eta$ and spatial coherence has previously been discussed qualitatively e.g.\ in Refs.\ \cite{Ginsberg:07, Riedl:12}. But we are not aware of a previous derivation of Eq.\ \eqref{eta-g1}.

For a pure BEC, $\eta(t)$ decays on a time scale in which $v_Rt$ reaches the sample size, as observed e.g.\ in Refs.\ \cite{Ginsberg:07, Riedl:12}. A much shorter decay time in a gas with a temperature slightly above the critical temperature $T_C$ for BEC has been observed in Ref.\ \cite{Ginsberg:07}. For nonzero temperatures below $T_C$, the coexistence of a BEC and an uncondensed fraction lead to the observation of a bimodal decay \cite{Riedl:12}, qualitatively agreeing with the expectation for the spatial coherence function. However, to our knowledge, quantitative agreement with any model for such a bimodal decay of the retrieval efficiency has not been reported yet.

Equation \eqref{eta-V} relates the fringe visibility $V(t)$ in Ramsey spectroscopy to the efficiency $\eta(t)$ in EIT-based storage and retrieval. Combination with Eq.\ \eqref{eta-g1} suggests that $V(t)$ should be related to $|g^{(1)}(v_Rt)|$. Indeed, it is well known that there is some relation between Ramsey spectroscopy and first-order spatial coherence. For example, Ref.\ \cite{Hagley:99} used photon recoil in a Ramsey experiment to study the first-order spatial coherence of a BEC and later Ref.\ \cite{Navon:15} quantitatively derived and experimentally studied the relation between $g^{(1)}(v_Rt)$ and the population transferred after two \emph{resonant} Ramsey pulses. In Ref.\ \cite{supplemental:model:experiment}, we derive the more general relation
\begin{align}
V(t)
= |g^{(1)}(v_R t)|
\end{align}
for the fringe visibility $V$ in Ramsey spectroscopy. This holds for a noninteracting gas as long as the dark-time Hamiltonian is purely kinetic and $u(\bm x)$ is a plane wave. It applies to arbitrary initial states.

\subsection{Raman-Nath Approximation}

\label{sec-Raman-Nath}

We turn to a situation in which atoms in states $|g\rangle$ and $|r\rangle$ experience different potentials $V_g(\bm x)$ and $V_r(\bm x)$. The corresponding single-atom dark-time Hamiltonians reads
\begin{align}
H_g
= \frac{\bm p^2}{2m} +V_g(\bm x)
,&&
H_r
= \frac{\bm p^2}{2m}+V_r(\bm x)
.\end{align}
In our experiment, the potentials $V_g(\bm x)$ and $V_r(\bm x)$ are light shifts created if the dipole trap is left on during the dark time, but in general this could also be other potentials, e.g.\ inhomogeneous Zeeman shifts. Even if both potentials are approximated as harmonic, this is a nontrivial problem, some parts of which have previously been addressed e.g.\ in Refs.\ \cite{Kuhr:05, Zhao:Kuzmich:08, Yang:11, Jenkins:12, Afek:17, Lampen:18}.

In principle, we could calculate $C(t)$ using Eqs.\ \eqref{C-def} and \eqref{Q-n-commuting}. However, the resulting calculation will typically become nontrivial because the time evolution of $|\psi_{r,n}(t)\rangle$ is not given by a trivial phase factor as in Eq.\ \eqref{psi-r-n(t)-simple}. Hence, while Eqs.\ \eqref{C-def} and \eqref{Q-n-commuting} can be useful for tackling the problem numerically, that approach will typically not produce a simple analytic result.

Instead, we consider Eq.\ \eqref{C-URUR} and apply the Raman-Nath approximation \cite{Raman:33} during the dark time. Conceptually, this approximation means that we ignore the distance which an atom travels during the dark time. Technically, this approximation consists in replacing the kinetic-energy operator by a constant real number during the dark time, see e.g.\ Ref.\ \cite{Adams:94}. Hence, Eq.\ \eqref{C-URUR} becomes
\begin{multline}
\label{C-Raman-Nath-temp}
C(t)
= \sum_n p_n
e^{i(E_{\text{kin},g,n}-E_{\text{kin},r,n})t/\hbar}
\\ \times
\langle\psi_{g,n}(0)|e^{iV_gt/\hbar}R e^{-iV_rt/\hbar}R^\dag|\psi_{g,n}(0)\rangle
,\end{multline}
where $E_{\text{kin},g,n}$ and $ E_{\text{kin},r,n}$ denote the expectation value of the kinetic-energy operator calculate for the states $|\psi_{g,n}(0)\rangle$ and $R^\dag|\psi_{g,n}(0)\rangle$, respectively. The remaining operators $V_g$, $V_r$, and $R$ are diagonal in the position representation so that \begin{multline}
C(t)
= \sum_n p_n
e^{i(E_{\text{kin},g,n}-E_{\text{kin},r,n})t/\hbar}
\\ \times
\mathcal V \int d^3x |v(\bm x)\psi_{g,n}(\bm x,0)|^2 e^{i[V_g(\bm x)-V_r(\bm x)]t/\hbar}
.\end{multline}
We assume that for those states $|\psi_{g,n}(0)\rangle$, which contribute noticeably to the thermal average, we can approximate $E_{\text{kin},g,n}= E_{\text{kin},r,n}$ because the typical kinetic energy in state $|g\rangle$ exceeds other effects, namely the kinetic energy associated with the total photon recoil and with the finite signal-beam waist because of the position-momentum uncertainty relation. Hence
\begin{align}
\label{C-Raman-Nath}
C(t)
= \mathcal V \int d^3x \varrho_g(\bm x,0) |v(\bm x)|^2 e^{i[V_g(\bm x)-V_r(\bm x)]t/\hbar}
,\end{align}
with $\varrho_g(\bm x,t)$ from Eq.\ \eqref{varrho}. The efficiency $\eta(t)$ is calculated by inserting $\mu(t)$ from Eq.\ \eqref{mu-def} and $C(t)$ from Eq.\ \eqref{C-Raman-Nath} into Eq.\ \eqref{eta-C}. The energy eigenstates of the potential $V_g(\bm x)$ no longer appear individually in Eqs.\ \eqref{mu-def} and \eqref{C-Raman-Nath}. Instead, only $\varrho_g(\bm x,t)$ appears, thus often allowing for a simple analytic solution.

To illustrate the plausibility of this result, we consider a Ramsey experiment in which the atoms are subject to a differential potential. For high enough temperature one can regard the atoms as having classical trajectories. For short enough dark time $t$ between the two Ramsey pulses, one can assume that the position of each atom is time independent. In this situation, an atom at position $\bm x$ will pick up a differential phase factor $e^{i[V_g(\bm x)-V_r(x)]t/\hbar}$. Averaging over the atomic positions yields a Ramsey pattern with fringe visibility $V(t)= |\int d^3x \varrho_g(\bm x)e^{i[V_g(\bm x)-V_r(x)]t/\hbar}|$. Obviously, this is closely related to Eq.\ \eqref{C-Raman-Nath} when keeping Eq.\ \eqref{V-C} in mind.

\subsection{Harmonic Potential and Gravitational Sag}

\label{sec-quadratic}

We now apply the Raman-Nath approximation to a situation in which both potentials $V_g(\bm x)$ and $V_r(\bm x)$ can be approximated as harmonic and in which $V_g(\bm x)$ is identical before and after storage and the system is in thermal equilibrium with $T$ far above quantum degeneracy before storage. As previously discussed in Sec.\ \ref{sec-exper-trap}, gravitational sag of the cloud in the trapping potential is a crucial aspect because it causes the position dependence of the differential light shift at cloud center to be linear, which leads to much faster decay of the retrieval efficiency compared to a purely quadratic differential potential.

We assume that both potentials $V_g(\bm x)$ and $V_r(\bm x)$ are light-shift potentials created by a travelling Gaussian light beam with wave vector along the $z$ axis used for dipole trapping the ground state. We obtain light-shift potentials \cite{grimm:00} $V_{j,\text{light}}(\bm x)= -\alpha_jI(\bm x)/2\epsilon_0c$, where $\alpha_j$ is the dynamical polarizability of state $j\in\{g,r\}$, $I(\bm x)$ the trapping-light intensity, $\epsilon_0$ the vacuum permittivity, and $c$ the vacuum speed of light. A harmonic approximation around the trap center yields $V_{j,\text{light}}(\bm x)= V_{j,0}+\kappa_j (x^2+y^2)/2$, where $V_{j,0}= V_{j,\text{light}}(0)$ is the peak value of the trapping potential, $\kappa_j= -4V_{j,0}/w_t^2$ the spring constant, and $w_t$ the beam waist of the dipole trapping beam. Here, we neglected the divergence of the Gaussian dipole-trapping beam.

Adding the gravitation potential $mgx$, we obtain the total potentials $V_j(\bm x)= \kappa_j[(x+x_{j,s})^2+y^2]/2$, where $x_{j,s}= mg/\kappa_j$ is the gravitational sag and we used an interaction picture to reset the zero of energy for each internal state individually. We now reset the coordinate origin along the $x$ axis to obtain
\begin{subequations}
\begin{align}
V_g(\bm x)
&
= \frac{\kappa_g}2 (x^2+y^2)
,\\
V_r(\bm x)
&
= \frac{\kappa_r}2 \left((x-x_0)^2+y^2\right)
,\end{align}
\label{V-quadratic}%
\end{subequations}
where $x_0= x_{g,s}-x_{r,s}$ is the differential gravitational sag. Note that an atom in state $|r\rangle$ localized at $\bm x=0$ experiences a force
\begin{align}
\label{F-kappa}
F
= \kappa_r x_0
= \frac{\kappa_r-\kappa_g}{\kappa_g} mg
\end{align}
along $x$.

We assume that before storage the system is in thermal equilibrium at temperature $T$. This requires $\kappa_g\geq 0$, whereas $\kappa_r$ may have either sign. We assume that $T$ is far above quantum degeneracy. Hence, the single-particle atomic density is
\begin{align}
\varrho_g(\bm x,0)
= \frac{1}{2\pi \sigma_x^2L_z}e^{-(x^2+y^2)/2\sigma_x^2}
\end{align}
with $\sigma_x= (\beta\kappa_g)^{-1/2}$.

We assume that the signal-beam profile is Gaussian
\begin{align}
\label{u-Gauss}
u(\bm x)
= \frac{1}{\sqrt{\mathcal V_G}}e^{-(x^2+y^2)/w^2}e^{ik_sz}
,\end{align}
where $w$ is the beam waist and $\mathcal V_G= \pi w^2L_z/2$ a normalization factor with the dimension of a volume. Here, we consider a cuboidal quantization volume $\mathcal V$ with edge lengths $L_x$, $L_y$, and $L_z$ and assumed $w\ll L_x=L_y$. In addition, we neglected the divergence of the Gaussian signal beam, the curvature of the wave fronts, and the Gouy phase, all based on the assumption that $z_R\ll L_z$, where $z_R =k_sw^2/2$ is the Rayleigh length.

Using the Raman-Nath approximation \eqref{C-Raman-Nath} in this situation and combining it with Eq.\ \eqref{eta-C-no-mu} yields
\begin{align}
\label{eta-quadratic-displace}
\frac{\eta(t)}{\eta_0}
= \frac{1}{|\zeta_1|^2} \exp\left( - \frac{t^2}{\tau_F^2} \frac{1}{|\zeta_1|^2} \right)
,\end{align}
where we abbreviated
\begin{align}
\label{w-r-tau-F}
\tau_F
= \frac{2\hbar}{w_r|F|}
,&&
w_r
= \left( \frac{1}{4\sigma_x^2} +\frac{1}{w^2}\right)^{-1/2}
\end{align}
and
\begin{align}
\label{tau-kappa}
\zeta_1(t)
= 1- i \frac{t}{\tau_\kappa}
,&&
\tau_\kappa
= \frac{4\hbar}{w_r^2|\kappa_g-\kappa_r|}
.\end{align}
$w_r$ is the radius of the part of the atom cloud which is transferred to the state $|r\rangle$. If $w$ and $2\sigma_x$ differ by a large factor, then $w_r$ equals the smaller of these quantities. For later reference, we use $\kappa_r/\kappa_g= \alpha_r/\alpha_g$ and Eq.\ \eqref{F-kappa} to rewrite
\begin{align}
\label{tau-F-tau-kappa-alpha}
\tau_\kappa
= \frac{4\hbar}{w_r\kappa_g|1-\frac{\alpha_r}{\alpha_g}|}
,&&
\tau_F
= \frac{2\hbar}{mgw_r|1-\frac{\alpha_r}{\alpha_g}|}
.\end{align}

Note that taking the finite value of $w_r$ into account is crucial here, because in the limit $w_r\to \infty$, Eq.\ \eqref{tau-F-tau-kappa-alpha} yields $\tau_\kappa\to 0$ and $\tau_F\to 0$. To make this plausible, note that coherent directed retrieval will be possible if the momentum spread of a pure single-atom state $|\psi_{r,n}(0)\rangle$ is larger than the differential change in momentum experienced during the dark time in the presence of the potential. Otherwise, the single-atom states $|\psi_{r,n}(t)\rangle$ and $R^\dag|\psi_{g,n}(t)\rangle$ have poor overlap in momentum space, causing $Q_n(t)$ and $C(t)$ to vanish according to Eqs.\ \eqref{Q-def} and \eqref{C-def}. For a finite $w_r$, the position-momentum uncertainty relation enforces a nonzero momentum spread $2\hbar/w_r$ of each state $|\psi_{r,n}(0)\rangle$, thus causing a nonzero time for the decay of $|C(t)|$. Note that only the differential potential $V_r(\bm x)-V_g(\bm x)$ is relevant for the directed retrieval because only this appears in Eq.\ \eqref{C-Raman-Nath}. In an experiment, the finite value of $w_r$ has contributions from the initial size $\sigma_x$ of the ground-state sample and from the transverse beam profiles of the signal and coupling beams. In EIT storage experiments, the signal beam waist typically is the smallest of these length scales because otherwise one cannot achieve high storage efficiency. Note that this is why we approximate the coupling beam as a plane wave throughout this work.

The Raman-Nath approximation is a good approximation as long as the distance which an atom travels during the dark time is small. In a harmonic differential potential, the dark time must be short compared to an oscillation period in the differential potential
\begin{align}
t
\ll 2\pi \sqrt{\frac{m}{|\kappa_r-\kappa_g|}}
.\end{align}
In addition, the Raman-Nath approximation neglects the distances travelled because of the initial thermal velocity, the net photon recoil, and the kinetic energy associated with the finite signal-beam waist because of the position-momentum uncertainty relation. This additionally requires $t\ll w_r/2\sigma_v$, $t\ll w_r/2v_R$, and $t\ll mw_r^2/4\hbar$.

If the gravitational sag is much smaller than the cloud size $x_{g,s}\ll w_r/2$, then the gravitational sag has negligible effect. In this case, the model can be simplified by setting $F=0$. In general, for $F=0$ we would expect the differential potential to excite the monopole mode, also known as the breathing mode, of the atomic cloud. If the spring constant should be negative, then the breathing mode would have imaginary frequency corresponding to exponential decay or growth instead of an oscillation. The short-time behavior of this is captured by the Raman-Nath approximation.

For $F= 0$ Eq.\ \eqref{eta-quadratic-displace}, which relies on the Raman-Nath approximation, yields an algebraic decay
\begin{align}
\label{eta-tau-kappa}
\frac{\eta(t)}{\eta_0}
= \frac{1}{1+{t^2}/{\tau_\kappa^2}}
.\end{align}
A previous analysis of the $F=0$ scenario by Kuhr et al.\ \cite{Kuhr:05} used a different approximation, which holds in a parameter regime different from ours and yields a different result, as detailed in Ref.\ \cite{supplemental:model:experiment}.

To make the decay time $\tau_\kappa$ from Eq.\ \eqref{tau-kappa} appearing in Eq.\ \eqref{eta-tau-kappa} plausible, we consider an atom in state $|r\rangle$ which has a certain momentum and a certain position with $y= 0$ and arbitrary $x$ at the beginning of the dark time. According to Eq.\ \eqref{V-quadratic}, it experiences a differential force along the $x$ axis of
\begin{align}
\label{F-r-F-g}
-\partial_x(V_r-V_g)
= (\kappa_g-\kappa_r)x+F
\end{align}
with $F$ from Eq.\ \eqref{F-kappa}. After the dark time $t$, the differential atomic momentum has changed by $(\kappa_g-\kappa_r)xt$, where we used $F=0$ and assumed that $x$ is unchanged because of the Raman-Nath approximation. Equating the modulus of this momentum change with the momentum width $2\hbar/w_r$ of state $|\psi_{r,n}(0)\rangle$ and replacing $x$ by the typical value $w_r/2$ yields $t= \tau_\kappa$.

Conversely, if we assume that the gravitational sag is much larger than the cloud size, then the harmonic part of the differential potential has little effect and we expect the differential potential to excite the dipole mode, also known as the sloshing mode, of the atomic cloud. Again, the short-time behavior of this is captured by the Raman-Nath approximation.

The condition that the gravitational sag is much larger than the cloud size $w_r/2\ll x_{g,s}$ is equivalent to $\tau_F\ll \tau_\kappa$. This is plausible because, if the cloud size $w_r/2$ is much smaller than the gravitational sag $x_{g,s}$, then the atomic cloud in state $|r\rangle$ essentially experiences a constant force $F$, i.e.\ the term $\propto x$ in Eq.\ \eqref{F-r-F-g} in negligible. If we consider $\tau_F\ll \tau_\kappa$, then the initial decay of $\eta(t)/\eta_0$ in Eq.\ \eqref{eta-quadratic-displace} from unity to a value much smaller than unity will be well approximated by
\begin{align}
\label{eta-tau-F}
\frac{\eta(t)}{\eta_0}
= \exp\left( - \frac{t^2}{\tau_F^2} \right)
.\end{align}
This is a good approximation except for the long-time tail of the decay, which is often of little interest because here $\eta(t)/\eta_0\ll 1$ anyway.

To make the decay time $\tau_F$ appearing in Eq.\ \eqref{eta-tau-F} plausible, we consider an atom in state $|r\rangle$ which has a certain momentum at the beginning of the dark time. After the dark time, its momentum has changed by $Ft$. Equating the modulus of this with the momentum width $2\hbar/w_r$ of state $|\psi_{r,n}(0)\rangle$ yields $t= \tau_F$.

For the parameters of our experiment, the gravitational sag is larger than the size of the cloud so that the linear potential dominates and the atoms hardly sample the curvature of the differential potential. Hence, Eq.\ \eqref{eta-tau-F} is applicable and the quadratic potential has negligible effect. Neglecting the quadratic potential from the start drastically simplifies the original problem and makes it possible to solve the problem analytically without resorting to the Raman-Nath approximation. In particular, this allows it to take the two-photon recoil $\hbar\bm k_R$ and the nonzero initial temperature $T$ of the atomic cloud into account, which we neglected in the Raman-Nath approximation. This is detailed in Ref.\ \cite{supplemental:model:experiment}.

We note that it might be tempting to associate the in-trap decoherence with a Markovian process driven by random motion of atoms in the trap. However, that ansatz would predict an exponential decay in time, which is in conflict with Eq.\ \eqref{eta-quadratic-displace}. The Markovian ansatz is not a good approximation because the atomic motion is not truly randomized on a short enough time scale.

\section{Conclusions}

To conclude, we studied the dark-time decay of the retrieval efficiency for light stored using Rydberg EIT. We experimentally demonstrated a $1/e$ time of 30 $\mu$s in free expansion at low atomic density and low temperature. Our experimental data, both inside the dipole trap and in free expansion, agree well with a model which we presented and that showed that it bears analogies to DLCZ single-photon sources and to the decay of fringe visibility in Ramsey spectroscopy. We also experimentally studied the decay of the degree to which the retrieved light is coherent. The model suggests that the trap-induced part of the decay of the retrieval efficiency should become negligible when moving from the present red-detuned 1064 nm dipole trap to a blue-detuned 532 nm dipole trap. This prediction is promising for future experiments aiming for a Rydberg cavity gate.

Another interesting perspective would be to extend the theoretical and experimental studies regarding the spatial coherence function and Ramsey spectroscopy beyond uncorrelated states. Thermalized correlated states might be particularly interesting, i.e.\ a noninteracting ensemble of identical fermions at zero temperature or a noninteracting thermalized gas of bosons or fermions at a nonzero temperature which is not far above quantum degeneracy.

\acknowledgments

We thank Johannes Otterbach, Thomas Pohl, and Richard Schmidt for discussions. This work was supported by Deutsche Forschungsgemeinschaft under Germany's excellence strategy via Nanosystems Initiative Munich and Munich Center for Quantum Science and Technology and under priority program 1929 GiRyd. T.S.\ acknowledges support from Studienstiftung des deutschen Volkes.

\appendix

\section{Atom-Light Interaction Hamiltonian}

\label{sec-app-Hamiltonian}

In this appendix, we present details regarding the atom-light interaction Hamiltonian $\mathcal V_{al}$ appearing in Eq.\ \eqref{V-al-matrix}. We assume that a signal (coupling) light field with angular frequency $\omega_s>0$ ($\omega_c>0$) and single-photon detuning $\Delta_s= \omega_s-\omega_{eg}$ ($\Delta_c= \omega_c-\omega_{re}$) is near resonant with the $|g\rangle \leftrightarrow |e\rangle$ ($|e\rangle \leftrightarrow |r\rangle$) transition with dipole matrix element $d_{eg}$ ($d_{re}$) and atomic resonance angular frequency $\omega_{eg}$ ($\omega_{re}$). We describe the coupling light field as a classical, plane wave with $E_c(\bm x,t)= \frac12 E_{c,0}e^{-i\omega_ct+i\bm k_c\cdot\bm x}+\cc$ with wave vector $\bm k_c$, complex amplitude $E_{c,0}$, and Rabi frequency $\Omega_c= -d_{re}E_{c,0}/\hbar$.

The signal light field, however, must be quantized to obtain a useful description of EIT-based storage because it is crucial that an atomic excitation from state $|g\rangle$ to $|e\rangle$ has a back action onto the signal light field, reducing its photon number by one. We include only a single monochromatic optical mode of the signal field in our model. The operator describing its electric field is $\hat E(\bm x)= \hat E^{(+)}(\bm x)+\Hc$ where $\hat E^{(+)}(\bm x)= E_{\omega_s} \sqrt{\mathcal V}u(\bm x)\hat a_s$ would become the positive-frequency component if one used the Heisenberg picture, $E_{\omega_s}= \sqrt{\hbar\omega_s/2\epsilon_0\mathcal V}$ is the field amplitude, $\mathcal V$ the quantization volume, $u(\bm x)$ the spatial mode function normalized to $\int_{\mathcal V} d^3x |u(\bm x)|^2= 1$, and $\hat a_s$ the annihilation operator for a photon in this mode with bosonic commutation relation $[\hat a_s,\hat a_s^\dag]= 1$. Using $E_{\omega_s}$ one introduces $g_R= -d_{eg}E_{\omega_s}/\hbar$, which is half the vacuum Rabi frequency.

We use an interaction picture and the rotating-wave approximation. We find that the Hamiltonian for the $i$th atom contains a modified internal-energy term $H_{\text{int},i}= \hbar \Delta_s |g_{i}\rangle\langle g_{i}| - \hbar \Delta_c |r_{i}\rangle\langle r_{i}|$ together with the potential $V_{al,i}$ describing the atom-light interaction with position representation
\begin{multline}
\label{V-al-i}
V_{al,i}(\bm x_i)
=
\hbar g_R \sqrt{\mathcal V} u(\bm x_{i}) \hat a_s |e_{i}\rangle\langle g_{i}|
+ \frac\hbar2 \Omega_c e^{i\bm k_c\cdot\bm x_{i}}|r_{i}\rangle\langle e_{i}|
\\
+\Hc
\end{multline}
The total atom-light interaction Hamiltonian is $\mathcal V_{al}= \sum_{i=1}^N V_{al,i}$. Spontaneous emission from state $|e\rangle$ into modes orthogonal to $u(\bm x)$ is not included in $\mathcal V_{al}$. In the following we always assume that $g_R$ and $\Omega_c$ are real and that $\Delta_s= \Delta_c= 0$. Reference \cite{Fleischhauer:02} uses the notation $\Omega$, where we use $\Omega_c/2$.

Had we considered a $\Lambda$-type level scheme instead of the ladder-type level scheme, i.e.\ had state $|r\rangle$ an energy lower than the energy of state $|e\rangle$, then we would have to replace $\Omega_c\mapsto \Omega_c^*$ and along with it $\bm k_c\mapsto -\bm k_c$ and $\Delta_c\mapsto -\Delta_c$. Everything else would remain unchanged.

Note that, strictly speaking, the positive-frequency component of the quantized electric field is $\hat E^{(+)}(\bm x)= \sum_m E_{\omega_m} \sqrt{\mathcal V} u_m(\bm x)\hat a_m$, where the $u_m(\bm x)$ form an orthonormal basis of mode functions. Each $u_m(\bm x)$ must be a monochromatic solution to the wave equation in classical electrodynamics. The corresponding angular frequency is $\omega_m$ and this yields $E_{\omega_m}= \sqrt{\hbar\omega_m/2\epsilon_0\mathcal V}$. Additionally, each $u_m(\bm x)$ must meet some boundary condition imposed by the quantization volume. Apart from that, one can choose the $u_m(\bm x)$ in an arbitrary fashion. The $\hat a_m$ are the corresponding bosonic photon annihilation operators.

As in Ref.\ \cite{Fleischhauer:02} we keep only one term in this sum, namely the one with $u(\bm x)$. The special treatment of this mode is justified because upon retrieval this term gives rise to directed emission because of constructive interference. The collection of all other modes gives rise to undirected retrieval. As we are not interested in the fate of those photons emitted in an undirected fashion, we can trace over them. This leaves us with a corresponding term in the dissipator of a quantum master equation, describing the atomic decay with rate coefficient $\Gamma_e$ which accompanies the spontaneous emission into those modes.

As a further approximation, we use the resulting Hamiltonian with just one term even when $u(\bm x)$ is a Gaussian beam which solves the wave equation only in paraxial approximation and also meets the boundary condition only in an approximate fashion. This is justified as long as the waist is much larger than the wavelength of the signal light because otherwise the paraxial approximation becomes poor.

\section{Retrieval Efficiency}

\label{sec-app-retrieval}

In this appendix, we derive the central result \eqref{eta-n-Q-plane-wave} for the retrieval efficiency for a separable pure initial state. For brevity, we drop the thermal averaging index $n$ from the notation throughout this appendix. We start by constructing an explicit expression for the projector $\mathcal P_u$. To this end, we consider an orthonormal basis of external states of the $i$th atom $(|\phi_{1,i}\rangle, |\phi_{2,i}\rangle, |\phi_{3,i}\rangle, ...)$. Hence, the product states $\bigotimes_{i=1}^N |\phi_{j_i,i}\rangle$ form an orthonormal basis of the Hilbert space $\mathcal H_\text{ext}$ of all external $N$-atom states. Similarly, the product states
\begin{align}
|\Phi_{g,j_1,j_2,...,j_N},1_s\rangle
= |1_s\rangle \bigotimes_{i=1}^N |\phi_{j_i,i},g_i\rangle
\end{align}
form an orthonormal basis of the subspace of states with one signal photon and zero atomic excitations. Likewise, the states
$|\phi_{j_i,i},r_i\rangle \bigotimes_{i'=1 , i'\neq i}^N |\phi_{j_{i'},i'},g_{i'}\rangle$ form an orthonormal basis of the subspace of states with one Rydberg excitation. Note that there are $N$ options for which of the atoms is excited. Since $R_i$ is unitary, we can alternatively use the states
\begin{align}
\label{single-excitation-subspace-R}
\left(R_i^\dag |\phi_{j_i,i},r_i\rangle \right) \bigotimes_{\substack{i'=1 \\ i'\neq i}}^N |\phi_{j_{i'},i'},g_{i'}\rangle
\end{align}
as an orthonormal basis of the subspace of states with one Rydberg excitation. Again, there are $N$ options for which of the atoms is excited.

For a given $u(\bm x)$, each of the states $|\Phi_{g,j_1,j_2,...,j_N},1_s\rangle$ has a 3d invariant subspace with a single excitation associated with it in analogy to Sec.\ \ref{sec-polariton}. Within each of these 3d subspaces, there is a symmetric Dicke state with one Rydberg excitation
\begin{align}
\label{Phi-r-j1-...-jN}
|\Phi_{r,j_1,j_2,...,j_N}\rangle
= \frac1{\sqrt N} \sum_{i=1}^N \left(R_i^\dag |\phi_{j_i,i},r_i\rangle \right) \bigotimes_{\substack{i'=1 \\ i'\neq i}}^N |\phi_{j_{i'},i'},g_{i'}\rangle
.\end{align}
The states $|\Phi_{r,j_1,j_2,...,j_N}\rangle$ form an orthonormal basis of the subspace $W_u$. Hence
\begin{align}
\label{P-u-Phi-Phi}
\mathcal P_u
= \sum_{j_1,j_2,...,j_N} |\Phi_{r,j_1,j_2,...,j_N}\rangle \langle\Phi_{r,j_1,j_2,...,j_N}|
.\end{align}
Note that the dimension of the subspace $W_u$ is a factor of $N$ smaller than the dimension of the subspace spanned by the states in Eq.\ \eqref{single-excitation-subspace-R} because among the $N$ options for which of the atoms is excited, only one superposition is realized, namely the symmetric Dicke state which couples to mode $u(\bm x)$.

To determine the retrieval efficiency based on Eq.\ \eqref{eta-P-Psi-r}, we first need to calculate $\mathcal P_u|\Psi_r(t)\rangle$. We facilitate this calculation by choosing the orthonormal basis $(|\phi_{1,i}\rangle, |\phi_{2,i}\rangle, |\phi_{3,i}\rangle, ...)$ in a way that is adapted to the problem. Specifically, we construct this basis such that
\begin{align}
\label{phi-1-i}
|\phi_{1,i}\rangle
= |\psi_{g,i}(t)\rangle
\end{align}
and that $R_i|\psi_{r,i}(t)\rangle$ lies in the two-dimensional (2d) subspace spanned by $|\phi_{1,i}\rangle$ and $|\phi_{2,i}\rangle$. We denote the corresponding expansion coefficients as $Q_{i}(t)$ and $G_{i}(t)$ defined by
\begin{align}
\label{Q-G-def}
R_i|\psi_{r,i}(t)\rangle
= Q_{i}(t)|\phi_{1,i}\rangle + G_{i}(t)|\phi_{2,i}\rangle
.\end{align}
Note that this agrees with the definition of $Q_{i}(t)$ in Eq.\ \eqref{Q-def}. In addition, as $R_i$ is unitary and $|\psi_{r,i}(t)\rangle$ is properly normalized
\begin{align}
|Q_{i}|^2 + |G_{i}|^2
= 1
.\end{align}
Combining Eqs.\ \eqref{Psi-r-n(t)}, \eqref{Phi-r-j1-...-jN}, \eqref{phi-1-i}, and \eqref{Q-G-def} yields
\begin{multline}
\langle\Phi_{r,j_1,j_2,...,j_N}|\Psi_r(t)\rangle
\\
= \frac1N \sum_{i=1}^N
\underbrace{\langle\phi_{j_i,i}|R_i|\psi_{r,i}(t)\rangle}_{= Q_i\delta_{1,j_i}+G_i\delta_{2,j_i}}
\prod_{\substack{i'=1 \\ i'\neq i}}^N \underbrace{\langle\phi_{j_{i'},i'}|\psi_{g,i'}(t)\rangle}_{=\delta_{1,j_{i'}}}
.\end{multline}
Insertion into Eq.\ \eqref{P-u-Phi-Phi} yields
\begin{multline}
\label{P-u-Psi-r(t)}
\mathcal P_u|\Psi_r(t)\rangle
\\
= |\Phi_{r,1,...,1}\rangle \frac1N \sum_{i=1}^N Q_{i}
+ \sum_{i=1}^N |\Phi_{r,1,...,1,j_i=2,1,...,1}\rangle \frac1N G_{i}
.\end{multline}
Combination with Eq.\ \eqref{eta-P-Psi-r} yields
\begin{align}
\frac{\eta(t)}{\eta_0}
= \frac1{N^2} \left| \sum_{i=1}^N Q_{i} \right|^2 + \frac1{N^2} \sum_{i=1}^N |G_{i}|^2
\end{align}
For $N\gg 1$ this yields Eq.\ \eqref{eta-n-Q-plane-wave} because $\frac1{N^2} \sum_{i=1}^N |G_{i}|^2\leq \frac1N \max_i |G_{i}|^2= O(\frac1N)$. As a result, the only term in $\mathcal P_u|\Psi_r(t)\rangle$ which gives a non-negligible contribution to $\eta(t)$ for $N\gg1$ comes from $|\Phi_{r,1,...,1}\rangle$, which equals $|\Phi(t)\rangle$ from Eq.\ \eqref{Phi-n}. Hence, for $N\gg 1$
\begin{align}
\label{P-Psi-r-Phi}
\mathcal P_u|\Psi_r(t)\rangle
= |\Phi(t)\rangle \langle \Phi(t) |\Psi_r(t)\rangle
.\end{align}
Inserting this into Eq.\ \eqref{eta-P-Psi-r} yields Eq.\ \eqref{eta-n-fidelity-Phi-Psi-r}.

In principle, the state in Eq.\ \eqref{P-u-Psi-r(t)} is entangled and it would remain entangled during the adiabatic passage from $\vartheta= \pi/2$ to 0, but for $N\gg1$ we obtain Eq.\ \eqref{P-Psi-r-Phi}, which becomes a product state $|\Phi_{g,1,...,1},1_s\rangle$ during the adiabatic passage from $\vartheta= \pi/2$ to 0. Hence, as a byproduct of our calculation, we obtained the final $N$-atom state after retrieval, at least for those cases in which retrieval occurs into mode $u(\bm x)$.

In our experiment, we use an optical fiber for transverse mode selection of the retrieved light. However, we do not select the longitudinal mode. Within our model, this lack of longitudinal mode selection has no effect onto the efficiency. To see this, we use the fact that $u(\bm x)$ has a plane-wave type dependence on $z$ and assume that $V_g(\bm x)$ and $V_r(\bm x)$ are independent of $z$. Hence, starting from thermal equilibrium, the initial atomic wave function $\psi_{g,n,i}(\bm x,0)$ has a plane-wave type dependence on $z$. Hence, the longitudinal mode along $z$ is unchanged when calculating $U_{g,i}^\dag(t) R_i U_{r,i}(t)R_i^\dag|\psi_{g,i}(0)\rangle$, which is relevant for Eq.\ \eqref{Q-def}. In essence, this reflects conservation of linear momentum along $z$ because the momentum added during the transition $|g\rangle\to |r\rangle$ is removed in the transition $|r\rangle\to |g\rangle$.

The following consideration will show why a state orthogonal to $W_u$ does not couple to the mode $u(\bm x)$. We start by considering the state $|\Psi_{g,\text{in}},1_s\rangle$ of Eq.\ \eqref{Psi-g-n(0)}. Taking photon recoil for a given $u(\bm x)$ into account, it is obvious that a transition of the $i$th atom from internal state $|g_i\rangle$ to $|r_i\rangle$ must change its external state from $|\psi_{g,i}(0)\rangle$ to $|\psi_{r,i}(0)\rangle$. When considering $N$ atoms, there is a corresponding $N$-dimensional subspace with one Rydberg excitation. It is spanned by the orthonormal basis $|\psi_{r,i}(0),r_{i}\rangle \bigotimes_{i'=1, i'\neq i}^N |\psi_{g,i'}(0),g_{i'}\rangle$ with $i\in\{1,2,\dots,N\}$. However, $\mathcal V_{al}$ couples the mode $u(\bm x)$ only to the symmetric Dicke state of Eq.\ \eqref{Psi-r-n(0)}. The ($N-1$)-dimensional subspace orthogonal thereto does not couple to the mode $u(\bm x)$. For an arbitrary initial state, this statement generalizes to the fact that a state orthogonal to $W_u$ does not couple to the mode $u(\bm x)$.

The state after storage $|\Psi_{r}(0)\rangle$ is an element of $W_u$. The dark-time evolution rotates this state inside the subspace spanned by the states in Eq.\ \eqref{single-excitation-subspace-R}. Whenever $\eta(t)< \eta_0$, this rotated state $|\Psi_{r}(t)\rangle$ has a nonzero component orthogonal to $W_u$. When the coupling light is turned back on for retrieval, this component causes spontaneous emission into modes orthogonal to $u(\bm x)$. The fate of those excitations is beyond the scope of the present paper. They are simply regarded as lost.

\section{Entangled Initial States}

\label{sec-app-entangled}

This appendix discusses a straightforward generalization of the model to arbitrary initial states. Again, the initial density matrix can be expanded as in Eq.\ \eqref{rho-N} but now each pure state $|\Psi_{g,n,\text{in}}\rangle$ can be entangled, i.e.\  nonseparable. This expansion is possible because $\rho_{N,\text{in}}$ is Hermitian so that it can be diagonalized.

Hence, Eq.\ \eqref{eta-eta-n} still holds but we need to generalize the calculation of $\eta_n(t)$ to entangled pure states. To this end, we expand the entangled pure state $|\Psi_{g,n,\text{in}}\rangle$ in an orthonormal set of separable pure states $|\Psi_{g,n,m,\text{in}}\rangle$ as
\begin{align}
\label{entangled-Psi-g-n-expand}
|\Psi_{g,n,\text{in}}\rangle
= \sum_m c_{n,m} |\Psi_{g,n,m,\text{in}}\rangle
\end{align}
with
\begin{align}
\label{entangled-orthonormal-Psi}
\langle\Psi_{g,n,m,\text{in}}|\Psi_{g,n,m',\text{in}}\rangle
= \delta_{m,m'}
.\end{align}
Hence, the complex expansion coefficients $c_{n,m}$ fulfill $\sum_m |c_{n,m}|^2= 1$. This expansion is possible because any Hilbert space has an orthonormal basis of separable pure states.

The treatment of storage and retrieval for each separable pure initial state $|\Psi_{g,n,m,\text{in}}\rangle$ is identical to the above. In particular, Eq.\ \eqref{P-Psi-r-Phi}, which holds for $N\gg1$, now simply picks up an additional index $m$, thus becoming
\begin{align}
\label{P-Psi-r-Phi-entangled}
\mathcal P_u|\Psi_{r,n,m}(t)\rangle
= |\Phi_{n,m}(t)\rangle \langle \Phi_{n,m}(t) |\Psi_{r,n,m}(t)\rangle
,\end{align}
where $|\Psi_{r,n,m}(t)\rangle$ and $|\Phi_{n,m}(t)\rangle$ are defined in analogy to the above, starting from the separable pure state $|\Psi_{g,n,m,\text{in}}\rangle$. For clarity, Eq.\ \eqref{P-Psi-r-Phi-entangled} explicitly includes the index $n$, which was omitted in appendix \ref{sec-app-retrieval} for brevity. Note that $R_i$ is unitary and that the dark time evolution is unitary. Hence, Eq.\ \eqref{entangled-orthonormal-Psi} implies
\begin{align}
\label{entangled-orthonormal-Phi}
\langle\Phi_{n,m}(t)|\Phi_{n,m'}(t)\rangle
= \delta_{m,m'}
.\end{align}

The discussion leading to Eq.\ \eqref{eta-P-Psi-r} relies on adiabatic following but it is irrelevant whether the initial pure state is entangled. Hence, Eq.\ \eqref{eta-P-Psi-r} is applicable here for calculating $\eta_n(t)$ for the possibly entangled pure state $|\Psi_{g,n,\text{in}}\rangle$. In addition, it is applicable for calculating $\eta_{n,m}(t)$ for the separable pure state $|\Psi_{g,n,m,\text{in}}\rangle$. Combining this with Eqs.\ \eqref{entangled-Psi-g-n-expand}, \eqref{P-Psi-r-Phi-entangled}, and \eqref{entangled-orthonormal-Phi} and with the fact that $\mathcal P_u$ is linear yields
\begin{align}
\label{entangled-average}
\eta_n(t)
= \sum_m |c_{n,m}|^2 \eta_{n,m}(t)
.\end{align}
This is the weighted average of the efficiencies $\eta_{n,m}(t)$.

For the special case of a thermalized initial state of noninteracting atoms, each $|\Psi_{g,n,\text{in}}\rangle$ is a bosonically symmetrized or fermionically anti-symmetrized version of one separable pure state. Hence, all the values of $\eta_{n,m}(t)$ for a given $n$ are independent of $m$ and the averaging in Eq.\ \eqref{entangled-average} becomes trivial.

There is a caveat here. If one considers a degenerate Fermi gas, which is an example of an entangled state, then Eq.\ \eqref{eta-P-Psi-r} might become a poor approximation for the retrieval efficiency. This is because, if just before retrieval an external single-atom state is already occupied by a ground-state atom, then Pauli blocking will prohibit the transition of another atom to this final state. This might result in a suppression of the directed retrieval.

\section{Beyond Plane Waves}

\label{sec-app-beyond-plane-waves}

In this appendix, we present details regarding the generalization beyond a plane-wave signal light field, which was only briefly summarized in Sec.\ \ref{sec-beyond-plane-waves}. The main difference will be the appearance of several nontrivial normalization factors. Apart from that, the treatment is largely analogous. The main reason for discussing this aspect in an appendix is that the main text becomes more transparent because the relevant physics is not obscured by a lengthy discussion of all the normalization factors.

We assume that the mode function is of the form $u(\bm x)= u_\perp(x,y)e^{ik_sz}/\sqrt{L_z}$, in which the longitudinal mode function remains that of a plane wave but there might be some nontrivial transverse mode function $u_\perp(x,y)$, such as the Gaussian of Eq.\ \eqref{u-Gauss}. Hence, the longitudinal properties of the signal light remain trivial.

We still define the single-particle states $|\psi_{e,n,i}(0)\rangle$ and $|\psi_{r,n,i}(0)\rangle$ by Eq.\ \eqref{psi-e-n-i(0)-psi-r-n-i(0)}. Hence, Eq.\ \eqref{psi-r-R-psi-g} still holds. But now these single-particle states are no longer properly normalized. Instead, calculating their norm squared yields the dimensionless real number
\begin{align}
\label{M-n-i(0)-app}
M_{n,i}(0)
&
= \langle\psi_{e,n,i}(0)|\psi_{e,n,i}(0)\rangle
= \langle\psi_{r,n,i}(0)|\psi_{r,n,i}(0)\rangle
\notag \\ &
= \mathcal V \int_{\mathcal V} d^3x |u(\bm x) \psi_{g,n,i}(\bm x,0)|^2
,\end{align}
which describes how well the mode $u(\bm x)$ overlaps with the atomic wave function before storage $\psi_{g,n,i}(\bm x,0)$.

Using these states, we can easily generalize two of the $N$-atom states of Eq.\ \eqref{3d-subspace} to
\begin{subequations}
\begin{multline}
|\Psi_{e,n}(0)\rangle
=
\\
\frac1{\sqrt{N_n(0)}} \sum_{i=1}^N |\psi_{e,n,i}(0),e_{i}\rangle
\bigotimes_{\substack{i'=1\\ i'\neq i}}^N |\psi_{g,n,i'}(0),g_{i'}\rangle
\end{multline}
and
\begin{multline}
\label{psi-r-n(0)-not-plane-app}
|\Psi_{r,n}(0)\rangle
=
\\
\frac1{\sqrt{N_n(0)}} \sum_{i=1}^N |\psi_{r,n,i}(0),r_{i}\rangle
\bigotimes_{\substack{i'=1\\ i'\neq i}}^N |\psi_{g,n,i'}(0),g_{i'}\rangle
.\end{multline}
\label{3d-subspace-no-plane-wave-app}%
\end{subequations}
The only difference from Eq.\ \eqref{3d-subspace} is the appearance of the dimensionless normalization factor $N_n(0)$ instead of the atom number $N$. The definition of $|\Psi_{g,n,\text{in}},1_s\rangle$ remains unchanged. The $N$-atom states $|\Psi_{e,n}(0)\rangle$ and $|\Psi_{r,n}(0)\rangle$ in Eq.\ \eqref{3d-subspace-no-plane-wave-app} are properly normalized because we choose
\begin{align}
\label{N-n(0)-app}
N_{n}(0)
= \sum_{i=1}^N M_{n,i}(0)
,\end{align}
which can be regarded as the effective number of atoms coupled to the EIT signal light.

Compared to a plane wave $u(\bm x)$, the situation here is conceptually a little more subtle because the 3d subspace spanned by states $(|\Psi_{g,n,\text{in}},1_s\rangle, \linebreak[1] |\Psi_{e,n}(0)\rangle, \linebreak[1] |\Psi_{r,n}(0)\rangle)$ is no longer invariant under application of the atom-light interaction potential $\mathcal V_{al}$. This is because a transition $|g\rangle\to |e\rangle$ is accompanied by multiplication with $u(\bm x)\sqrt{\mathcal V}$ whereas the reverse transition $|e\rangle\to |g\rangle$ is accompanied by multiplication with $u^*(\bm x)\sqrt{\mathcal V}$. If and only if $u(\bm x)$ is a plane wave, then these factors cancel which makes the 3d subspace invariant. As argued in appendix \ref{sec-app-beyond-subspace}, we can safely ignore this subtlety and restrict our considerations to only this 3d subspace.

The state after storage is again $|\Psi_{r,n}(0)\rangle$. The dark time evolution proceeds as in the plane-wave case. In particular, Eq.\ \eqref{psi-g-n-i(t)-psi-r-n-i(t)} still holds and Eq.\ \eqref{Psi-r-n(t)} becomes
\begin{align}
|\Psi_{r,n}(t)\rangle
&
= \mathcal U_d(t) |\Psi_{r,n}(0)\rangle
\notag \\ &
= \frac1{\sqrt{N_n(0)}} \sum_{i=1}^N |\psi_{r,n,i}(t),r_{i}\rangle
\bigotimes_{\substack{i'=1\\ i'\neq i}}^N |\psi_{g,n,i'}(t),g_{i'}\rangle
.\end{align}
Again, this differs from the plane-wave case only in the normalization factor $N_{n}(0)$. Note that the unitary time evolution during the dark time implies that the norm of each $|\psi_{r,n,i}(t)\rangle$ is time independent, which is why the normalization factor $N_n(0)$ appearing in $|\Psi_{r,n}(t)\rangle$ is also time independent.

The retrieval is also treated as in the plane-wave case. In particular, Eqs.\ \eqref{eta-P-Psi-r} and \eqref{eta-n-fidelity-Phi-Psi-r} remain unchanged and the Dicke state of Eq.\ \eqref{Phi-n} is generalized to
\begin{multline}
|\Phi_n(t)\rangle
\\
= \frac1{\sqrt{N_n(t)}} \sum_{i=1}^N \left(R_i^\dag|\psi_{g,n,i}(t),r_{i}\rangle \right)
\bigotimes_{\substack{i'=1\\ i'\neq i}}^N |\psi_{g,n,i'}(t),g_{i'}\rangle
.\end{multline}
Again, the only difference from the plane-wave case is that the single-atom state $R_i^\dag|\psi_{g,n,i}(t)\rangle $ is not normalized, which yields the normalization factor for the $N$-atom state
\begin{align}
N_{n}(t)
= \sum_{i=1}^N M_{n,i}(t)
\end{align}
with the dimensionless real number
\begin{align}
M_{n,i}(t)
&
= \langle\psi_{g,n,i}(t)|R_iR_i^\dag|\psi_{g,n,i}(t)\rangle
\notag \\ &
= \mathcal V \int_{\mathcal V} d^3x |u(\bm x) \psi_{g,n,i}(\bm x,t)|^2
.\end{align}
In the limit $t\to0$, the last three equations coincide with Eqs.\ \eqref{psi-r-n(0)-not-plane-app}, \eqref{N-n(0)-app}, and \eqref{M-n-i(0)-app}. Hence Eq.\ \eqref{eta-n-fidelity-Phi-Psi-r} yields $\eta_n(0)/\eta_0= 1$. In addition, note that $0\leq M_{n,i}(t)\leq 1$. For the special case in which $u(\bm x)$ or $\psi_{g,n,i}(\bm x,t)$ is a plane wave, we obtain $M_{n,i}(t)= 1$ and if this applies to all $i$, then $N_{n}(t)= N$.

We also find that Eq.\ \eqref{eta-n-Q-plane-wave} is generalized to
\begin{align}
\label{eta-n-Q-app}
\frac{\eta_n(t)}{\eta_0}
= \frac{1}{N_n(0)N_n(t)} \left| \sum_{i=1}^N Q_{n,i}(t) \right|^2
\end{align}
and Eq.\ \eqref{Q-def} remains unchanged. In particular, for $t= 0$ we combine Eqs.\ \eqref{psi-r-R-psi-g}, \eqref{Q-def}, and \eqref{M-n-i(0)-app} to obtain
\begin{align}
\label{Q(0)-M(0)-app}
Q_{n,i}(0)
= M_{n,i}(0)
,\end{align}
which again yields $\eta_n(0)/\eta_0= 1$.

For a pure BEC, Eq.\ \eqref{eta-n-Q-app} simplifies to
\begin{align}
\label{eta-BEC-app}
\frac{\eta_\text{BEC}(t)}{\eta_0}
= \frac{|Q(t)|^2}{M(0)M(t)}
.\end{align}
We turn to the uncorrelated state of Eq.\ \eqref{rho-factorize}. Wanting to average $\eta_n(t)$ over $n$, we are facing the difficulty that in $N_n(t)= \sum_{i=1}^N M_{n,i}(t)$ the sum over $i$ appears in the denominator in Eq.\ \eqref{eta-n-Q-app}. To solve this problem, we note that $N_n(t)= \sum_{i=1}^N M_{n,i}(t)$ for $N\gg1$ is the sum over a large number $N$ of uncorrelated random variables $M_{n,i}(t)$ which each have the same mean value
\begin{align}
\mu(t)
= \sum_n p_{n,i} M_{n,i}(t)
= \sum_n p_{n,1} M_{n,1}(t)
\end{align}
and the same standard deviation because the particles are identical. The central limit theorem states that $\frac1N N_n(t)$ is a normally-distributed, random variable with mean value $\mu(t)$ and a standard deviation of order $O(N^{-1/2})$. We consider $N\gg 1$ and neglect this standard deviation, i.e\ we replace $N_n(t)$ by $N\mu(t)$ and $N_n(0)$ by $N\mu(0)$. Hence, Eq.\ \eqref{eta-C-plane-wave} generalizes to
\begin{align}
\label{eta-C-app}
\frac{\eta(t)}{\eta_0}
= \frac{|C(t)|^2}{\mu(0)\mu(t)}
,\end{align}
where Eq.\ \eqref{C-def} remains unchanged. The quantities $C(t)$ and $\mu(t)$ are obtained from $Q_n(t)$ and $M_n(t)$ by averaging over $n$. As mentioned below Eq.\ \eqref{rho-single-atom-diagonalized}, a pure BEC is an example of an uncorrelated state. Hence, Eq.\ \eqref{eta-C-app} reproduces Eq.\ \eqref{eta-BEC-app} if the initial state is a pure BEC. Note that Eq.\ \eqref{Q(0)-M(0)-app} yields
\begin{align}
C(0)
= \mu(0)
.\end{align}

If Eq.\ \eqref{Hg-rho-commute} holds, i.e.\ typically because the Hamiltonian is identical before and after storage, then $M_{n,i}(t)= M_{n,i}(0)$ so that $\mu(t)= \mu(0)$ and
\begin{align}
\label{eta-C-no-mu-app}
\frac{\eta(t)}{\eta_0}
= \left| \frac{C(t)}{C(0)}\right|^2
\end{align}
and Eqs.\ \eqref{C-def}, \eqref{Q-n-commuting}, and \eqref{Q-n-simple} remain unchanged, now with $\langle\psi_{r,n,i}(0)|\psi_{r,n,i}(0)\rangle= M_n(0)$ according to Eq.\ \eqref{M-n-i(t)}.

\section{Non-Invariance of the Subspace}

\label{sec-app-beyond-subspace}

As mentioned in appendix \ref{sec-app-beyond-plane-waves}, if $u(\bm x)$ is not a plane wave, then the 3d subspace spanned by the states $(|\Psi_{g,n,\text{in}},1_s\rangle, \linebreak[1] |\Psi_{e,n}(0)\rangle, \linebreak[1] |\Psi_{r,n}(0)\rangle)$ is no longer invariant under application of the atom-light interaction potential $\mathcal V_{al}$. In this appendix, we argue why this is not a major concern.

As an example, we consider application of $\mathcal V_{al}$ to any vector in this 3d subspace. This will create a vector in a 4d subspace spanned e.g.\ by the above three vectors combined with
\begin{multline}
|\Psi_{g,n}^{(1)},1_s\rangle
= \frac{1}{\sqrt{N_{g,n,i}^{(1)}(0)}} |1_s\rangle
\sum_{i=1}^N R_iR_i^\dag|\psi_{g,n,i}(0), g_{i}\rangle
\\
\bigotimes_{\substack{i'=1\\i'\neq i}}^N |\psi_{g,n,i'}(0), g_{i'}\rangle
\end{multline}
with some normalization factor $N_{g,n,i}^{(1)}(0)$. Applying $\mathcal V_{al}$ to any vector in this 4d subspace yields a vector in a 5d subspace etc. In this way, we obtain an infinite hierarchy of subspaces.

If we consider the 4d subspace rather than the 3d subspace, then we obtain a negligible correction to the final result $\eta_n(t)$ of our calculation because the single-atom state of only one atom in $|\Psi_{g,n}^{(1)},1_s\rangle$ differs from $|\Psi_{g,n,\text{in}},1_s\rangle$. All other single-atom states are identical. Assuming that the number of coupled atoms is large $N_n(0)\gg 1$, this difference has negligible effect. The same applies if we extend the model to the 5d subspace and so forth, unless the number $k$ of applications of $V_{eg}= \sum_{i=1}^N \hbar g_R \sqrt{\mathcal V} u(\bm x_{i}) \hat a_s |e_{i}\rangle\langle g_{i}|+\Hc$ becomes comparable to $N_n(0)$.

Now, the number $k$ of applications of $V_{eg}$ that we need to consider is set by the duration $t_s$ of the storage pulse. The typical value of $k$ which needs to be taken into account is twice the number of $|g\rangle \leftrightarrow |e\rangle$ Rabi flops which the Bloch vector can undergo during time $t_s$. Let $\varphi_R= \sqrt{N_n(0)} g_Rt_s$ denote the pulse area. Then the number of Rabi flops is $\varphi_R/2\pi$. All states obtained by a much larger number of applications of $V_{eg}$ have negligible amplitude in the state after storage. To achieve adiabatic following in the dark state during the storage process, we need $1\ll \varphi_R/\pi$. But typically $N_n(0)$ is large enough that this leaves enough room to choose $t_s$ such that $1\ll \varphi_R/\pi \ll N_n(0)$. We assume that such a choice was made. Hence, it suffices to restrict $k$ to $k\approx \varphi_R/\pi \ll N_n(0)$. A similar argument applies to the adiabatic passage during retrieval.

\clearpage

\begin{widetext}
\section*{Supplemental Material}
\end{widetext}

\renewcommand{\thefigure}{S\arabic{figure}}
\setcounter{figure}{0}
\setcounter{section}{19}

\section*{I. Experimental Details}

\subsection*{A. Possible Reasons for the Offset in Fig.\ \ref{fig-tau-vs-T}}

Here, we list a few obvious candidates of physical effects that might, in principle, explain the offset observed in Fig.\ \ref{fig-tau-vs-T} together with reasons why these candidates can probably be excluded in our experiment. A possible deviation from the perfectly counterpropagating beam geometry would change the slope of the straight-line fit in Fig.\ \ref{fig-tau-vs-T}, not its offset. Differential light shifts should vanish because all light fields are off during the dark time. Phase fluctuations of the EIT coupling light and homogenous stray electric fields should not affect the dark-time decay of $\eta$. Interactions of the Rydberg atom with a surrounding ground-state atom can be excluded because we experimentally verified that $\tau$ does not improve when lowering the density of ground-state atoms even further.

After this qualitative discussion, we now turn to quantitative arguments. According to Eqs.\ \eqref{tau-exp-approx} and \eqref{eta-tau-exp-2d}, the release of the atomic gas from a 2d harmonic trap shortly before storage is expected to cause a decay of $\eta$ with $1/e$ time $\sqrt{e-1}\,\tau_\text{rel} \approx 1.3\, \tau_\text{rel} $ where $\tau_\text{rel} = w\sqrt{m/k_BT}= 1.8$ ms at $T= 0.2$ $\mu$K. In the zero-temperature limit, Eq.\ \eqref{tau-exp} yields $\tau_\text{rel}= \tau_w= mw^2/\hbar= 88$ ms. This decay is much too slow to explain the offset observed in Fig.\ \ref{fig-tau-vs-T}. The radiative $1/e$ population lifetimes of the Rydberg states in a room-temperature environment are expected to range from 65 $\mu$s for the $50S$ state to 130 $\mu$s for the $70S$ state \cite{Saffman:05}. According to Eq.\ \eqref{eta-gamma-rg}, this causes an exponential decay of $\eta(t)$ with the same $1/e$ time, which is too long to explain the offset observed in Fig.\ \ref{fig-tau-vs-T}. With an efficiency of $\eta(0)$ of typically 10 to 20\%, typically one incoming photon on average, and a length of the medium of typically 0.4 mm, we expect the probability that two Rydberg excitations are stored at a relative distance short enough that they can significantly interact with each other to be negligible.

Finally, any effects relating to Rydberg level shifts from stray electric fields, interactions of a Rydberg atom with another atom in the ground or Rydberg state, or the radiative population lifetime of the Rydberg state should all exhibit a strong dependence on $n$. But the experimental data show no discernable dependence on $n$.

\subsection*{B. Photoionization by the Dipole Trapping Light}

In principle, the faster decay of $\eta$ in the presence of the dipole trap observed in Fig.\ \ref{fig-trap} could also be caused by photoionization of the Rydberg state by the trapping light. However, we will show now that this is not the case for the parameters of our experiment.

According to Ref.\ \cite{Saffman:05} the photoionization cross section decreases with increasing principal quantum number $n$ and has a non-monotonic dependence on the orbital angular momentum with the maximum occurring for $D$ states. As fairly small electric fields might cause mixing between orbital angular momentum states, we use $\sigma_{50D}=1.2$ pm$^2$ \cite{Saffman:05} as a worst-case value for the cross section which will drastically overestimate the actual 1064-nm photoionization cross section for the data in Fig.\ \ref{fig-trap}. For the parameters of our 1064 nm dipole trap with a trap depth of $k_B\times18$ $\mu$K, we estimate that photoionization alone should cause a $1/e$ lifetime of 1.3 ms for population in the internal state $|r\rangle$. According to Eq.\ \eqref{eta-gamma-rg}, $\eta(t)$ should decay exponentially with the same $1/e$ time. This effect is negligible in Fig.\ \ref{fig-trap}.

In principle, photoionization caused by the 532 nm light sheets could also be relevant, despite the fact that the atoms are located at positions where the intensity of this light is small. To test this, we repeated the free-expansion measurements of Fig.\ \ref{fig-trap} with the 532-nm light sheets on. This had no effect onto the observed $1/e$ decay time showing that the 532 nm light sheets are not the issue here.

\subsection*{C. Density Dependence of the Visibility}

The measurement of the visibility in Fig.\ \ref{fig-visibility} differs from Ref. \cite{Tiarks:19} in a variety of aspects. For example, the atomic density is much lower. The atomic ensemble is much longer. The reference light is not stored in an atomic ground-state memory. The data are taken after release from the dipole trap. The EIT signal input light pulse is longer. The intensity and beam waist of the EIT coupling light are different. The principal quantum number $n=70$ differs slightly from $n=69$ in Ref. \cite{Tiarks:19}. We improved the frequency locks for the lasers producing the EIT signal and coupling light. In principle, the big improvement in the visibility measured in Fig.\ \ref{fig-visibility} compared to Ref. \cite{Tiarks:19} could depend on many of these aspects. We will show now that the atomic density is crucial.

To this end, we measured the visibility as a function of the peak atomic density for storage in the 70$S$ state. To change the density, we varied the separation of the centers of the light sheets between 0.09 and 0.43 mm. The atom number hardly varied, lying between $1.6\times10^5$ and $2.0\times10^5$. The temperature was fixed at 0.6 $\mu$K. Each light sheet was operated at a power between 190 and 250 mW. We estimate that this caused $L$ to vary between 0.05 and 0.39 mm. The dark time between storage and retrieval was chosen to be 5 $\mu$s. This choice was made because except for really low density, the retrieval efficiency oscillates as a function of dark time because of atom-atom interactions \cite{Baur:phd, Mirgorodskiy:17}. For the 70$S$ state the first revival of the efficiency occurs for $t=5$ $\mu$s.

\begin{figure}[!t]
\includegraphics[width=\columnwidth]{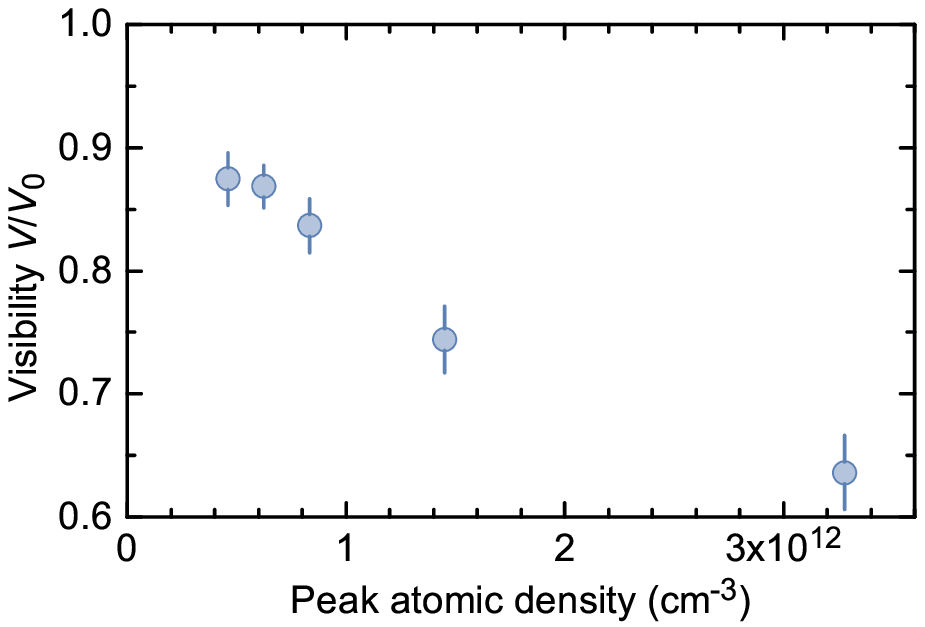}
\caption{Density dependence of the visibility for storage in the 70$S$ state. The experimental data were taken at a dark time of $t=5$ $\mu$s. Clearly, the visibility deteriorates for high density.}
\label{fig-visibility-vs-n}
\end{figure}

Results are shown in Fig.\ \ref{fig-visibility-vs-n}. Clearly, the visibility deteriorates when increasing the density. For comparison, note that our measurements in Ref.\ \cite{Tiarks:19} yielded $V= 66(2)\%$ at a peak density of $\varrho_0\approx 2\times10^{12}$ cm$^{-3}$, which agrees fairly well with the data in Fig.\ \ref{fig-visibility-vs-n}. This clearly shows that low atomic density is a necessary condition for obtaining high visibility, as in Fig.\ \ref{fig-visibility}.

\section*{II. Modelling Details}

\subsection*{A. Decay of Coherence between the Rydberg and Ground States}

Here we consider a decay of the off-diagonal element $\rho_{rg}(t)$ of the single-atom density matrix according to
\begin{align}
\label{gamma-rg}
\partial_t \rho_{rg}(t)
= -\frac12 \gamma_{rg} \rho_{rg}(t)
\end{align}
with rate coefficient $\gamma_{rg}$. This might be the result of population decay or of dephasing. We study how this causes a decay of $\eta(t)$. For a somewhat related treatment see e.g.\ Ref.\ \cite{Mewes:05}.

To do so, we first rewrite this in terms of a quantum master equation for the $N$-atom density matrix $\rho_N(t)$ with dissipator in Lindblad form
\begin{multline}
\label{master-equation}
\partial_t \rho_N(t)
=
\\
\frac12 \sum_k \gamma_k (2A_k \rho_N(t) A_k^\dag- A_k^\dag A_k \rho_N(t) - \rho_N(t) A_k^\dag A_k)
.\end{multline}
Here, the $A_k$ are Lindblad operators and the $\gamma_k$ decay rate constants. We ignore that there might be other contributions to the time evolution.

We first study dephasing. We assume that the sum in Eq.\ \eqref{master-equation} contains $N$ terms, one for each atom, with $A_i= |r_i\rangle\langle r_i|$ and that all rate coefficients $\gamma_i$ are identical. Comparison with Eq.\ \eqref{gamma-rg} yields $\gamma_i= \gamma_{rg}$.

We abbreviate the $N$-atom state $|\StateJ_{n,i}\rangle= |\psi_{r,n,i}(0),r_{i}\rangle \linebreak[1] \bigotimes_{i'=1,i'\neq i}^N |\psi_{g,n,i'}(0), g_{i'}\rangle$. According to Eq.\ \eqref{3d-subspace}, the $N$-atom state after storage is $|\Psi_{r,n}(0)\rangle= \sum_{i=1}^N |\StateJ_{n,i}\rangle/ \sqrt N$. Hence, the $N$-atom density matrix takes on the form $\rho_N(t)= \sum_n P_n \rho_{N,n}(t)$, where each component $\rho_{N,n}(t)$ separately evolves according to Eq.\ \eqref{master-equation} with the initial condition $\rho_{N,n}(0)= |\Psi_{r,n}(0)\rangle\langle \Psi_{r,n}(0)|$. One easily obtains $A_i |\StateJ_{n,i'}\rangle= \delta_{ii'} |\StateJ_{n,i'}\rangle$. We combine this with Eq.\ \eqref{master-equation} and $A_i^2= A_i^\dag= A_i$ to obtain
\begin{align}
\label{dt-StateJ}
\partial_t |\StateJ_{n,i}\rangle\langle \StateJ_{n,i'}|
= \gamma_{rg} (\delta_{ii'}-1) |\StateJ_{n,i}\rangle\langle \StateJ_{n,i'}|
\end{align}
and
\begin{align}
\rho_{N,n}(t)
= \frac1N \sum_{i=1}^N \sum_{i'=1}^N e^{(\delta_{ii'}-1)\gamma_{rg}t} |\StateJ_{n,i}\rangle \langle \StateJ_{n,i'}|
.\end{align}
Equation \eqref{eta-n-fidelity-Psi-r(0)-Psi-r(t)} generalizes to
\begin{align}
\frac{\eta_n(t)}{\eta_0}
= \langle \Psi_{r,n}(0)|\rho_{N,n}(t)|\Psi_{r,n}(0)\rangle
,\end{align}
which is, again, the fidelity of the initial and final state. We use $\langle \StateJ_{n,i}|\StateJ_{n,i'}\rangle= \delta_{ii'}$ and obtain
\begin{align}
\frac{\eta_n(t)}{\eta_0}
= \frac1N+ \left(1-\frac1N\right) e^{-\gamma_{rg}t}
.\end{align}
This is independent of $n$ because the dephasing mechanism acts only on the internal state, which is independent of $n$. We neglect terms of order $O(1/N)$ and perform thermal averaging over the index $n$. Hence
\begin{align}
\label{eta-gamma-rg}
\frac{\eta(t)}{\eta_0}
= e^{-\gamma_{rg}t}
.\end{align}

An analogous calculation for population decay can be based on $A_i= |f_i\rangle\langle r_i|$, where the population lost in state $|r_i\rangle$ reappears in a final internal state $|f_i\rangle$, which might equal $|g_i\rangle$ or $|e_i\rangle$ but is typically more likely to be yet another internal state. In any case, for population decay we assume $\langle r_i|f_i\rangle= 0$. Combination with Eq.\ \eqref{gamma-rg} again yields $\gamma_i= \gamma_{rg}$. In addition to the decay of $\rho_{rg}(t)$, this Lindblad operator causes decay of the population $\rho_{rr}(t)$ according to
\begin{align}
\partial_t \rho_{rr}(t)
= -\gamma_{rg} \rho_{rr}(t)
.\end{align}
A calculation largely analogous to the above yields instead of Eq.\ \eqref{dt-StateJ}
\begin{align}
\partial_t |\StateJ_{n,i}\rangle\langle \StateJ_{n,i'}|
= \gamma_{rg} (\delta_{ii'}|\StateF_{n,i}\rangle\langle \StateF_{n,i'}| -|\StateJ_{n,i}\rangle\langle \StateJ_{n,i'}|)
\end{align}
with the $N$-atom state $|\StateF_{n,i}\rangle= |\psi_{r,n,i}(0),f_{i}\rangle \linebreak[1] \bigotimes_{i'=1,i'\neq i}^N \linebreak[1] |\psi_{g,n,i'}(0), g_{i'}\rangle$. This calculation also produces Eq.\ \eqref{eta-gamma-rg}.

\subsection*{B. Linear Potential and Finite Signal Beam Waist}

For the parameters of our experiment, the gravitational sag is larger than the size of the cloud so that the linear potential dominates and the atoms hardly sample the curvature of the differential potential. Hence, Eq.\ \eqref{eta-tau-F} is applicable and the quadratic potential has negligible effect. Neglecting all quadratic terms in Eq.\ \eqref{V-quadratic}, we obtain
\begin{align}
\label{H-linear}
V_g(\bm x)
= 0
,&&
V_r(\bm x)
= - \bm F\cdot\bm x
,\end{align}
where $\bm F$ is a position-independent force.

The point which we address in the present section is that neglecting the quadratic potential from the start makes it possible to obtain an analytic solution without resorting to the Raman-Nath approximation. In particular, this allows it to take the two-photon recoil $\hbar\bm k_R$ and the nonzero initial temperature $T$ of the atomic cloud into account, both of which were neglected in the Raman-Nath approximation.

As discussed in Sec.\ \ref{sec-quadratic}, it is crucial that the finite spatial size $w_r$ of the cloud transferred to state $|r\rangle$ is taken into account and that in an experiment, this is typically dominantly limited by the signal beam waist $w$. Hence, this situation is well approximated when assuming that the atomic density distribution is homogeneous. That is equivalent to neglecting the transverse trapping potential before storage. Hence, the ground-state Hamiltonian is identical before and after storage, i.e.\ Eq.\ \eqref{Hg-rho-commute} holds. As a result, we can calculate $\eta(t)$ from Eqs.\ \eqref{C-def}, \eqref{Q-n-commuting}, and \eqref{eta-C-no-mu}. We assume that the signal beam has the Gaussian transverse mode of Eq.\ \eqref{u-Gauss}.

As we neglect the trapping potential before storage, the initial external states $|\psi_{g,n}(0)\rangle$ are the plane waves of Eq.\ \eqref{psi-g-plane-wave} with wave vector $\bm k_n$. But in contrast to Sec.\ \ref{sec-recoil} $|\psi_{r,n}(0)\rangle$ is now not an eigenstate of $H_r$. In momentum space, its time evolution can still be expressed relatively easily. This is because the Schr\"odinger equation in momentum representation $i\hbar\partial_t \psi(\bm k,t)= (\hbar^2\bm k^2/2m -i\bm F\cdot \nabla_{\bm k})\psi(\bm k,t)$ with $\nabla_{\bm k}=(\partial_{k_x},\partial_{k_y},\partial_{k_z})$ can be solved easily after changing coordinates from $(\bm k,t)$ to $(\bm k-\bm k_F,t)$ with $\bm k_F(t)= \bm Ft/\hbar$. This yields $\psi(\bm k,t)= \psi(\bm k-\bm k_F,0)\exp[-(k^2-\bm k\cdot\bm k_F+k_F^2/3)i\hbar t/2m]$ so that
\begin{multline}
\label{chi-ri-k}
\psi_{r,n}(\bm k,t)
= 2\sqrt{\frac{\mathcal V_G}{\mathcal V}}e^{-(k^2-\bm k\cdot\bm k_F)i\hbar t/2m+i\varphi_F}
\delta_{k_z,\nu_{n,z}+k_{F,z}}
\\ \times
e^{-w^2[(k_x-\nu_{n,x}-k_{F,x})^2+(k_y-\nu_{n,y}-k_{F,y})^2]/4}
,\end{multline}
where $\delta_{i,j}$ is the Kronecker symbol and we abbreviated $\bm \nu_n= \bm k_n+\bm k_R$ and $\varphi_F= -\hbar k_F^2t/6m$. Note that we explicitly use discrete values of $k_z$ here because of the finite quantization volume.

For $k_{F,z}\neq 0$ we obtain $\langle \psi_{r,n}(0)|\psi_{r,n}(t)\rangle= 0$ in this model because of the Kronecker symbol in Eq.\ \eqref{chi-ri-k}. This is because the cloud transferred to state $|r\rangle$ does not have a finite spatial size along $z$ resulting in an infinitely narrow initial momentum distribution along $z$. Hence, we consider $F_z= 0$.

To simplify the notation and without loss of generality, we rotate the coordinate system in the $xy$ plane to obtain $F_y=0$. Based on the assumption that $L_x= L_y$ are large, we approximate $k_x$ and $k_y$ as continuous and obtain
\begin{multline}
\!\!
\langle \psi_{r,n}(0)|\psi_{r,n}(t)\rangle
= \frac1{\zeta_0} e^{i\varphi_F}e^{-i\hbar \nu_{n,z}^2t/2m-w^2\nu_{n,y}^2(\zeta_0-1)/2\zeta_0}
\\ \times
e^{-w^2[\nu_{n,x}^2+(\nu_{n,x}+k_{F,x})^2]/4+w^2[2\nu_{n,x}+k_{F,x}(1+it/\tau_w)]^2/4\zeta_0}
,\end{multline}
where we abbreviated $\zeta_0(t)= 1+it/\tau_w$ and
\begin{align}
\label{tau-w}
\tau_w
= \frac{mw^2}\hbar
.\end{align}
Inserting this into Eqs.\ \eqref{C-def} and \eqref{Q-n-commuting} yields $C(t)$. The sum over $n$ is converted into an integral, as in Sec.\ \ref{sec-recoil}. A lengthy but straightforward calculation solves this integral and produces
\begin{multline}
C(t)
= \frac1{\zeta} e^{i\varphi_F} e^{-i\hbar {k_R,z}^2t/2m}
\\ \times
\exp\left( -\frac12 k_{R,z}^2\sigma_v^2t^2 -\frac18 w^2k_{F,x}^2\left(1-\frac{it}{\tau_w}\right)\right)
\\ \times
\exp\left(-\frac{\zeta-1}{2\zeta} w^2\left(k_{R,y}^2+(k_{R,x}+k_{F,x})^2\right) \right)
,\end{multline}
where we abbreviated
\begin{align}
\zeta(t)
= 1+ \frac{it}{\tau_w} + \frac{\sigma_v^2t^2}{w^2}
.\end{align}
Insertion into Eq.\ \eqref{eta-C-no-mu} yields
\begin{multline}
\label{C-sq-linear}
\frac{\eta(t)}{\eta_0}
= \frac{1}{|\zeta|^2} \exp\left(-t^2k_{R,z}^2\sigma_v^2-\frac14 w^2k_{F,x}^2\right)
\\ \times
\exp\left(-w^2\left(k_{R,y}^2+(k_{R,x}+k_{F,x})^2\right)\Re\frac{\zeta-1}\zeta \right)
.\end{multline}

In many experiments, the temperature is high enough that $w\sigma_k\gg 1$. This yields
\begin{align}
\label{zeta-high-T}
|\zeta(t)|
= \left(1+\frac{\sigma_v^2}{w^2} t^2\right)\left[1+O\left(\frac1{w^2\sigma_k^2}\right)\right]
\end{align}
and $\Re\frac{\zeta-1}{\zeta}= \frac{|\zeta|-1}{|\zeta|}[1+O(\frac1{w^2\sigma_k^2})]$.

As an aside, we note that a special case of Eq.\ \eqref{C-sq-linear} with $w\sigma_k\gg 1$ can be compared to the literature. To this end, consider the special case $\bm F=0$, which yields
\begin{align}
\label{Jenkins}
\frac{\eta(t)}{\eta_0}
= \frac{1}{|\zeta|^2} \exp\left( -t^2\sigma_v^2 \left(k_{R,z}^2+\frac1{|\zeta|}(k_{R,x}^2+k_{R,y}^2)\right)\right)
\end{align}
with $|\zeta|$ from Eq.\ \eqref{zeta-high-T}. For the special case $k_{R,z}^2\ll k_{R,x}^2+k_{R,y}^2$, Eq.\ \eqref{Jenkins} was previously derived in Ref.\ \cite{Jenkins:12}.

Our experiment is performed in a different regime, namely for $k_{R,x}= k_{R,y}= 0$ and $w\sigma_k\gg 1$. Hence, Eq.\ \eqref{C-sq-linear} simplifies to
\begin{align}
\label{C-sq-linear-kz=0}
\frac{\eta(t)}{\eta_0}
= \frac{1}{|\zeta|^2} \exp\left(-t^2k_{R,z}^2\sigma_v^2-\frac{t^2}{\tau_{F,\infty}^2} \left(5-\frac{4}{|\zeta|}\right)\right)
\end{align}
with $|\zeta|$ from Eq.\ \eqref{zeta-high-T}. Here, we abbreviated $\tau_{F,\infty}= 2\hbar/w|F|$.

In our experiment, we additionally have $\tau_{F,\infty}\ll w/\sigma_v$ and $\tau_{F,\infty}\ll 1/\sigma_vk_{R,z}$ so that the dominant decay in Eq.\ \eqref{C-sq-linear-kz=0} is obtained when approximating $\sigma_v= 0$ yielding
\begin{align}
\label{eta-tau-F-infty}
\frac{\eta(t)}{\eta_0}
= \exp\left(-\frac{t^2}{\tau_{F,\infty}^2} \right)
,&&
\tau_{F,\infty}
= \frac{2\hbar}{w|F|}
.\end{align}
This reproduces Eqs.\ \eqref{w-r-tau-F} and \eqref{eta-tau-F} with $\sigma_x\to \infty$ so that $w\to w_r$.

The added value of this calculation compared to Sec.\ \ref{sec-quadratic} consist in not having to use the Raman-Nath approximation to arrive at Eq.\ \eqref{eta-tau-F-infty}. In addition, the solution Eq.\ \eqref{C-sq-linear} contains other limiting cases. For example, if the fastest time scale in the decay is $\tau_w$, then the decay is governed by
\begin{align}
\label{eta-tau-w}
\frac{\eta(t)}{\eta_0}
= \frac1{1+t^2/\tau_w^2}
\end{align}
with $\tau_w$ from Eq.\ \eqref{tau-w}.  To make this time scale plausible, we consider the momentum width $2\hbar/w$ of the state $|\psi_{r,n}(0)\rangle$. The typical distance travelled after the dark time is $2\hbar t/w$. Equating this with the signal-beam rms radius $w/2$ yields $t= \tau_w/4$. This is the time scale on which the atoms disperse out of the signal beam because of the kinetic energy associated with the position-momentum uncertainty relation.

Another limit of Eq.\ \eqref{C-sq-linear} is obtained if $w/\sigma_v$ is the fastest time scale in the decay. Here, see also Ref.\ \cite{Jenkins:12}
\begin{align}
\label{C-sigma-v-w}
\frac{\eta(t)}{\eta_0}
= \frac1{(1+\sigma_v^2t^2/w^2)^2}
.\end{align}
The associated time scale obviously describes how atoms leave the signal beam because of thermal motion.

\subsection*{C. Relation to the Result of Kuhr et al.}

In this section, we discuss how Eq.\ \eqref{eta-tau-kappa} with $\tau_\kappa$ from Eq.\ \eqref{tau-kappa} relates to the result of Kuhr et al.\ \cite{Kuhr:05}. The conclusion will be that the Raman-Nath result \eqref{eta-tau-kappa} is applicable at high temperature, thus representing the semiclassical limit, whereas the result of Kuhr et al.\ requires $|\kappa_g-\kappa_r|/\kappa_g\ll 1$ combined with an intermediate temperature regime characterized by Eq.\ \eqref{Kuhr-intermediate-T}. Note that Kuhr et al.\ considered Ramsey spectroscopy, but as explained above the treatment of Ramsey spectroscopy is largely equivalent to EIT-based storage and retrieval.

Throughout this discussion, we assume that the photon recoil $\bm k_R$ vanishes and that the beam waist is infinite $w\to\infty$. The combination of these two assumptions is equivalent to
\begin{align}
\label{Kuhr-T}
R
= \mathbbm1
.\end{align}
In addition, we assume
\begin{align}
\label{Kuhr-beta-hbar-omega}
\hbar\omega_g
\ll k_BT
,\end{align}
where $\omega_g= \sqrt{\kappa_g/m}$ and $\omega_r= \sqrt{\kappa_r/m}$ are the angular frequency of the harmonic potentials experienced by atoms in states $|g\rangle$ and $|r\rangle$.

\subsubsection*{1. Approach of Kuhr et al.}

Kuhr et al.\ \cite{Kuhr:05} studied a 3d harmonic potential whereas Eq.\ \eqref{eta-tau-kappa} applies to a 2d harmonic potential. Hence, we start by generalizing the approach of Kuhr et al.\ to $d$ dimensions. In this process, we express their approach in our formalism.

The basic assumption in the argument of Kuhr et al.\ is that when making the transition from $|g\rangle$ to $|r\rangle$ the total energy of an atom changes according to
\begin{align}
\label{Kuhr-E-ri}
E_{r,n}
= \sqrt{\frac{\kappa_r}{\kappa_g} }E_{g,n}
.\end{align}
We will provide a justification for this assumption below. Using $R= \mathbbm1$ and approximating the Hamiltonians $H_g$ and $H_r$ by their expectation values $E_{g,n}$ and $E_{r,n}$, Eq.\ \eqref{Ramsey-C} for Ramsey spectroscopy simplifies to
\begin{align}
\label{Kuhr-C-e^E}
C(t)
= \sum_n p_n e^{i(E_{g,n}-E_{r,n})t/\hbar}
,\end{align}
which is identical to the combination of Eqs.\ \eqref{C-def} and \eqref{Q-n-simple} for EIT-based storage and retrieval.

Using $\hbar\omega_g\ll k_BT$ from Eq.\ \eqref{Kuhr-beta-hbar-omega}, we can approximate the sum over $n$ as an integral over $E_{g}$
\begin{align}
\label{Kuhr-C-E-gi}
C(t)
= \int_0^\infty dE_{g} p(E_{g})
\exp\left( \frac{iE_{g}t}{\hbar} \left(1-\sqrt{\frac{\kappa_r}{\kappa_g}}\right)\right)
,\end{align}
where we abbreviated
\begin{align}
\label{Kuhr-chi-2-dist-E-gi}
p(E_{g})
= \beta \frac{(\beta E_{g})^{d-1}}{\Gamma(d)} e^{-\beta E_{g}}
.\end{align}
Here $\Gamma$ denotes the gamma function. Obviously $p(E_{g})$ is the thermal distribution of $E_{g}$ for a $d$ dimensional harmonic trap.

As an aside, note that the replacement $\beta\to 1/2$ would convert $p(E_{g})$ into the $\chi^2$ distribution with $2d$ degrees of freedom. This is not a coincidence. The $\chi^2$ distribution with $\nu$ degrees of freedom describes a random variable of the form $\sum_{n=1}^\nu X_n^2$ where the $X_n$ are uncorrelated random variables which each have a standardized Gaussian distribution, where standardized means zero mean value and unit variance. The thermal distribution of the variable $2\beta E_{g}$ in $d$ dimensions is the sum of $d$ random variables $2\beta\frac m2\omega_g^2x_j^2$ and $d$ random variables $2\beta\frac1{2m}p_j^2$ which are uncorrelated and each have a standardized Gaussian distribution. Here, $j$ is the Cartesian coordinate index. This is why $p(E_{g})$ is identical to a $\chi^2$ distribution with $2d$ degrees of freedom when replacing $\beta\to 1/2$.

The integral in Eq.\ \eqref{Kuhr-C-E-gi} is easily solved, yielding
\begin{align}
\label{Kuhr-C-Kuhr}
C(t)
= \left(1-i\frac{t}{K}\right)^{-d}
\end{align}
with the typical time scale
\begin{align}
\label{Kuhr-K-kappa}
K
= \beta\hbar \left(1-\sqrt{\frac{\kappa_r}{\kappa_g}}\right)^{-1}
= \beta\hbar \frac{\omega_g}{\omega_g-\omega_r}
.\end{align}
Hence $|C(t)|^2= (1+t^2/K^2)^{-d}$. Following Kuhr et al.\ we abbreviate $\eta= (\kappa_g-\kappa_r)/\kappa_g$. Assuming $|\eta|\ll 1$, we obtain $(1-\sqrt{1-\eta})^{-1}\approx 2/\eta$ so that
\begin{align}
\label{Kuhr-K}
K
\approx
2\tau_\kappa
,&&
\tau_\kappa
= \beta\hbar \frac{\kappa_g}{\kappa_g-\kappa_r}
.\end{align}
For $d=3$, Eqs.\ \eqref{Kuhr-C-Kuhr} and \eqref{Kuhr-K} agree with the results of Kuhr et al.

Note that the dependence of Eq.\ \eqref{Kuhr-C-Kuhr} on $d$ is an example for the possibility to multiply 1d results to obtain results for higher dimensions, see Eq.\ \eqref{separate}.

\subsubsection*{2. Raman-Nath Approach}

To achieve a useful comparison with the Raman-Nath based result Eq.\ \eqref{eta-tau-kappa} we generalize the Raman-Nath approach to $d$ dimensions and restrict it to $w\to\infty$. In analogy to Eq.\ \eqref{eta-tau-kappa}, we obtain
\begin{align}
\label{Kuhr-Raman-Nath}
C(t)
= \left(1-i\frac{t}{\tau_\kappa}\right)^{-d/2}
\end{align}
with $\tau_\kappa$ from Eq.\ \eqref{Kuhr-K}. Hence $|C(t)|^2= (1+t^2/\tau_\kappa^2)^{-d/2}$, which agrees with Eqs.\ \eqref{tau-kappa} and \eqref{eta-tau-kappa} with $d=2$ and $w\to\infty$. In Eq.\ \eqref{tau-kappa} we used $|\kappa_g-\kappa_r|$ instead of $\kappa_g-\kappa_r$ to make $\tau_\kappa$ positive but that is only a matter of convention and not essential.

However, Eq.\ \eqref{Kuhr-Raman-Nath} obviously differs from Eqs.\ \eqref{Kuhr-C-Kuhr} and \eqref{Kuhr-K} obtained from the approach of Kuhr et al. Hence, even after generalizing the two approaches to $d$ dimensions their results still do not agree with each other.

Interestingly, the $1/e$ decay times for $|C(t)|^2$ obtained from the two approaches tend to be fairly similar. Eqs.\ \eqref{Kuhr-C-Kuhr} and \eqref{Kuhr-K} yield a $1/e$ time of $2\tau_\kappa\sqrt{e^{1/d}-1}$, whereas Eq.\ \eqref{Kuhr-Raman-Nath} yields $\tau_\kappa\sqrt{e^{2/d}-1}$. For example, for $d= 3$ we obtain $1.26\tau_\kappa$ and $0.97\tau_\kappa$. For $d=2$ or $d=1$ the two $1/e$ times are even more similar.

To reveal the origin of the difference between the two approaches, we first present an alternative semiclassical approach which will reproduce the Raman-Nath result. We choose a formulation which makes the analogy to our above formulation of the approach of Kuhr et al.\ clear. For simplicity, we discuss the example of Ramsey spectroscopy.

Our central assumption in this approach is that each Ramsey pulse is so short that the position and the momentum of the atom cannot change during a pulse. Note that this also makes use of our above assumption of zero photon recoil, see Eq.\ \eqref{Kuhr-T}. We decompose the total energies of the atom immediately before and after the first pulse according to $E_{g,n}= E_{\text{kin},g,n}+V_{g,n}$ and $E_{r,n}= E_{\text{kin},r,n}+V_{r,n}$ into the kinetic parts $E_{\text{kin},g,n}$ and $E_{\text{kin},r,n}$ and the potential parts $V_{g,n}$ and $V_{r,n}$. The unchanged atomic momentum implies $E_{\text{kin},r,n}= E_{\text{kin},g,n}$ and the unchanged atomic position implies $V_{r,n}= V_{g,n}{\kappa_r}/{\kappa_g}$. Together this reads
\begin{align}
\label{Kuhr-T-V-gi}
E_{\text{kin},r,n}
= E_{\text{kin},g,n}
,&&
V_{r,n}
= \frac{\kappa_r}{\kappa_g} V_{g,n}
\end{align}
Note how this differs from Eq.\ \eqref{Kuhr-E-ri}.

The rest of the calculation is strictly analogous to the above description of the approach of Kuhr et al. We insert the resulting expressions for $E_{g,n}$ and $E_{r,n}$ into Eq.\ \eqref{Kuhr-C-e^E}, which again relies on $R=\mathbbm 1$. The kinetic energies cancel, which is reminiscent of the Raman-Nath approximation. Again, using $\hbar\omega_g\ll k_BT$ from Eq.\ \eqref{Kuhr-beta-hbar-omega} we approximate the sum over $n$ as an integral over $V_{g}$ and obtain
\begin{align}
\label{Kuhr-C-V-gi}
C(t)
= \int_0^\infty dV_{g} p(V_{g})
\exp\left( \frac{iV_{g}t}{\hbar} \left(1-\frac{\kappa_r}{\kappa_g}\right)\right)
,\end{align}
where we abbreviated
\begin{align}
p(V_{g})
= \beta \frac{(\beta V_{g})^{\frac d2-1}}{\Gamma(\frac d2)} e^{-\beta V_{g}}
.\end{align}
Obviously $p(V_{g})$ is the thermal distribution of $V_{g}$ for a $d$ dimensional harmonic trap. Note that the thermal distribution of the variable $2\beta V_{g}$ is a $\chi^2$ distribution with $d$ degrees of freedom. The difference in the numbers of degrees of freedom compared to Eq.\ \eqref{Kuhr-chi-2-dist-E-gi} is easily understood because $E_{g}$ contains kinetic and potential energy terms yielding a total of $2d$ terms, whereas $V_{g}$ contains only potential energy terms yielding a total of $d$ terms.

The integral in Eq.\ \eqref{Kuhr-C-V-gi} is easily solved. It reproduces Eq.\ \eqref{Kuhr-Raman-Nath}. Hence, we re-derived the Reman-Nath result Eq.\ \eqref{Kuhr-Raman-Nath} with a method which is closely analogous to our above description of the approach of Kuhr et al. The difference boils down to starting from Eq.\ \eqref{Kuhr-E-ri} versus starting from Eq.\ \eqref{Kuhr-T-V-gi}. The rest of the method is identical. While we presented a reasonable justification for Eq.\ \eqref{Kuhr-T-V-gi}, we have not yet presented any argument supporting Eq.\ \eqref{Kuhr-E-ri}.

\subsubsection*{3. Refinement of the Approach of Kuhr et al.}

We will now give two arguments in support of the result of Kuhr et al. One will be a brief, intuitive but rough argument. The other will be a solid mathematical derivation, which will clarify in which parameter regime the approach of Kuhr et al.\ is applicable,

For simplicity, we restrict the discussion to the 1d harmonic potential. For higher dimensions one can use Eq.\ \eqref{separate}. Let $|\StateSingle_{a,n}\rangle$ denote the $n$th energy eigenstate of a 1d harmonic oscillator with oscillator length $a= \sqrt{\hbar/m\omega}$. The corresponding energy eigenvalues are $\hbar\omega(n+1/2)$. Note that the oscillator lengths $a_g$ and $a_r$ for states $|g\rangle$ and $|r\rangle$ typically differ. Let $\Delta a= a_r-a_g$ and assume $|\Delta a|\ll a_g$.

We now give an intuitive but rough argument in support of Eq.\ \eqref{Kuhr-E-ri}. Obviously if $\Delta a$ would vanish exactly, then $n$ would be conserved in the transition from $|g\rangle$ to $|r\rangle$. If we postulate without much justification that the conservation of $n$ holds to a good approximation even for small nonzero values of $\Delta a$ and use this to calculate the energies $E_{g,n}= \hbar\omega_g(n+1/2)$ and $E_{r,n}= \hbar\omega_r(n'+1/2)$, then we will immediately obtain Eq.\ \eqref{Kuhr-E-ri}.

We now turn to a solid mathematical derivation of Eq.\ \eqref{Kuhr-C-Kuhr}. It will reveal that the result of Kuhr et al.\ is applicable in an intermediate temperature regime.

As we assumed $R=\mathbbm 1$ we obtain
\begin{align}
\label{chi-ag-n}
|\psi_{r,n}(0)\rangle
= |\psi_{g,n}(0)\rangle
= |\StateSingle_{a_g,n}\rangle
.\end{align}
This result is exact, unlike Eqs.\ \eqref{Kuhr-E-ri} and \eqref{Kuhr-T-V-gi} which are obviously approximations because they describe classical particles.

As the $|\StateSingle_{a_r,n'}\rangle$ form an orthonormal basis of energy eigenstates of $H_r$, we can expand $|\psi_{r,n}(0)\rangle$ in this basis and time evolution simply yields
\begin{align}
|\psi_{r,n}(t)\rangle
= \sum_{n'=0}^\infty e^{-i\omega_r t(n'+1/2)} |\StateSingle_{a_r,n'}\rangle\langle\StateSingle_{a_r,n'}|\psi_{r,n}(0)\rangle
.\end{align}
Combination with Eqs.\ \eqref{C-def}, \eqref{Q-n-commuting}, and \eqref{chi-ag-n} yields
\begin{align}
\label{Kuhr-C-n-n'}
C(t)
= C_0 \sum_{n,n'=0}^\infty p_n
e^{it(\omega_g n-\omega_r n')} |\langle\StateSingle_{a_g,n}|\StateSingle_{a_r,n'}\rangle|^2
\end{align}
with $C_0= e^{it(\omega_g-\omega_r)/2}$, $p_n= (1-q) q^n$, and $q= e^{-\beta\hbar\omega_g}$. Hence, we need to determine $|\langle\StateSingle_{a_g,n}|\StateSingle_{a_r,n'}\rangle|^2$.

Using standard methods, one can show that
\begin{multline}
\partial_a |\StateSingle_{a,n'}\rangle
= \frac{\sqrt{(n'+1)(n'+2)}}{2a}|\StateSingle_{a,n'+2}\rangle
\\
-\frac{\sqrt{(n'-1)n'}}{2a}|\StateSingle_{a,n'-2}\rangle
.\end{multline}
From here, one easily obtains a similar expansion of $\partial_a^2 |\StateSingle_{a,n'}\rangle$ in the orthonormal basis of the $|\StateSingle_{a,n}\rangle$. Combining these results, we obtain the leading terms of the Taylor series for $|\StateSingle_{a_g+\Delta a,n'}\rangle$ neglecting terms of order $O(\Delta a^3)$. Inserting this series into $|\langle \StateSingle_{a_g,n}|\StateSingle_{a_g+\Delta a,n'}\rangle|^2$ we obtain
\begin{multline}
\label{phi-phi}
\left| \langle \StateSingle_{a_g,n}|\StateSingle_{a_g+\Delta a,n'}\rangle \right|^2
= \delta_{n,n'} +\frac{\Delta a^2}{4a_g^2}
\Big((n-1)n\delta_{n,n'+2}
\\
-(n^2+n+1)\delta_{n,n'}-(n+1)(n+2)\delta_{n,n'-2}\Big)
+O\left(\frac{n^3\Delta a^3}{a_g^3}\right)
.
\!
\end{multline}

We insert Eq.\ \eqref{phi-phi} into Eq.\ \eqref{Kuhr-C-n-n'} and use the Kronecker symbols to collapse the sum over $n'$. Hence
\begin{multline}
\label{Kuhr-C-n}
C(t)
= C_0 (1-q) \sum_{n=0}^\infty e^{-n\xi}
\Biggl( 1 + \frac{\Delta a^2}{4a_g^2}
\Big((n-1)n e^{2i\omega_rt}
\\
-(n^2+n+1)-(n+1)(n+2) e^{-2i\omega_rt}\Big)
+O\left(\frac{n^3\Delta a^3}{a_g^3}\right)\Biggr)
,\end{multline}
where we abbreviated
\begin{align}
\xi(t)
= \beta\hbar\omega_g-it(\omega_g-\omega_r)
= \beta\hbar\omega_g\left(1-i\frac{t}K\right)
\end{align}
and used $K$ from Eq.\ \eqref{Kuhr-K-kappa}.

As an aside, we consider the terms oscillating with frequencies $\pm2\omega_r$ and note that higher-order terms not shown here oscillate with integer multiples of these frequencies. All these terms are caused by the monopole mode with angular frequency $2\omega_r$, also known as the breathing mode, excited by the transfer to state $|r\rangle$ because of the different trapping frequencies \cite{Yang:11}. This breathing mode gives rise to collapse and revival of $|C(t)|^2$. This is not to be confused with the collapse and revival of $|C(t)|^2$ caused by the dipole mode with angular frequency $\omega_r$, also known as the sloshing mode, created even for $\Delta a= 0$ by nonzero net photon recoil \cite{Afek:17, Lampen:18} or by creating the spin wave not in the trap center \cite{Yang:11}.

We use $\hbar\omega_g\ll k_BT$ from Eq.\ \eqref{Kuhr-beta-hbar-omega} to obtain $1-q\approx \beta\hbar\omega_g$ and to approximate $\sum_{n=0}^\infty$ as $\int_0^\infty dn$. Hence
\begin{multline}
C(t)
= C_0 \frac{1}{1-i\frac{t}K}
\Biggl(1 + \frac{\Delta a^2}{4a_g^2 \xi^2}
\Big( (2-\xi) e^{2i\omega_rt}
\\
-(2+\xi+\xi^2)
-(2+3\xi+2\xi^2) e^{-2i\omega_rt}\Big)
+O\left(\frac{\Delta a^3}{a_g^3\xi^3}\right)\Biggr)
.\end{multline}
The factor $(1-it/K)^{-1}$ decays with a typical decay time $K$. Throughout the rest of this section, we ignore the long-time tail of this decay, i.e.\ we ignore times with $t\gg K$ because there $|C(t)/C(0)|\ll 1$. Combination with $\beta\hbar\omega_g\ll 1$ from Eq.\ \eqref{Kuhr-beta-hbar-omega} yields $|\xi|\ll 1$. In particular, this implies $C_0= e^{-\Im\xi/2}\approx 1$. Hence
\begin{multline}
\label{Kuhr-C-xi-no-sum}
C(t)
\approx \frac{1}{1-i\frac{t}K}
\\ \times
\Biggl(1 - \frac{\Delta a^2}{2a_g^2 \xi^2} \Bigl(1-2i\sin(2\omega_rt)\Bigr) +O\left(\frac{\Delta a^3}{a_g^3\xi^3}\right)\Biggr)
.\end{multline}
If terms of order $O(\Delta a^2)$ are negligible, then Eq.\ \eqref{Kuhr-C-xi-no-sum} will reproduce Eq.\ \eqref{Kuhr-C-Kuhr} with $d= 1$, which was the goal of this calculation.

Since $1\leq |1-2i\sin(2\omega_rt)|\leq \sqrt5$ terms of order $O(\Delta a^2)$ are negligible in Eq.\ \eqref{Kuhr-C-xi-no-sum} if and only if $|\Delta a/a_g \xi|\ll 1$. As we ignore the long-time tail $t\gg K$, we obtain $|\xi|\approx \beta\hbar\omega_g$ so that the condition is equivalent to $|\Delta a|/a_g\ll \beta\hbar\omega_g$. Combination with $\beta\hbar\omega_g\ll 1$ from Eq.\ \eqref{Kuhr-beta-hbar-omega} yields
\begin{align}
\label{Kuhr-intermediate-T}
\hbar\omega_g
\ll k_BT
\ll \hbar\omega_g \frac{a_g}{|\Delta a|}
.\end{align}
This defines an intermediate-temperature regime, in which Eq.\ \eqref{Kuhr-C-Kuhr} is applicable. Note that comparing the leftmost and rightmost expressions in Eq.\ \eqref{Kuhr-intermediate-T} yields $|\Delta a|/a_g\ll 1$, which is equivalent to $|\kappa_g-\kappa_r|/\kappa_g\ll 1$.

Above Eq.\ \eqref{chi-ag-n}, we gave an intuitive but rough argument in support of Eq.\ \eqref{Kuhr-E-ri}, which relies on the approximation that the quantum number $n$ is conserved. This requires the quantum states to be discrete. This is not compatible with the high-temperature limit because that can usually be derived without referring to the discreteness of quantum states. This agrees with the fact that the result holds only in the intermediate temperature regime of Eq.\ \eqref{Kuhr-intermediate-T}. In the high-temperature limit, the Raman-Nath result \eqref{Kuhr-Raman-Nath} is applicable. The Raman-Nath result does not rely on discrete quantum states, as is obvious from Eq.\ \eqref{C-Raman-Nath}.

\subsection*{D. Free Expansion from a Harmonic Trap}

Another simple example is obtained when considering free expansion from a harmonic trap. Before storage the system is in thermal equilibrium in a trapping potential. After storage, the trap is off $V_g(\bm x)= V_r(\bm x)= 0$. For simplicity, we consider a 1d model with a harmonic potential along $x$ before storage $V_{g,\text{before}}(x)= m\omega^2 x^2/2$ with angular trapping frequency $\omega$. We assume that the signal-beam profile along $x$ is a Gaussian. We assume that the net photon recoil $\bm k_R$ vanishes and obtain
\begin{align}
\label{v-Gauss-1d}
v(x)
= \left(\frac{2}{\pi w^2}\right)^{1/4}e^{-x^2/w^2}
,\end{align}
where $w$ is the beam waist and we consider a 1d version of the quantization volume with length $L_x\gg w$. A generalization to higher dimensions using Eq.\ \eqref{separate} is straightforward.

\subsubsection*{1. Pure BEC}

First, we consider a BEC at $T=0$. As all atoms initially occupy the same spatial wave function, we drop the indices $n,i$ from the notation. Free expansion of the initial external state yields
\begin{align}
\psi_g(x,t)
= \frac1{\pi^{1/4}\sqrt{a_0(1+i\omega t)}}  e^{-x^2/2a_0^2(1+i\omega t)}
,\end{align}
where
\begin{align}
a_0
= \sqrt{\frac\hbar{m\omega}}
\end{align}
is the harmonic-oscillator length. Similarly, after storage at time $t=0$, the normalized external state of the atoms transferred to $|r\rangle$ evolves into
\begin{align}
\frac{\psi_r(x,t)}{\sqrt{M(0)}}
= \frac1{\pi^{1/4}\sqrt{b_0(1+i\omega ta_0^2/b_0^2)}} e^{-x^2/2b_0^2(1+i\omega ta_0^2/b_0^2)}
\end{align}
where $b_0= (2/w^2+1/a_0^2)^{-1/2}$. Equations \eqref{Q-def} and \eqref{eta-BEC} yield
\begin{align}
\label{eta-expanding-BEC-temp}
\frac{\eta_\text{BEC}(t)}{\eta_0}
= \frac{L_x}{M(t)} \left|\int_{-\infty}^\infty dx v^*(x) \psi_g^*(x,t) \frac{\psi_r(x,t)}{\sqrt{M(0)}}
\right|^2
.\end{align}
We obtain $M(t)= L_x/\sqrt{\pi[a_0^2(1+\omega^2 t^2)+w^2/2]}$ from Eq.\ \eqref{M-n-i(t)}. Hence
\begin{align}
\label{eta-expanding-BEC}
\frac{\eta_\text{BEC}(t)}{\eta_0}
= \sqrt{\frac{w^4(2a_0^2+w^2)(2a_0^2+w^2+2a_0^2\omega^2t^2)} {(2a_0^2+w^2)^2(w^2+a_0^2\omega^2t^2)^2+4a_0^8\omega^2t^2}}
.\end{align}

For $a_0\ll w$ this simplifies to
\begin{align}
\frac{\eta_\text{BEC}(t)}{\eta_0}
\approx \frac{\sqrt{1+2t^2/\tau_a^2}}{1 +t^2/\tau_a^2}
\approx \frac1{\sqrt{1+t^2/2\tau_a^2}}
,&&
a_0\ll w
,\end{align}
where we abbreviated
\begin{align}
\label{tau-a}
\tau_a
= \frac{w}{a_0\omega}
= \frac{ma_0w}\hbar
.\end{align}
This time scale is plausible because the position-momen\-tum uncertainty relation implies that the finite initial spatial size of the BEC $a_0$ causes a 1d rms width of the velocity distribution of $\hbar/\sqrt2ma_0$ for the atomic gases in states $|g\rangle$ and $|r\rangle$. For $t= \tau_a$, the typical distance travelled because of this velocity during the dark time equals $w/\sqrt2$ so that the atoms in states $|g\rangle$ and $|r\rangle$ have reduced spatial overlap with the signal-beam mode $u(\bm x)$. Note that in the limit $w\to\infty$ we obtain $\tau_a\to\infty$ so that there is no decay at all. This reflects the fact that in this limit $R=\mathbbm 1$. Combined with $H_g= H_r$ during the dark time no decay is expected, as discussed in the context of Eq.\ \eqref{R=1-Hg=Hr}.

Conversely for $w\ll a_0$, Eq.\ \eqref{eta-expanding-BEC} becomes
\begin{align}
\label{tau-w-BEC}
\frac{\eta_\text{BEC}(t)}{\eta_0}
\approx \frac1{\sqrt{1+ t^2/\tau_w^2}}
,&&
w\ll a_0
\end{align}
with $\tau_w= mw^2/\hbar$ from Eq.\ \eqref{tau-w}. To compare with the related Eq.\ \eqref{eta-tau-w} note that the two results become identical when taking into account that the 2d result Eq.\ \eqref{eta-tau-w} can be factored into 1d components according to Eq.\ \eqref{separate}. This is interesting because Eq.\ \eqref{tau-w-BEC} applies to a BEC, whereas Eq.\ \eqref{eta-tau-w} was obtained by thermal averaging at high temperature. However, as both results are obtained by assuming that the dominant kinetic energy comes from the Gaussian waist, the relevant quantity is identical for all initial states which contribute noticeably to the thermal average so that the averaging has no effect. This explains why the thermal average yields the same result as a BEC here.

\subsubsection*{2. Simplified Model for High Temperature}

To obtain an approximate, simple, analytic model for a gas with a temperature $T$ far above quantum degeneracy, we perform two steps in an \emph{ad hoc} fashion. First, we replace all complex quantities in Eq.\ \eqref{eta-expanding-BEC-temp} by their moduli, thus setting their phases to zero. Second, we apply the result to a situation with $T$ far above quantum degeneracy. The justification for this approach will come from a numerical calculation detailed below. We obtain
\begin{align}
\label{eta-release-high-T}
\frac{\eta(t)}{\eta_0}
= \frac{L_x}{\mu(t)} \left| \int_{-\infty}^{\infty} dx |v(x)| \sqrt{\varrho_g(x,t)\varrho_r(x,t)} \right|^2
\end{align}
with $\mu(t)$ from Eq.\ \eqref{mu-def}, $\varrho_g(x,t)$ from Eq.\ \eqref{varrho} and $\varrho_r(x,t)= \sum_n p_n |\psi_{r,n}(x,t)|^2/M_n(0)$, also normalized to $\int_{\mathcal V} d^3x \varrho_g(\bm x,t)= 1$. For a pure BEC $\mu(t)$, $\varrho_g(x,t)$, and $\varrho_r(x,t)$ would simplify to $M(t)$, $|\psi_g(x,t)|^2$, and $|\psi_r(x,t)|^2/M(0)$, respectively, which suggests that we picked appropriate generalizations when moving from the pure BEC to a high-temperature gas. A corresponding approximation neglecting phases for calculating $\mu(t)$ is not needed because $\mu(t)$ in Eq.\ \eqref{mu-def} is insensitive to phases anyway. If the signal beam is a plane wave, Eq.\ \eqref{eta-release-high-T} will simplify to $\eta(t)/\eta_0= |\int_{-\infty}^\infty dx \sqrt{\varrho_g(x,t)\varrho_r(x,t)}|^2$, the 3d version of which was previously used e.g.\ in Ref.\ \cite{Yang:11}.

This model only takes the decay of spatial overlap into account. Depending on the physical effect under investigation, this can be a good or a terrible approximation. For example, the decay of the retrieval resulting from photon recoil combined with thermal atomic motion in a homogeneous gas discussed in Sec.\ \ref{sec-recoil} of the paper would not produce any decay of $\eta(t)$ within this approximation.

For a high-temperature gas released from a 1d harmonic trap, we use
\begin{align}
\varrho_g(x,t)
= \frac1{\sqrt{2\pi}\sigma_g} e^{-x^2/2\sigma_g^2}
,&&
\sigma_g(t)
= \sqrt{\sigma_x^2+\sigma_v^2t^2}
\end{align}
with $\sigma_v= \sqrt{k_BT/m}$ and $\sigma_x= \sigma_v/\omega$ and
\begin{align}
\varrho_r(x,t)
= \frac1{\sqrt{2\pi}\sigma_r} e^{-x^2/2\sigma_r^2}
,&&
\sigma_r(t)
= \sqrt{\sigma_r^2(0)+\sigma_{v,r}^2t^2}
\end{align}
with
\begin{align}
\sigma_r(0)
= \left(\frac1{\sigma_x^2}+\frac4{w^2}\right)^{-1/2}
,&&
\sigma_{v,r}
= \sqrt{\sigma_v^2+\frac{\hbar^2}{m^2w^2}}
.\end{align}
This expression for $\sigma_{v,r}$ takes into account that the finite signal beam waist contributes to the momentum spread of the atoms transferred to internal state $|r\rangle$ according to the position-momentum uncertainty relation.

Equation \eqref{eta-release-high-T} yields
\begin{align}
\label{eta-release-thermal-approx}
\frac{\eta(t)}{\eta_0}
= \sqrt{\frac1{4\sigma_g^2(t)}+\frac1{w^2}} \frac1{\sigma_r(t)}
\frac1{\frac1{4\sigma_g^2(t)}+\frac1{4\sigma_r^2(t)}+\frac1{w^2}}
.\end{align}
For $w\ll \sigma_x$ this simplifies to
\begin{align}
\label{eta-tau-exp}
\frac{\eta(t)}{\eta_0}
\approx \frac{\sqrt{1+4\frac{t^2}{\tau_\text{rel}^2}}}{1+2\frac{t^2}{\tau_\text{rel}^2}}
\approx \frac1{\sqrt{1+\frac{t^2}{\tau_\text{rel}^2}}}
\end{align}
with
\begin{align}
\label{tau-exp}
\tau_\text{rel}
= \frac{w}{\sqrt{\sigma_v^2+\frac{\hbar^2}{m^2w^2}}}
.\end{align}
If additionally $\sigma_v\ll \hbar/mw$, then $\tau_\text{rel}\approx \tau_w$ with $\tau_w$ from Eq.\ \eqref{tau-w}. In this limit, Eq.\ \eqref{eta-tau-exp} becomes identical to Eq.\ \eqref{tau-w-BEC}, i.e.\ the atoms disperse out of the signal beam because of the kinetic energy associated with the position-momentum uncertainty relation.

Conversely, if additionally $\hbar/mw\ll \sigma_v$, then
\begin{align}
\label{tau-exp-approx}
\tau_\text{rel}
\approx \frac w{\sigma_v}
.\end{align}
This time scale describes how atoms leave the signal beam because of thermal motion. If the release occurs from a 2d harmonic trap, then Eq.\ \eqref{separate} is applicable, converting Eq.\ \eqref{eta-tau-exp} into
\begin{align}
\label{eta-tau-exp-2d}
\frac{\eta(t)}{\eta_0}
\approx \frac{1+4\frac{t^2}{\tau_\text{rel}^2}}{\left(1+2\frac{t^2}{\tau_\text{rel}^2}\right)^2}
\approx \frac1{1+\frac{t^2}{\tau_\text{rel}^2}}
.\end{align}

\subsubsection*{3. Numerical Calculation for High Temperature}

To test whether Eq.\ \eqref{eta-release-high-T} is a good approximation, we additionally perform a numerical calculation. Again, we consider a 1d situation. We start by explicitly writing Eq.\ \eqref{psi-g-n-i(t)-psi-r-n-i(t)} in the position representation
\begin{align}
\psi_{r,n}(x,t)
= \int_{-\infty}^\infty dx_0 U(x,x_0,t) \psi_{r,n}(x_0,0)
\end{align}
and
\begin{align}
\psi_{g,n}(x,t)
= \int_{-\infty}^\infty dx_1 U(x,x_1,t) \psi_{g,n}(x_1,0)
,\end{align}
where
\begin{align}
U(x_a,x_b,t)
= \frac1{\sqrt{2\pi iA}} e^{-(x_a-x_b)^2/2iA}
,&&
A
= \frac{\hbar t}{m}
\end{align}
is the position representation of the time-evolution operator $U(t)= \exp(-ik^2A/2)$ describing the dark time. Using this, the definition of $Q_n(t)$ in Eq.\ \eqref{Q-def} yields
\begin{align}
Q_n(t)
= L_x \int_{-\infty}^\infty dx_0 v(x_0) \psi_{g,n}(x_0,0) I_1(x_0)
\end{align}
with
\begin{align}
I_1(x_0)
= \int_{-\infty}^\infty dx_1 \psi_{g,n}^*(x_1,0) I_2(x_0,x_1)
\end{align}
and
\begin{align}
I_2(x_0,x_1)
= \int_{-\infty}^\infty dx v^*(x) U^*(x,x_1,t) U(x,x_0,t)
.\end{align}
Using $v(x)$ from Eq.\ \eqref{v-Gauss-1d} we obtain
\begin{multline}
I_2(x_0,x_1)
= v_0 \frac{w}{2\sqrt\pi A} \exp\left(\frac{w^2}{2A^2}x_0x_1\right)
\\ \times
\exp\left(-\left(\frac{w^2}{4A^2}-i\frac{1}{2A}\right)x_0^2
-\left(\frac{w^2}{4A^2}+i\frac{1}{2A}\right)x_1^2\right)
.\end{multline}

The initial states are the energy eigenstates of the harmonic oscillator
\begin{align}
\psi_{g,n}(x)
= \psi_{0,n} e^{-x^2/2a_0^2} H_n\left(\frac x{a_0}\right)
\end{align}
with
\begin{align}
\psi_{0,n}
= \frac1{\pi^{1/4}\sqrt{a_0}} \frac1{\sqrt{2^nn!}}
,\end{align}
where $H_n$ is a Hermite polynomial. We use that for $a,b,c\in\mathbbm C,n\in\mathbbm Z_0^+$ with $\Re a> 0, c^2\neq a$ (to prove this, calculate $\partial_b$ and use mathematical induction with respect to $n$ combined with the recurrence relation and differential property of the Hermite polynomials)
\begin{multline}
\int_{-\infty}^\infty dx e^{-a x^2+bx} H_n(cx)
\\
= \sqrt{\frac\pi a} e^{b^2/4a} \left(1-\frac{c^2}a\right)^{n/2}H_n\left(\frac{c}{2a\sqrt{1-\frac{c^2}a}}\, b\right)
\end{multline}
and obtain
\begin{multline}
I_1(x_0)
= \psi_{0,n} v_0 \frac{a_0w}{2A} \frac{1}{\sqrt{\xi_1}}
\xi_3^n H_n\left(\xi_4 \frac{x_0}{a_0}\right)
\\ \times
\exp\left(-\left(\frac{w^2}{4A^2}-i\frac{1}{2A}-\frac{\xi_2}{a_0^2} \right)x_0^2\right)
\end{multline}
where we abbreviated the dimensionless complex parameters
\begin{align}
\xi_1
= \frac12+\left(\frac{a_0w}{2A}\right)^2+i\frac{a_0^2}{2A}
,&&
\xi_2
= \left(\frac{a_0w}{2A}\right)^4 \frac{1}{\xi_1}
\end{align}
and
\begin{align}
\xi_3
= \sqrt{1-\frac{1}{\xi_1}}
,&&
\xi_4
= \frac{1}{\xi_1\xi_3} \left(\frac{a_0w}{2A}\right)^2
.\end{align}

We now use that for $a\in\mathbbm C,n\in\mathbbm Z_0^+$ with $\Re a>0$
\begin{align}
\label{int-x^n-Gauss}
\int_{-\infty}^\infty dx \, x^n e^{-ax^2}
&
=
\begin{cases}
a^{-(n+1)/2} \Gamma\left(\frac{n+1}2\right), & n \text{ even} \\
0, & n \text{ odd} \\
\end{cases}
\notag \\ &
= \sqrt{\frac{\pi}{a}} \, \frac{H_n(0)}{(-4)^{n/2}} \, \frac1{a^{n/2}}
.\end{align}
We expand $H_n(x)= \sum_{r=0}^n d_{n,r} x^r$ with coefficients $d_{n,r}$ and obtain for $a,b,c\in\mathbbm C, n\in\mathbbm Z_0^+$ with $\Re a>0$
\begin{align}
\int_{-\infty}^\infty dx e^{-ax^2} H_n(bx)H_n(cx)
= \sqrt{\frac\pi a} \sum_{r,s=0}^n g_{n,r,s} \frac{b^r c^s}{a^{(r+s)/2}}
,\end{align}
where we abbreviated
\begin{align}
g_{n,r,s}
= d_{n,r} d_{n,s} \frac{d_{r+s,0}}{(-4)^{(r+s)/2}}
.\end{align}
Hence
\begin{align}
Q_n(t)
= L_x v_0^2 \frac{a_0w}{2A} \frac{1}{\sqrt{\xi_1}}
\frac1{\sqrt{\xi_5}}
\frac1{2^nn!} \xi_3^n \sum_{r,s=0}^n g_{n,r,s} \frac{\xi_4^r}{\xi_5^{(r+s)/2}}
\end{align}
with
\begin{align}
\xi_5
= \xi_1^*-\xi_2+\frac{a_0^2}{w^2}
.\end{align}

From Eq.\ \eqref{M-n-i(t)} we obtain in an analogous way
\begin{align}
M_n(t)
= L_x v_0^2 \frac{a_0w}{2\sqrt2A} \frac{1}{\sqrt{\tilde\xi_1}}
\frac1{\sqrt{\tilde\xi_5}}
\frac1{2^nn!} \tilde\xi_3^n \sum_{r,s=0}^n g_{n,r,s} \frac{\tilde\xi_4^r}{\tilde\xi_5^{(r+s)/2}}
\end{align}
with $\tilde \xi_1= \frac12+(\frac{a_0w}{2\sqrt2A})^2+i\frac{a_0^2}{2A}$, $\tilde \xi_2= (\frac{a_0w}{2\sqrt2A})^4 \frac{1}{\tilde \xi_1}$, $\tilde \xi_3= \sqrt{1-\frac{1}{\tilde \xi_1}}$, $\tilde \xi_4
= \frac{1}{\tilde \xi_1\tilde \xi_3} (\frac{a_0w}{2\sqrt2A})^2$, and $\tilde\xi_5
= \tilde\xi_1^*-\tilde\xi_2$.

As this expression for $M(t)$ is difficult to evaluate for $t=0$, we additionally calculate from Eq.\ \eqref{M-n-i(t)}
\begin{align}
M_n(0)
= L_x v_0^2 \frac{1}{\sqrt{\xi_6}} \frac1{2^nn!}  \sum_{r=0}^n U_{n,r} \frac{1}{\xi_6^r}
\end{align}
where we expanded $H_n^2(x)= \sum_{r=0}^n u_{n,r} x^{2r}$ with coefficients $u_{n,r}$ and abbreviated $\xi_6= 1+\frac{2a_0^2}{w^2}$ and $U_{n,r}= u_{n,r}\frac{d_{2r,0}}{(-4)^r}$ and used that for $a,b\in\mathbbm C, n\in\mathbbm Z_0^+$ with $\Re a>0$
\begin{align}
\int_{-\infty}^\infty dx e^{-ax^2} H_n^2(bx)
= \sqrt{\frac\pi a} \sum_{r=0}^n U_{n,r} \frac{b^{2r}}{a^r}
,\end{align}
which follows from Eq.\ \eqref{int-x^n-Gauss}.

\begin{figure}[!t]
\includegraphics[width=\columnwidth]{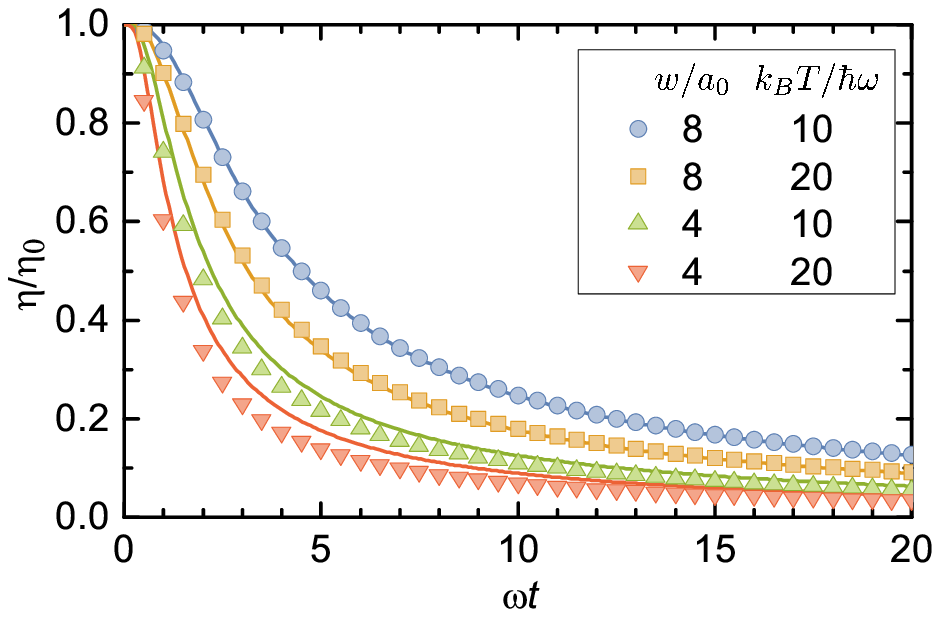}
\caption{Release of a gas with a temperature far above quantum degeneracy from a 1d harmonic trap. Numerical results (dots) agree fairly well with Eq.\ \eqref{eta-release-thermal-approx} of the simplified model (lines). In the numerical calculation, the sums over $n$ are truncated to $n\leq 40$.}
\label{fig-numerics}
\end{figure}

Eventually, we calculate $\eta(t)$ from Eq.\ \eqref{eta-C} using $C(t)$ from Eq.\ \eqref{C-def} and $\mu(t)$ from Eq.\ \eqref{mu-def}. In doing so, we truncate the sums in Eqs.\ \eqref{C-def} and \eqref{mu-def} to $n\leq n_\text{max}$ with $n_\text{max}= 40$ and perform a numerically calculation. Numerical results for different values of $k_BT/\hbar\omega$ and $w/a_0$ are shown in Fig.\ \ref{fig-numerics}. Note that the condition $k_BT/\hbar\omega\gg 1$ is necessary for $T$ to be far above quantum degeneracy. The result Eq.\ \eqref{eta-release-thermal-approx} of the simplified analytic model (lines) agrees fairly well with the numerical results. Note that for the parameters shown in the figure, Eq.\ \eqref{eta-release-thermal-approx} is fairly well approximated by both the middle and the rightmost expression in Eq.\ \eqref{eta-tau-exp}.

In our experiment with $T= 0.2$ $\mu$K, $\omega/2\pi= 96$ Hz, and $w=8$ $\mu$m, we obtain $k_BT/\hbar\omega\approx 43$ and $w/a_0\approx 7.3$, not too far from the values studied in Fig.\ \ref{fig-numerics}. In the numerical calculation, one should choose $n_\text{max}$ such that $k_BT/\hbar\omega\ll n_\text{max}$ because states up to $n\approx k_BT/\hbar\omega$ are noticeably populated. When increasing $n_\text{max}$ beyond 40, our numerical implementation not detailed here is no longer stable, which is why we chose to restrict our numerical calculations to $k_BT/\hbar\omega\leq 20$.

\subsection*{E. Ramsey Spectroscopy}

\label{sec-supp-Ramsey}

Here, we show that the fringe visibility in an appropriately chosen Ramsey experiment is given by Eq.\ \eqref{V-C} with $C(t)$ from Eq.\ \eqref{C-URUR}. Throughout this appendix, we will use a notation which is chosen to make the analogy to EIT-based storage and retrieval obvious by often using the same symbols for quantities which correspond to each other.

\subsubsection*{1. Single-Atom Ramsey Spectroscopy}

Compared to EIT, Ramsey spectroscopy can be implemented in a much simpler system, namely in a single atom with only two internal atomic states. Hence, we first consider Ramsey spectroscopy performed on a single two-level atom. We denote the ground and excited internal states as $|g\rangle$ and $|r\rangle$. We assume that these states are connected by an electric-dipole transition with matrix element $d_{rg}$ and resonance angular frequency $\omega_{gr}>0$, which is driven by a classical electric field $E(\bm x,t)= \frac12 E_0 \sqrt{\mathcal V}v(\bm x)e^{-i\omega_\text{Ram}t}+\cc$, where $\omega_\text{Ram}>0$ is the angular frequency and $E_0$ the complex amplitude. The spatial mode function $v(\bm x)$ is normalized to $\int_{\mathcal V} d^3x |v(\bm x)|^2=1$, where $\mathcal V$ is a quantization volume.

We denote the Rabi frequency as $\Omega_R= -d_{rg}E_0/\hbar$, the detuning as $\Delta_R= \omega_\text{Ram}-\omega_{rg}$, use an interaction picture and the rotating-wave approximation and obtain a Hamiltonian which contains a modified internal-state energy term $H_\Delta$ together with a potential $V_R$ given by
\begin{subequations}
\begin{align}
\label{Ramsey-V-R}
H_\Delta
&
= \hbar \Delta_R |g\rangle\langle g|
,&
V_R
&
= \tfrac12 \hbar \Omega_R S_r^\dag+\Hc
,\\
\label{Ramsey-S-def}
S_r
&
= R \otimes |g\rangle\langle r|
,&
R(\bm x)
&
= \sqrt{\mathcal V}v^*(\bm x)
.\end{align}
\end{subequations}
Note that an analogous atom-light interaction potential will be obtained for a two-photon transition if the intermediate state can be eliminated adiabatically because of a large single-photon detuning. Another situation described by an analogous potential is an rf transition between two atomic ground states connected by a magnetic-dipole transition. In the latter two cases, it is quite possible that $\bm k_R\cdot\bm x$ is essentially constant across the sample, which means that in these cases one can approximate $\bm k_R= 0$.

Ramsey spectroscopy consists in preparing the atom in the internal state $|g\rangle$, applying two light pulses separated by a dark time $t$, and finally measuring the probability $P_r$ that the atom is found in state $|r\rangle$. The pulse area $\varphi_\text{Ram}= \Omega_Rt_p$ is typically chosen to be identical for both pulses. $t_p$ is the duration of one pulse. Note that $\varphi_\text{Ram}$ might be complex.

We simplify the calculation by assuming that the Rabi frequency $\Omega_R$ is so large that all terms in the Hamiltonian other than $V_R$ have negligible effect during each pulse. In particular $\Delta_R$ is assumed to have negligible effect during the pulse. Hence, the time-evolution operator describing the effect of one pulse is $U_p= \exp(-iV_Rt_p/\hbar)= d_a\mathbbm1+(d_b|r\rangle\langle g|-\Hc)$ with operators $d_a$ and $d_b$ which act onto the external degree of the atom. Their position representations are $d_a(\bm x)= \cos|\varphi_\text{Ram} \sqrt{\mathcal V}v(\bm x)/2|$ and $d_b(\bm x)= -i \frac{\varphi_\text{Ram} v(\bm x)}{|\varphi_\text{Ram} v(\bm x)|} \sin|\varphi_\text{Ram} \sqrt{\mathcal V}v(\bm x)/2|$. The appearance of the position-dependent mode function $v(\bm x)$ inside the trigonometric functions makes it difficult to draw analogies to EIT-based storage and retrieval. But this expression can be much simplified if we consider pulses which are short enough that $|\varphi_\text{Ram}|\ll 1$, which yields
\begin{align}
\label{Ramsey-U-p-approx}
U_p
= c_a\mathbbm1+c_b S_r^\dag-c_b^* S_r
,\end{align}
where $c_a= 1$, $c_b= -i\varphi_\text{Ram}/2$, and terms of order $O(\varphi_\text{Ram}^2)$ were neglected. The point is that $c_a$ and $c_b$ are position-independent complex number. They do not act on the atomic externa state. However, $U_p$ acts on the internal and external state, because of the appearance of $S_r$ but that appearance is simple enough that one can make a connection to EIT-based storage and retrieval.

The appearance of a position dependence inside a trigonometric function is also avoided if $|v(\bm x)|^2$ is independent of $\bm x$, which implies that the mode is a plane wave. In that case, $R$ is unitary and one finds that Eq.\ \eqref{Ramsey-U-p-approx} still holds but now with $c_a= \cos|\varphi_\text{Ram}/2|$ and $c_b= -i\frac{\varphi_\text{Ram}}{|\varphi_\text{Ram}|}\sin|\varphi_\text{Ram}/2|$. Specifically, for the much used $\varphi_\text{Ram}= \pi/2$ we obtain $c_a= 1/\sqrt2$ and $c_b= -i/\sqrt2$. In the following, we will assume that Eq.\ \eqref{Ramsey-U-p-approx} holds. Hence, our formalism is applicable for an arbitrary mode function and small pulse area or for a plane-wave mode with arbitrary pulse area.

Typical Ramsey experiments use $\varphi_\text{Ram}= \pi/2$ and a plane wave, both to maximize the number of atoms transferred for $t=0$. But typical experiments on EIT-based storage and retrieval use a signal beam waist much smaller than the sample size to achieve high storage efficiency. To clarify the connection to EIT-based retrieval, we keep the possibility of a small signal beam waist in our calculation by not making any assumptions about $c_a$ and $c_b$.

During the dark time, the Rabi frequency vanishes and other effects will dominate the dynamics. We assume that the time evolution during the dark time is described by $U_{d,\Delta}= U_d U_\Delta= U_\Delta U_d$, where $U_\Delta= \exp(-iH_\Delta t)$ with $H_\Delta$ from Eq.\ \eqref{Ramsey-V-R} and $U_d= \exp(-iH_dt/\hbar)$ with a dark-time Hamiltonian of the form
\begin{align}
\label{Ramsey-H-d}
H_d
= H_g \otimes |g\rangle\langle g|+H_r\otimes |r\rangle\langle r|
,\end{align}
where $H_g$ and $H_r$ are operators acting on the external atomic state. This implies $U_{d,\Delta}= U_{g,\Delta}\otimes|g\rangle\langle g|+U_r\otimes|r\rangle\langle r|$, with $U_r= \exp(-iH_rt/\hbar)$, $U_{g,\Delta}= e^{-i\Delta_R t}U_g$, and $U_g= \exp(-iH_gt/\hbar)$. Hence, the internal state is unchanged during the dark time. The total time-evolution operator for the full sequence consisting of both pulses separated by the dark time reads $U_t= U_pU_{d,\Delta} U_p$.

While we assume that the initial internal state is $|g\rangle$, the initial external state is typically a mixed state described by a density matrix $\rho_\text{in}$, yielding the initial single-atom density matrix $\widetilde \rho_\text{in}= \rho_\text{in} \otimes |g\rangle\langle g|$. The density matrix after the second Ramsey pulse is $U_t \widetilde \rho_\text{in} U_t^\dag$. Hence, the probability of finding the atom in internal state $|r\rangle$ after the second Ramsey pulse is $P_r= \tr[U_t \widetilde \rho_\text{in} U_t^\dag (\mathbbm1\otimes|r\rangle\langle r|)] = \tr(\rho_\text{in} U_{rg}^\dag U_{rg})$, where we abbreviated an operator $U_{rg}= \langle r|U_t|g\rangle$, which acts only on the external state. We easily obtain $U_{rg}= c_ac_b(U_rR^\dag+R^\dag U_{g,\Delta})$ and
\begin{multline}
\label{Ramsey-P-r-URUR}
P_r
= |c_ac_b|^2 [\tr(\rho_\text{in} RR^\dag) + \tr(\rho_\text{in} U_{g,\Delta}^\dag R R^\dag U_{g,\Delta})
\\
+2\Re\tr(\rho_\text{in} U_{g,\Delta}^\dag R U_r R^\dag)]
.\end{multline}
Now we assume
\begin{align}
\label{Ramsey-URUR-drop-U}
\tr(\rho_\text{in} RR^\dag)
= \tr(\rho_\text{in} U_{g,\Delta}^\dag R R^\dag U_{g,\Delta})
.\end{align}
There are two important scenarios in which Eq.\ \eqref{Ramsey-URUR-drop-U} holds. The first scenario is if $v(\bm x)$ is a plane wave because then $R$ is unitary and each side of Eq.\ \eqref{Ramsey-URUR-drop-U} simplifies to $\tr(\rho_\text{in})= 1$. The second scenario is if the initial external state $\rho_\text{in}$ commutes with $H_g$
\begin{align}
\label{Ramsey-Hg-rho-commute}
[H_g,\rho_\text{in}]
= 0
,\end{align}
which is the case e.g.\ if the initial state is in thermal equilibrium and the Hamiltonians before and after the first Ramsey pulse are identical. Equation \eqref{Ramsey-Hg-rho-commute} implies that $U_{g,\Delta}$ commutes with $\rho_\text{in}$ and that yields Eq.\ \eqref{Ramsey-URUR-drop-U}.

We insert Eq.\ \eqref{Ramsey-URUR-drop-U} and $U_{g,\Delta}= e^{-i\Delta_R t}U_g$ from above into Eq.\ \eqref{Ramsey-P-r-URUR} and use that $e^{-i\Delta_R t}$ is only a scalar factor. We obtain
\begin{align}
\label{Ramsey-P-r-C}
P_r(t)
= 2|c_ac_b|^2 \Bigl( C(0)+\Re[e^{i\Delta_R t} C(t)] \Bigr)
,\end{align}
where we abbreviated a complex number
\begin{align}
\label{Ramsey-C}
C(t)
= \tr[\rho_\text{in} U_g^\dag(t)RU_r(t)R^\dag]
.\end{align}
This is identical to Eq.\ \eqref{C-URUR} for the coherence $C(t)$ in EIT-based storage and retrieval.

We rewrite Eq.\ \eqref{Ramsey-P-r-C} as
\begin{align}
\label{Ramsey-P-r-V}
P_r(t)
= \frac{P_{r,0}}2 [1+ V \cos(\Delta_R t+\vartheta_\text{Ram})]
,\end{align}
where we abbreviated $P_{r,0}= 4|c_ac_b|^2 C(0)$, used that $C(0)$ is always real, and abbreviated
\begin{align}
\label{Ramsey-V-def}
V(t)
= \left|\frac{C(t)}{C(0)}\right|
,&&
\vartheta_\text{Ram}(t)
= \arg C(t)
,\end{align}
where $\arg z$ with $-\pi< \arg z\leq \pi$ denotes the argument of a complex number $z= |z|e^{i\arg z}$. In a Ramsey experiment, one typically chooses $\Delta_R$ so large that $C(t)$ varies slowly compared to $\Delta_Rt$. Hence, $V$ and $\vartheta_\text{Ram}$ have the interpretation of a slowly-varying fringe visibility $V= \frac{\max(P_r)-\min(P_r)}{\max(P_r)+\min(P_r)}$ and a slowly-varying phase shift of the otherwise sinusoidal fringe pattern in Eq.\ \eqref{Ramsey-P-r-V}. As the fringe visibility $V$ is given by Eq.\ \eqref{Ramsey-V-def} we derived Eq.\ \eqref{V-C} with $C_1(t)= C(t)$.

The result obtained here is fairly generic because only few assumptions were needed to derive it. First, the time evolution during each pulse is described by Eq.\ \eqref{Ramsey-U-p-approx}, which is the case e.g.\ if the signal beam is a plane wave or the pulse area is small, second, the dark-time Hamiltonian has the form of Eq.\ \eqref{Ramsey-H-d}, and, third, Eq.\ \eqref{Ramsey-URUR-drop-U} holds, which is the case e.g.\ if the signal beam is a plane wave or the initial external state $\rho_\text{in}$ commutes with $H_g$.

\subsubsection*{2. Atomic-Ensemble Ramsey Spectroscopy}

We now turn to Ramsey spectroscopy performed on an ensemble of noninteracting, identical atoms. As the atoms do not interact, all equations valid for single-atom Ramsey spectroscopy remain valid. The only difference is that many quantities therein now pick up an index $i$ for the $i$th atom. In particular, Eq.\ \eqref{Ramsey-C} becomes
\begin{align}
\label{Ramsey-C-i}
C_i(t)
= \tr[\rho_{N,\text{in}} U_{d,i}^\dag(t)S_{r,i}U_{d,i}(t)S_{r,i}^\dag]
\end{align}
with the initial $N$-body density matrix $\rho_{N,\text{in}}$ in which all internal states are $|g\rangle$. Note, however, that our above assumptions leading to Eq.\ \eqref{Ramsey-U-p-approx} imply that $c_{a,i}$ and $c_{b,i}$ are independent of $i$.

The average number of atoms detected in internal state $|r\rangle$ is $\langle N_r\rangle= \sum_{i=1}^N P_{r,i}(t)$ with $P_{r,i}(t)$ from Eq.\ \eqref{Ramsey-P-r-C}. $\langle N_r\rangle$ exhibits a sinusoidal pattern with fringe visibility
\begin{align}
\label{Ramsey-V-i}
V(t)
= \left| \frac{\sum_{i=1}^N C_i(t)}{\sum_{i=1}^N C_i(0)} \right|
= \left| \frac{C_1(t)}{C_1(0)} \right|
,\end{align}
where for the last step, we used that the indistinguishability of the atoms implies that $C_i(t)$ is independent of $i$ for all $t$.

For the uncorrelated state of Eq.\ \eqref{rho-factorize}, $C_1(t)$ from Eq.\ \eqref{Ramsey-C-i} simplifies to
\begin{align}
C_i(t)
= \tr[\rho_\text{in} U_{g,i}^\dag(t)R_iU_{r,i}(t)R_i^\dag]
,\end{align}
which equals $C(t)$ from Eq.\ \eqref{C-URUR}. Hence, for an atomic ensemble in an uncorrelated state, we derived that $C_1(t)$ equals $C(t)$ from Eq.\ \eqref{C-URUR}.

The results \eqref{Ramsey-C-i} and \eqref{Ramsey-V-i} for Ramsey spectroscopy apply for arbitrary initial states if the signal beam is a plane wave. Alternatively, they apply if $\rho_{N,\text{in}}$ commutes with the dark-time Hamiltonian and the pulse area of the Ramsey pulses is small. Hence, we derived Eq.\ \eqref{V-C}. Our study of the correspondence to the efficiency in EIT-based storage and retrieval, however, is restricted to uncorrelated initial states or single-atom Ramsey spectroscopy. Extending the study of this correspondence to other initial states is beyond the present scope.

\subsection*{F. Relation to the Spatial Coherence Function}

Here, we establish the relation to the spatial first-order coherence function. We assume that the signal-light field is a plane wave and that the Hamiltonian after storage contains only kinetic energy. However, we assume neither that the initial atomic external states are product states nor that the atomic sample is homogeneous. For most of the calculation, we do not even assume that the system is thermalized before storage.

\subsubsection*{1. Retrieval Efficiency}

To put things into perspective, we first rewrite some of our previous results in terms of the operators
\begin{subequations}
\begin{align}
S_{e,i}^\dag
&
= R_{e,i}^\dag \otimes |e_i\rangle\langle g_i|
,&
\mathcal S_e
&
= \sum_{i=1}^N S_{e,i}
,\\
S_{c,i}^\dag
&
= R_{c,i}^\dag \otimes |r_i\rangle\langle e_i|
,&
\mathcal S_c
&
= \sum_{i=1}^N S_{c,i}
,\\
S_{r,i}^\dag
&
= R_i^\dag \otimes |r_i\rangle\langle g_i|
,&
\mathcal S_r
&
= \sum_{i=1}^N S_{r,i}
\end{align}
\label{spatial-S-def}%
\end{subequations}
where $R_{e,i}^\dag$ and $R_{c,i}^\dag$ act on the external state of the $i$th atom and have position representations
\begin{align}
R_{e,i}^\dag(\bm x)
= \sqrt{\mathcal V}u(\bm x)
,&&
R_{c,i}^\dag(\bm x)
= e^{i\bm k_c\cdot\bm x}
\end{align}
in analogy to $R_i^\dag$ in Eq.\ \eqref{R-def}. The operators $S_{e,i}$, $S_{c,i}$, and $S_{r,i}$ act on the external and internal state of the $i$th atom. The operators $\mathcal S_e$, $\mathcal S_c$, and $\mathcal S_r$ act on an $N$-atom state.

Comparison with Eq.\ \eqref{V-al-i} shows that we can write $\mathcal V_{al}$ in the representation-free form
\begin{align}
\label{spatial-V-al}
\mathcal V_{al}
= \hbar g \hat a_s \mathcal S_e^\dag + \frac\hbar2 \Omega_c \mathcal S_c^\dag +\Hc
\end{align}
Comparison with Eqs.\ \eqref{3d-subspace}, \eqref{Psi-r-n(t)}, and \eqref{Phi-n} yields
\begin{subequations}
\begin{align}
|\Psi_{e,n}(0)\rangle
&
= \frac1{\sqrt N}\mathcal S_e^\dag |\Psi_{g,n,\text{in}}\rangle
,\\
|\Psi_{r,n}(0)\rangle
&
= \frac1{\sqrt N}\mathcal S_r^\dag |\Psi_{g,n,\text{in}}\rangle
,\\
|\Psi_{r,n}(t)\rangle
&
= \frac1{\sqrt N}\mathcal U_d(t)\mathcal S_r^\dag |\Psi_{g,n,\text{in}}\rangle
,\\
|\Phi_{n}(t)\rangle
&
= \frac1{\sqrt N}\mathcal S_r^\dag \mathcal U_d(t)|\Psi_{g,n,\text{in}}\rangle
.\end{align}
\label{spatial-Psi-e-n-S}%
\end{subequations}
Here, applying the operator $\mathcal S_e^\dag$ ($\mathcal S_r^\dag$) to $|\Psi_{g,n,\text{in}}\rangle$ creates a spin wave, i.e.\ a Dicke state, with exactly one excitation in internal state $|e\rangle$ ($|r\rangle$). Hence, Eq.\ \eqref{eta-n-fidelity-Phi-Psi-r} for a separable pure initial state and $N\gg 1$ can be written as
\begin{align}
\label{spatial-eta-n-S}
\frac{\eta_n(t)}{\eta_0}
&
= \left| \frac1N \langle\Psi_{g,n,\text{in}}|\mathcal U_d^\dag(t)\mathcal S_r \mathcal U_d(t)\mathcal S_r^\dag |\Psi_{g,n,\text{in}}\rangle\right|^2
\notag \\ &
= \left| \frac1N \langle\Psi_{g,n,\text{in}}|\mathcal S_r(t) \mathcal S_r^\dag(0) |\Psi_{g,n,\text{in}}\rangle\right|^2
,\end{align}
where we wrote
\begin{align}
\label{spatial-S-r(t)}
\mathcal S_r(t)
= \mathcal U_d^\dag(t) \mathcal S_r \mathcal U_d(t)
\end{align}
for the operator in the Heisenberg picture. Whenever referring to operators in the Heisenberg picture, we include an explicit time argument. Next, we use that in the Heisenberg picture
\begin{align}
\label{spatial-S-r-i(t)}
S_{r,i}(t)
= U_{g,i}^\dag(t) R_i U_{r,i}(t) \otimes |g_i\rangle \langle r_i|
.\end{align}
Hence, Eq.\ \eqref{C-URUR} for an uncorrelated initial state can be written as
\begin{subequations}
\begin{align}
\label{spatial-C-S}
C(t)
&
= \tr[ \widetilde\rho_\text{in} S_r(t) S_r^\dag(0) ]
\\
\label{spatial-C-S-N}
&
= \frac1N \tr[ \rho_{N,\text{in}} \mathcal S_{r}(t) \mathcal S_{r}^\dag(0) ]
,\end{align}
\end{subequations}
where $\widetilde\rho_\text{in}= \rho_\text{in}\otimes |g\rangle\langle g|$ includes the internal state in the single-atom density matrix as in Eq.\ \eqref{rho-factorize} and we dropped the index $i$ from the single-particle operator $S_{r,i}$ because $S_{r,i}= S_{r,1}$. The factor $1/N$ appearing in Eq.\ \eqref{spatial-C-S-N} comes from the fact that according to Eq.\ \eqref{spatial-S-def} $\mathcal S_{r}(t)$ and $\mathcal S_{r}^\dag(0)$ each contain a sum over $i$ combined with the fact that in the resulting double sum the internal state makes all off-diagonal terms vanish and that all $N$ diagonal terms are identical because the particles are indistinguishable.

After rewriting these previous results, we now turn to the relation to the spatial coherence function. To define the spatial coherence function, we describe the atomic state in second quantization. The field operator $\hat\Psi_j(\bm x)$ annihilates an atom in internal state $j\in\{g,e,r\}$ at position $\bm x$ with commutation relations $[\hat\Psi_j(\bm x),\hat\Psi_{j'}^\dag(\bm x')]_\pm= \delta_{j,j'} \delta^{(3)}(\bm x-\bm x')$ and $[\hat\Psi_j(\bm x),\hat\Psi_{j'}(\bm x')]_\pm= 0$. The upper (lower) sign in the commutator is applicable if the atoms are fermions (bosons). In addition, $[\hat\Psi_j(\bm x),\hat a_s]= [\hat\Psi_j(\bm x),\hat a_s^\dag]= 0$. The corresponding atom-number operators are $\hat N_j= \int d^3x \linebreak[1] \hat\Psi_j^\dag(\bm x)\hat\Psi_j(\bm x)$. Operators in second quantization are represented by a hat.

The spatial first-order coherence function is defined as \cite{naraschewski:99}
\begin{align}
G^{(1)}(\bm x_1,\bm x_2)
= \langle \hat\Psi_g^\dag(\bm x_1)\hat\Psi_g(\bm x_2)\rangle
\end{align}
with positions $\bm x_1$ and $\bm x_2$. It is often useful to decompose these coordinates into the center-of-mass coordinate $\bm x'= (\bm x_1+\bm x_2)/2$ and relative coordinate $\bm x= \bm x_1-\bm x_2$. Integration over the center-of-mass coordinate $\bm x'$ yields \cite{naraschewski:99}
\begin{align}
\label{spatial-G1-x-x'}
G^{(1)}(\bm x)
= \int d^3x' \langle \hat\Psi_g^\dag(\bm x'+\tfrac12\bm x)\hat\Psi_g(\bm x'-\tfrac12\bm x)\rangle
.\end{align}
The normalized version thereof is \cite{naraschewski:99}
\begin{align}
\label{spatial-g1-def}
g^{(1)}(\bm x)
= \frac{G^{(1)}(\bm x)}{\int d^3x G^{(1)}(\bm x,\bm x)}
= \frac{G^{(1)}(\bm x)}{\langle \hat N_g\rangle}
.\end{align}

Next, we turn to general operators for noninteracting particles. We consider a single-particle operator $A$ and denote the version of it which acts on the $i$th atom as $A_i$. We assume that the $|u_k\rangle$ form an orthonormal basis of states. The general procedure for rewriting the $N$-atom operator $\mathcal A= \sum_i A_i$ in second quantization is $\hat A= \sum_{k,k'} \hat a_k^\dag \langle k|A_i|k'\rangle \hat a_{k'}$, where $\hat a_k^\dag$ creates an atom in state $|u_k\rangle$. We apply this to the $N$-atom operators $\mathcal S_e^\dag$, $\mathcal S_c^\dag$, and $\mathcal S_r^\dag$ of Eq.\ \eqref{spatial-S-def}. Here, the set of states $|\bm x,j\rangle$ with $j\in\{g,e,r\}$ and position $\bm x$ form an orthonormal basis and $\hat\Psi_j^\dag(\bm x)$ creates an atom in internal state $j$ at position $\bm x$. Hence, in second quantization the operators $\mathcal S_e^\dag$, $\mathcal S_c^\dag$, and $\mathcal S_r^\dag$ become
\begin{subequations}
\begin{align}
\hat S_e^\dag
= \int d^3x \hat\Psi_e^\dag(\bm x) e^{i\bm k_s\cdot\bm x} \hat\Psi_g(\bm x)
,\\
\hat S_c^\dag
= \int d^3x \hat\Psi_r^\dag(\bm x) e^{i\bm k_c\cdot\bm x} \hat\Psi_e(\bm x)
,\\
\hat S_r^\dag
= \int d^3x \hat\Psi_r^\dag(\bm x) e^{i\bm k_R\cdot\bm x} \hat\Psi_g(\bm x)
.\end{align}
\end{subequations}
Similarly, the atom-light interaction Hamiltonian $\mathcal V_{al}$ of Eq.\ \eqref{spatial-V-al} becomes
\begin{align}
\hat V_{al}
= \hbar g \hat a_s \hat S_e^\dag + \frac\hbar2 \Omega_c \hat S_c^\dag +\Hc
\end{align}

We now turn to the storage process. As mentioned above, the signal beam mode $u(\bm x)$ is assumed to be a plane wave. Hence, $v(\bm x)= e^{i\bm k_R\cdot\bm x}/\sqrt{\mathcal V}$ with $\bm k_R= \bm k_s+\bm k_c$. We consider an initial pure state of the form
\begin{align}
|\Theta_{g,n,\text{in}},1_s\rangle
= |1_s\rangle\otimes |\Theta_{g,n,\text{in}}\rangle
,\end{align}
where $|\Theta_{g,n,\text{in}}\rangle$ describes a state with exactly $N$ atoms, all initially in internal state $|g\rangle$, i.e.
\begin{subequations}
\begin{align}
\hat N_g|\Theta_{g,n,\text{in}}\rangle
&
= N|\Theta_{g,n,\text{in}}\rangle
,\\
\hat N_e|\Theta_{g,n,\text{in}}\rangle
&
= \hat N_r|\Theta_{g,n,\text{in}}\rangle
= 0
.\end{align}
\label{spatial-N-psi-gn}%
\end{subequations}
In analogy to Eq.\ \eqref{spatial-Psi-e-n-S}, we define states
\begin{subequations}
\begin{align}
|\Theta_{e,n}(0)\rangle
= \frac1{\sqrt N}\hat S_e^\dag(0) |\Theta_{g,n,\text{in}}\rangle
,\\
|\Theta_{r,n}(0)\rangle
= \frac1{\sqrt N}\hat S_r^\dag(0) |\Theta_{g,n,\text{in}}\rangle
.\end{align}
\end{subequations}
These are properly normalized because $u(\bm x)$ is a plane wave. If the initial $N$-atom state $|\Theta_{g,n,\text{in}}\rangle$ happens to be a product state, then the states $|\Theta_{g,n,\text{in}},1_s\rangle$, $|\Theta_{e,n}(0)\rangle$, and $|\Theta_{r,n}(0)\rangle$ simplify to the states in Eq.\ \eqref{3d-subspace}. It is straightforward to show that the set of orthonormal $N$-atom states $|1_s,\Theta_{g,n,\text{in}}\rangle$, $|\Theta_{e,n}(0)\rangle$, and $|\Theta_{r,n}(0)\rangle$ spans a 3d subspace, which is invariant under application of $\hat V_{al}$. Hence, we can again model the storage process simply as an adiabatic passage from $|1_s,\Theta_{g,n,\text{in}}\rangle$ to $|\Theta_{r,n}(0)\rangle$.

The dark-time Hamiltonian of Eqs.\ \eqref{H-d-i} and \eqref{H-kin} describing only kinetic energy becomes
\begin{align}
\hat H_d
= \sum_{j\in\{e,g,r\}} \int d^3x \hat\Psi_j^\dag(\bm x)\frac{-\hbar^2\nabla^2}{2m} \hat\Psi_j(\bm x)
.\end{align}
As the internal state $|e\rangle$ is never populated during the dark time, we might as well drop it from this sum. The resulting time-evolution operator is $\hat U_d(t)= e^{-i\hat H_dt/\hbar}$.

Using a calculation in the momentum representation, one finds that the spin-wave operator in the Heisenberg picture has the form
\begin{align}
\hat S_r^\dag(t)
&
= \hat U_d^\dag(t) \hat S_r^\dag(0) \hat U_d(t)
\notag \\ &
= e^{-i\omega_R t} \int d^3x e^{i\bm k_R\cdot\bm x} \hat\Psi_r^\dag(\bm x-\bm v_Rt)\hat\Psi_g(\bm x)
,\end{align}
where $\bm v_R= \hbar\bm k_R/m$ is the recoil velocity associated with $\bm k_R$ as in Eq.\ \eqref{tau-R-v-R} and $\omega_R= \hbar\bm k_R^2/2m$ is the corresponding recoil angular frequency. Based on this, one finds
\begin{multline}
\label{spatial-S-S}
\hat S_r(t)\hat S_r^\dag(0)|\Theta_{g,n,\text{in}}\rangle
\\
= e^{-i\omega_R t} \int d^3x \hat\Psi_g^\dag(\bm x+\tfrac12\bm v_Rt)\hat\Psi_g(\bm x-\tfrac12\bm v_Rt) |\Theta_{g,n,\text{in}}\rangle
.\end{multline}

When rewriting Eq.\ \eqref{spatial-eta-n-S} in second quantization, $\mathcal S_{r}(t)$ again becomes $\hat S_{r}(t)$. Hence, for a separable pure state
\begin{align}
\label{spatial-2nd-eta-n-S}
\frac{\eta_n(t)}{\eta_0}
= \left|\frac1N \langle \Theta_{g,n,\text{in}}|\hat S_r(t) \hat S_r^\dag(0)|\Theta_{g,n,\text{in}}\rangle\right|^2
.\end{align}
Combining this with Eqs.\ \eqref{spatial-G1-x-x'}, \eqref{spatial-g1-def}, \eqref{spatial-N-psi-gn}, and \eqref{spatial-S-S}, we obtain for a separable pure state
\begin{align}
\label{spatial-eta-g1-n}
\frac{\eta_n(t)}{\eta_0}
= |g^{(1)}_n(\bm v_Rt)|^2
,\end{align}
where the subscript $n$ in $g^{(1)}_n(\bm v_Rt)$ refers to taking the expectation values in $G^{(1)}$ and $\langle \hat N_g\rangle$ in Eqs.\ \eqref{spatial-G1-x-x'} and \eqref{spatial-g1-def} with respect to $|\Theta_{g,n,\text{in}}\rangle$.

Now, we turn to an uncorrelated state. Rewriting Eq.\ \eqref{spatial-C-S-N} in second quantization, $\mathcal S_{r}(t)$ again becomes $\hat S_{r}(t)$. Hence, for an uncorrelated state
\begin{align}
\label{spatial-C-g1}
C(t)
= \frac1N\tr\left( \rho_{N,\text{in}} \hat S_{r}(t) \hat S_{r}^\dag(0) \right)
= e^{-i\omega_R t}g^{(1)}(\bm v_Rt)
,\end{align}
where for the second step we used  Eqs.\ \eqref{spatial-G1-x-x'}, \eqref{spatial-g1-def}, \eqref{spatial-N-psi-gn}, and \eqref{spatial-S-S}. We insert this into Eq.\ \eqref{eta-C-plane-wave} and obtain for an uncorrelated state
\begin{align}
\label{spatial-eta-g1}
\frac{\eta(t)}{\eta_0}
= |g^{(1)}(\bm v_Rt)|^2
,\end{align}
where $g^{(1)}(\bm v_Rt)$ without a subscript refers to taking the expectation values in $G^{(1)}$ and $\langle \hat N_g\rangle$ in Eqs.\ \eqref{spatial-G1-x-x'} and \eqref{spatial-g1-def} with respect to $\rho_{N,\text{in}}$.

If the system is isotropic, e.g.\ because the atomic sample is homogeneous, then $|g^{(1)}(\bm x)|^2$ will depend only on $|\bm x|$. In that case, Eqs.\ \eqref{spatial-eta-g1-n} and \eqref{spatial-eta-g1} are derivations of Eq.\ \eqref{eta-g1}, one for separable pure states, the other for uncorrelated states.

For the separable initial state of Eq.\ \eqref{rho-N}, which might be mixed and might be correlated, we obtain $\eta(t)= \sum_n P_n \eta_n(t)$ according to Eq.\ \eqref{eta-eta-n}. Combination with Eq.\ \eqref{spatial-eta-g1-n} yields
\begin{align}
\label{spatial-eta-g1-sum}
\frac{\eta(t)}{\eta_0}
= \sum_n P_n |g^{(1)}_n(\bm v_Rt)|^2
.\end{align}
For comparison, we consider
\begin{align}
\label{spatial-eta-g1-simplified}
|g^{(1)}(\bm v_Rt)|^2
= \left| \sum_n P_n g^{(1)}_n(\bm v_Rt)\right|^2
.\end{align}
Interestingly, the righthand sides of Eqs.\ \eqref{spatial-eta-g1-sum} and \eqref{spatial-eta-g1-simplified} typically differ. This suggest that for separable initial states, which are neither pure nor uncorrelated, there might be deviations from Eq.\ \eqref{eta-g1}. However, a detailed study of such deviations is beyond the present scope.

\subsubsection*{2. Ramsey Spectroscopy}

To discuss the relation between Ramsey spectroscopy and the spatial coherence function, we start by similarly rewriting the previous results on atomic-ensemble Ramsey spectroscopy. In particular, using $C_i(t)$ from Eq.\ \eqref{Ramsey-C-i}, we obtain
\begin{align}
\sum_{i=1}^N C_i(t)
= \tr[\rho_{N,\text{in}} \mathcal S_r(t)\mathcal S_r^\dag(0)]
,\end{align}
where we used $\mathcal S_r$ and $\mathcal S_r(t)$ from Eqs.\ \eqref{spatial-S-def} and \eqref{spatial-S-r(t)} with $\mathcal U_d(t)= \exp(-i \mathcal H_d t/\hbar)$, $\mathcal H_d= \sum_{i=1}^N H_{d,i}$, and $H_{d,i}$ from Eq.\ \eqref{Ramsey-H-d}. In analogy to the transition from Eq.\ \eqref{spatial-C-S} to Eq.\ \eqref{spatial-C-S-N}, we used that each $\mathcal S_r$ contains a sum over $i$ and that the internal state makes all off-diagonal terms of the resulting double sum vanish.

As in Eq.\ \eqref{Ramsey-V-i}, we now use $C_i(t)= C_1(t)$ because the particles are indistinguishable. Hence, we can drop the index $i$ from $C_i(t)$. Thus, we obtain Eq.\ \eqref{spatial-C-S-N}, now for Ramsey spectroscopy. Note that this equation holds for Ramsey spectroscopy for arbitrary initial states, because in this section we assume that the signal beam is a plane wave. This is in contrast to our derivation of Eq.\ \eqref{spatial-C-S-N} for EIT-based retrieval, which holds only for uncorrelated states.

After rewriting the above results on Ramsey spectroscopy, we now discuss the connection to the spatial coherence function. To this end, we use second quantization, so that Eq.\ \eqref{spatial-C-S-N} becomes Eq.\ \eqref{spatial-C-g1}. As we assumed that $u(\bm x)$ is a plane wave, $C_i(0)= 1$ for all $i$ according to Eq.\ \eqref{C(0)=1-plane-wave}. Hence, Eq.\ \eqref{Ramsey-V-i} simplifies to $V(t)= |C(t)|$ and inserting Eq.\ \eqref{spatial-C-g1} yields for arbitrary initial external states
\begin{align}
V(t)
= |g^{(1)}_n(\bm v_Rt)|
.\end{align}

\end{document}